# Automating IETF Insights generation with AI

Jaime Jiménez

jaime.jimenez@ericsson.com

Ericsson

Jorvas, Finland

## ABSTRACT

This paper presents the *IETF Insights* project, an automated system that streamlines the generation of comprehensive reports on the activities of the Internet Engineering Task Force (IETF) Working Groups. The system collects, consolidates, and analyzes data from various IETF sources, including meeting minutes, participant lists, drafts and agendas. The core components of the system include data preprocessing code and a report generation module that produces high-quality documents in LaTeX or Markdown. By integrating large Language Models (LLMs) for summaries based on the data as ground truth, the *IETF Insights* project enhances the accessibility and utility of IETF records, providing a valuable overview of the IETF's activities and contributions to the community.

## 1 INTRODUCTION

The Internet Engineering Task Force (IETF), established in 1986, is the foremost standards body for the Internet, dedicated to producing influential technical documents. It operates on principles of openness, technical competence, and rough consensus, inviting participation from all interested parties. Working Groups (WGs) are central to the IETF's mission, serving as the primary means for developing Internet standards. These groups, guided by charters, focus on specific technical areas and make decisions through rough consensus rather than formal voting.

The IETF's primary output is a series of technical documents called Request for Comments (RFCs). These RFCs, often developed within WGs, describe the Internet's technical foundations, including addressing, routing, transport or security technologies. Example output of this process are protocol specifications like TLS [16], QUIC [9], or WebRTC [1]. This structure ensures that the IETF's standards reflect the collective expertise and real-world experience of the global Internet community.

During this process, the IETF produces extensive documentation, such as minutes, participant lists, drafts, and agendas, making it challenging to maintain an overview of activities. This paper introduces the *IETF Insights* project, an automated system that streamlines report generation on IETF WG activities. By utilizing advanced data processing and large language models (LLMs), the system consolidates and analyzes data, enhancing the accessibility and utility of IETF records. This project aims to provide a comprehensive overview of IETF activities, supporting transparency and informed decision-making [10].

The rest of the paper is structured as follows: the **Introduction** presents the IETF's mission and the *IETF Insights* project. The **Background** explores IETF public records and LLMs' role in data processing. The **Components** section outlines the project's architecture and workflow. The **Discussion** assesses LLM accuracy, performance, and integration challenges. The **Conclusion** highlights the project's contributions and enhanced utility of IETF records. An **Appendix** provides a sample generated report for IETF 119.

**Research Questions**

The *IETF Insights* project aims to automate the generation of broad and comprehensive reports on IETF working group activities using AI. To understand the effectiveness and impact of this automation, we pose the following research questions:

The *IETF Insights* project aims to automate the generation of broad and comprehensive reports on IETF working group activities using AI. To achieve this goal, we have established the following design objectives:

- **DO1** - Develop a system that can accurately generate structured data and summaries from IETF meetings, utilizing minutes, participant lists, drafts, and agendas as input sources.

- **DO2** - Analyze the performance of different sizes and types of LLMs (e.g., Ollama [14], OpenAI [15], Anthropic [2]) for processing IETF data, considering factors such as accuracy, speed, and resource utilization.

- **DO3** - After implementation, identify potential challenges and limitations of the IETF Insights project.

## 2 BACKGROUND
### 2.1 IETF Public Records

The IETF maintains an Open Records policy that makes a wide range of data publicly available. This data includes document-based information such as email messages and meeting minutes, structured data about Working Groups, and raw data like meeting attendance statistics [6].

The IETF Mail Archive Website [5] provides access to an extensive archive of all IETF mailing lists, which contain discussions, decisions, and historical context of various working groups. This archive offers a searchable interface for users.

For programmatic access to IETF-related data, the organization provides Application Programming Interfaces (APIs). The Datatracker API [3] allows retrieval of detailed information about documents, meetings, and working groups. It supports RESTful operations and returns data in JSON format.

The IETF utilizes GitHub for version control and collaborative development. The IETF's GitHub Organization [4] hosts repositories for various working groups and projects, facilitating global collaboration on drafts and issue management.

For file transfer and synchronization, the IETF supports the use of rsync. The rsync IETF Server [7] maintains backups of Internet-Drafts (I-Ds), RFCs, meeting materials, and other files within the IETF's FTP directory.

IETF meetings and sessions are recorded and made available on platforms such as YouTube, often accompanied by transcripts [8].

### 2.2 Large Language Models (LLMs)

Large Language Models (LLMs) have revolutionized Natural Language Processing (NLP) by enabling machines to *understand* and generate human-like text. These models employ mechanisms to capture long-range dependencies and contextual relationships in text. Pre-trained on vast corpora, LLMs can be fine-tuned for specific tasks, demonstrating versatility and efficiency. LLMs can be categorized based on their deployment method: those running locally and those accessed via APIs. LLMs can be further divided into large and small models based on their parameter size and computational requirements.

**Local Models** provide greater control and privacy, but often requires quantization to reduce size and computational demands. This quantization can lead to some limitations in performance and accuracy. Using Ollama, a platform designed for efficient deployment and management of machine learning models [14], we have tested various locally-run LLMs.

**Commercial Models** are accessed through APIs and are typically proprietary, meaning they are developed and maintained by specific organizations and are not fully open-source. For example, popular commercial models, such as GPT-4o and Claude 3.5 Sonnet, offer advanced capabilities and are optimized for performance and efficiency.

| Model | Parameters | Size | Category |
|---|---|---|---|
| codestral:latest | 22.2B | 12 GB | Large |
| llama3:70b-instruct | 70.6B | 39 GB | Large |
| command-r:latest | 35B | 20 GB | Large |
| mixtral:latest | 47B | 26 GB | Large |
| gemma2 | 9.2B | 5.4 GB | Small |
| phi3 | 3.8B | 2.4 GB | Small |
| llama3 | 8B | 4.7 GB | Small |

Table 1: Comparison of Local LLM Models

*GPT-4o* ("o" for "omni") is a state-of-the-art model that supports multimodal inputs, including text, audio, image, and video, and generates outputs in text, audio, and image formats. It is designed for rapid response times, with audio input processing as fast as 232 milliseconds.

*Claude 3.5 Sonnet* is part of the Claude model family and is known for its high intelligence and efficiency. Claude 3.5 Sonnet is available for free and boasts a large context window of 200K tokens.

## 3  COMPONENTS

### 3.1  System Architecture

The code has two main components, each composed of multiple Python scripts for preprocessing and report generation. These components can be broadly divided into data preparation and report generation.

**Data Preprocessing** This component includes scripts for retrieving, consolidating, and preparing IETF data. It extracts data from various sources, cleans it, and organizes it for further processing. The main task is to consolidate data from different sources. It uses `rsync` to synchronize data, it processes meeting attendance data using the `pandas` library for data manipulation and uses fuzzy matching with the `fuzzywuzzy` library to match participant names, maintaining data consistency even with slight variations in affiliation or names. The script uses vectorized operations to speed up data processing.

**Report Generation** This component is dedicated to generating detailed reports about the IETF working groups during a specific session. It takes the prepared data and produces comprehensive reports. The reports are compiled using LaTeX, allowing for quality document formatting. This component leverages LLMs for generating summaries from a vectorized database of IETF data. It uses the selected LLMs recursively on a per-WG basis, correlating the contents form the data sources. In addition to the plain documents in text, it also uses GraphRAG [12], a retrieval-augmented generation framework that combines graph-based data retrieval with language models, enhancing the quality and depth of the reports. The code integrates data from various sources and allows for easy extension and customization to accommodate new report formats, data sources or LLMs.

To improve the accuracy and relevance of the LLM outputs, we use custom `structured prompts`. These prompts guide the LLM to focus on specific aspects of the data, ensuring that the generated content is aligned with the requirements of the IETF reports. By providing clear and detailed instructions within the prompts, we can achieve more precise and contextually appropriate summaries and insights.

The system architecture of the project is visualized in the following PlantUML diagram:

### 3.2  Workflow

The typical workflow for generating an IETF compiled report involves:

(1) Updating the datastores and indexing with the latest IETF meeting data.

(2) Running the report generation code with appropriate parameters (meeting, output format, working groups, etc).

(3) Generating individual reports for each specified working group.

(4) Compiling the individual reports into a single LaTeX document if LaTeX output is selected.

(5) Producing the final comprehensive LaTeX document, which can be further processed into PDF format.

This implementation allows for flexible, automated generation of detailed IETF working group reports, leveraging both structured data analysis and advanced natural language processing capabilities.

## 4  DISCUSSION

We now address some of the objectives posed in the introduction based on the analysis conducted. Due to constraints in time and resources, we have not conducted a statistically significant evaluation for each LLM. Instead, our conclusions are drawn from our practical experiences as developers and users. The experiments and development were conducted on a MacBok with 64 GB RAM and an Apple M3 Max processor.

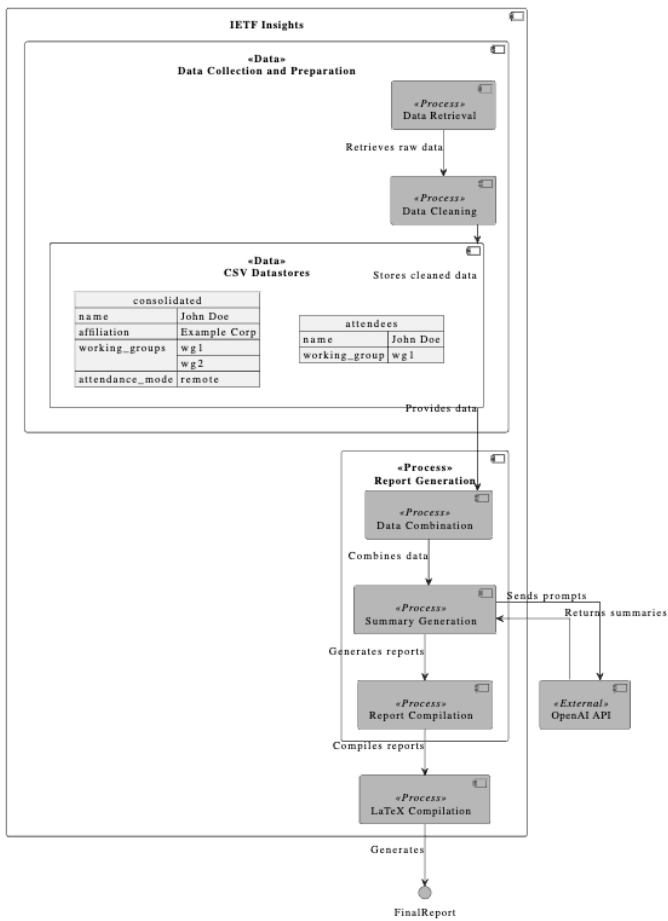

Figure 1: IETF Insights System Architecture

**DO1: How accurately can LLMs generate structured data and summaries from IETF meeting minutes, participant lists, and agendas?**

Our findings indicate that smaller LLMs often struggle with context size limitations and do not produce high-quality outputs. These models, while being freely available, lack the capacity to handle complex and extensive input data effectively. We tried modifying the default context size in Ollama but still there were quite a lot of hallucinations, chunking the text also meant that the context provided was not kept in between AI API interactions.

Larger models such as sonnet-3.5 from Anthropic and GPT-4o from OpenAI have demonstrated superior performance (see Appendix A as reference). These models can generate structured data and summaries with a high degree of accuracy by leveraging their extensive context size and advanced architecture. The structured prompting seems to help provide some common output pattern per WG, which helps readability and consistency.

For each working group, the LLM generates a structured summary that includes an overview of the main companies involved, the number of attendees, the key topics discussed, and the main decisions made. When mentioned in the minutes, the summaries also indicate the individuals presenting or discussing. The summaries are consistently formatted, facilitating comparison and analysis of different working groups' activities. Instances of hallucinations are minimal, and inline links to the drafts are provided.

That said, the output itself is not `insightful` as such, in the sense that it cannot make connections between the draft and the discussion or provide at the same time an `opinionated and correct` view of the work.

**DO2: What are the performance implications of using different sizes and types of LLMs for processing IETF data?**

The performance implications of using different LLMs are significant. Lightweight models like Phi3 [13] offer comparable processing times to API calls to larger models and require fewer computational resources, however they are simply not useful for processing large contexts. In contrast, larger models like Llama3 [11] provide more accurate and nuanced outputs but at the cost of increased processing time and resource consumption, while still not being on par with the larger commercial models.

One future approach could be that of delegating some simple subtasks to local LLMs and more complex ones to the API models, but this would require a more complex orchestration and data management, which we have not implemented.

**DO3: What are the challenges and limitations of integrating LLMs into the IETF Insights project, and how can they be mitigated?**

The main challenges are not necessarily related to the LLM integration, as API interfaces are very consistant and simple to use. The preprocessing of the attendance was challenging as affiliations and names were not consistant over time and had to be normalized. Another challenge was using the IETF transcripts, which did not work for non-native speakers, so much so that we discarded those as input data as they were not reliable and only added confusion to the LLM processing part.

From the LLM point of view, the main challenge was processing the data from multiple sources which was complicated. Ensuring data integrity and consistency is crucial for accurate report generation. Integration LLMs over APIs requires handling of API limits and error management to avoid disruptions in the report generation process. Additionally, compiling multiple LaTeX documents into a single comprehensive report requires management of document structure and formatting. These challenges can be mitigated by implementing robust data validation and error handling mechanisms, optimizing data processing pipelines, and using modular and extensible scripts to accommodate new data sources and report formats.

## 5  CONCLUSION

This project aims at showing the potential of leveraging data processing techniques and LLMs to automate the generation of comprehensive reports on IETF Working Group activities. By consolidating and analyzing vast amounts of data from various IETF sources, the system enhances the accessibility and utility of IETF records, providing valuable insights into the organization's activities and contributions.

Our findings indicate that smaller LLMs struggle with context limitations and are not reliable enough while larger models like GPT-4o and Claude 3.5 Sonnet offer sufficient performance in generating accurate and structured summaries. The primary challenges lied in preprocessing inconsistent data and managing the integration of multiple data sources.

## 6  ACKNOWLEDGMENTS

We'd like to thank Ericsson for their support of this work. We also appreciate the valuable discussions and insights provided by Péter Mátray, Jari Arkko, Carsten Bormann, Patrik Salmela, Marco Tiloca, Lorenzo Corneo and Ignacio Castro.

# A    APPENDIX A

# IETF119 Meeting Report

Generated by IETF Reporter, by Jaime Jiménez

26th of March 2024

# Contents

# 1 6lo Working Group (6lo)

## 1.1 Attendees Overview

### 1.1.1 Representation and Attendance

The meeting was attended by 22 participants, representing key institutions such as Apple, Cisco, Huawei Technologies, and ETRI. Notable attendees included Stuart Cheshire from Apple and Éric Vyncke from Cisco.

The discussions were rich and centered around the latest advancements in IPv6 over Low power Wireless Personal Area Networks (6LoWPAN). The group reviewed several Internet Engineering Task Force (IETF) draft documents, providing a platform for dialogue on technical specifications and future work. References to these drafts were made available through hyperlinks, such as [draft-ietf-6lo-path-aware-semantic-addressing-04](draft-ietf-6lo-path-aware-semantic-addressing-04).

Meeting materials can be accessed directly via the [IETF 119 Materials](IETF 119 Materials) page.

## 1.2 Meeting Discussions

### 1.2.1 Path-Aware Semantic Addressing for LLNs

Luigi Iannone presented the latest on Path-Aware Semantic Addressing for Low-Power and Lossy Networks (LLNs), emphasizing the need for reviews from both the 6lo group and other areas before proceeding to Working Group Last Call (WGLC). The document in question can be found at [draft-ietf-6lo-path-aware-semantic-addressing-04](draft-ietf-6lo-path-aware-semantic-addressing-04).

### 1.2.2 Generic Address Assignment Option for 6LoWPAN ND

The discussion on the Generic Address Assignment Option for 6LoWPAN Neighbor Discovery (ND) highlighted the intention to update the document and continue discussions on the mailing list. Luigi Iannone led the presentation, with the draft available at [draft-iannone-6lo-nd-gaao-02](draft-iannone-6lo-nd-gaao-02).

### 1.2.3 Reliability Considerations of PASA

The reliability aspects of Path-Aware Semantic Addressing (PASA) were also presented by Luigi Iannone. It was suggested that the document might have fulfilled its purpose and could remain as is without progressing to RFC publication. The draft is located at [draft-li-6lo-pasa-reliability-03](draft-li-6lo-pasa-reliability-03).

### 1.2.4 Transmission of SCHC-Compressed Packets over IEEE 802.15.4

Carles Gomez presented on the transmission of Static Context Header Compression (SCHC)-compressed packets over IEEE 802.15.4 networks. The draft detailing this topic is found at [draft-ietf-6lo-schc-15dot4-05](draft-ietf-6lo-schc-15dot4-05).

### 1.2.5 Transmission of IPv6 Packets over Short-Range OWC

Younghwan Choi discussed the transmission of IPv6 packets over short-range Optical Wireless Communications (OWC). The draft for this topic is available at [draft-choi-6lo-owc-02](draft-choi-6lo-owc-02). A call for working group adoption of this draft was supported by a show of hands, with a follow-up review call to be initiated on the 6lo mailing list.

The meeting concluded with a consensus on the importance of these discussions for the evolution of 6LoWPAN technologies and their applications. The next steps include further reviews, updates to drafts, and continued collaboration on the mailing list to refine the technical direction and strategy of the working group's efforts.

# 2 IPv6 Maintenance Working Group (6MAN)

## 2.1 Attendees Overview

### 2.1.1 Prominent Attendees and Total Attendance

The 6MAN Working Group meeting saw participation from key players in the industry, including representatives from Google, Apple, Cisco, Huawei, and APNIC. The meeting was attended by a total of 105 individuals, indicating a strong interest in the ongoing development and maintenance of IPv6.

### 2.1.2 Summary of Discussions

The meeting discussions revolved around several draft documents and proposals aimed at improving IPv6 functionality and deployment. Key topics included the handling of Network Resource Partition (NRP) information, signaling DHCPv6 Prefix Delegation availability, and the deprecation of the IPv6 Router Alert Option. Attendees engaged in technical debates that contextualized the need for these changes within the broader scope of IPv6 deployment challenges. For instance, the [draft-ietf-6man-enhanced-vpn-vtn-id](draft-ietf-6man-enhanced-vpn-vtn-id) sparked discussions on the complexity of data plane operations and the necessity for simplicity in extension headers.

Meeting materials can be accessed directly via the [IETF 119 Materials](IETF 119 Materials) page.

## 2.2 Meeting Discussions

### 2.2.1 Carrying NRP Information in IPv6 Extension Header

The presentation on carrying NRP information in IPv6 Extension Header highlighted the readiness for a working group last call and a request for IANA code point early allocation. The discussion emphasized the need for simplicity in the extension header to ensure successful hardware processing at each node along the path.

### 2.2.2 Signaling DHCPv6 Prefix Delegation Availability

The debate on DHCPv6 Prefix Delegation availability focused on the potential inconsistencies between multiple Prefix Information Options (PIOs) and the proposal of using an RA flag instead of a P flag. The discussion concluded with an agreement to add more text to the draft to clarify these concerns.

### 2.2.3 Deprecation of IPv6 Router Alert Option

The session on deprecating the IPv6 Router Alert Option was skipped due to the presenter's absence.

### 2.2.4 Entering IPv6 Zone Identifiers into User Interfaces

The conversation on IPv6 zone identifiers in user interfaces revealed a divergence of opinions on the necessity and implementation of the proposed standard. The discussion highlighted the need for further clarification of terminology and use cases.

### 2.2.5 Stub Router Flag in ICMPv6 Router Advertisement Messages

The Stub Router Flag presentation led to a consensus on the importance of adhering to RFC4861 guidelines and the need for a "best effort" approach when dealing with configuration changes that could affect router advertisements.

### 2.2.6 Triggering Unsolicited Router Advertisements Upon Configuration Changes

A request for working group adoption was made for the draft concerning unsolicited router advertisements upon configuration changes. The ensuing discussion raised questions about enterprise router behavior and the potential need to adjust the interval between advertisements.

### 2.2.7 RFC 6296bis IPv6-to-IPv6 Network Prefix Translation

The RFC 6296bis update prompted a lively debate on the merits and drawbacks of IPv6-to-IPv6 Network Prefix Translation (NPTv6). Participants discussed whether to adopt the draft as informational, with opinions ranging from support for publication to concerns about the potential negative impact on applications and users.

### 2.2.8 Considerations for DetNet Extensions for IPv6

The DetNet extensions presentation was met with clarification that there is no consensus beyond existing QoS marking, and that network partitioning discussions are outside the scope of the DetNet working group.

# 3 Authentication and Authorization for Constrained Environments (ACE)

## 3.1 Attendees Overview

### 3.1.1 Attendance Summary

The meeting was attended by 15 participants representing various organizations, including Ericsson, Inria, RISE Research Institutes of Sweden, ACS, Aiven, Nokia, SECOM, cyberstorm.mu, and others. The diversity of attendees from prominent companies and institutions highlights the broad interest and collaborative efforts within the ACE working group.

### 3.1.2 Meeting Summary

The ACE working group meeting focused on discussing several Internet Engineering Task Force (IETF) draft documents, addressing comments and feedback, and outlining the next steps for each draft. The discussions aimed to refine the drafts and move them closer to completion, with particular attention to addressing comments from the last working group calls and ensuring alignment with RFC 9200 requirements. The meeting materials can be accessed via the following link: Meeting Materials.

## 3.2 Meeting Discussions

### 3.2.1 Message from Chairs

The meeting began with an introduction by Tim Hollebeek and a brief discussion on the agenda, including a rearrangement to move the presentation on the pub-sub profile to the end of the session.

### 3.2.2 draft-ietf-ace-oscore-gm-admin

Marco Tiloca presented the outcomes of the working group last call for the draft-ietf-ace-oscore-gm-admin, highlighting the resolution of comments from various reviewers. The next step is to address remaining comments and prepare for submission to the IESG. The presentation slides are available at draft-ace-oscore-gm-admin slides.

### 3.2.3 draft-ietf-ace-workflow-and-params

Marco Tiloca also presented updates to the draft-ietf-ace-workflow-and-params, including changes to the formulation of two profile requirements from RFC 9200. A notable discussion point was the token upload workflow and its implications for clients and resource servers. The issue was taken up for further discussion on the mailing list. The presentation slides can be found at draft-ace-workflow-and-params slides.

### 3.2.4 draft-ietf-ace-group-oscore-profile

Rikard Höglund presented updates to the draft-ietf-ace-group-oscore-profile, emphasizing the separation of access control within OSCORE groups and adherence to NIST zero-trust principles. The updates included editorial improvements and clarifications on the processing of proof-of-possession evidence. Future work includes considering an access token covering multiple OSCORE groups and aligning with RFC 9200 profile requirements. The presentation slides are available at draft-group-oscore-profile slides.

### 3.2.5 draft-ietf-ace-est-oscore

Mališa Vučinić presented the latest updates to the draft-ietf-ace-est-oscore, addressing reviews and closing related issues. Open issues were discussed, including the support of Content-Formats and the use of static DH keys for one-time signing. The draft aims to address the remaining open issues and solicit reviews before proceeding to the working group last call. The presentation slides can be accessed at draft-est-oscore slides.

### 3.2.6 draft-ace-edhoc-oscore-profile

Göran Selander presented the draft-ace-edhoc-oscore-profile, focusing on the workflow and the conveyance of the access token within the EDHOC protocol. The discussion included the use of unique identifiers for credentials and the potential use of the EDHOC Reverse message flow. The presentation slides are available at [draft-edhoc-oscore-profile slides](draft-edhoc-oscore-profile slides).

### 3.2.7 draft-ietf-ace-pubsub-profile

Due to time constraints, the draft-ietf-ace-pubsub-profile was not presented. Further discussions are expected to take place on the mailing list.

# 4 Automated Certificate Management Environment (ACME)

## 4.1 Attendees Overview

### 4.1.1 Attendance Summary

The meeting was attended by representatives from prominent companies and institutions such as Verisign, DigiCert, Let's Encrypt, Google, Microsoft, and Akamai, with a total attendance of 65 participants.

The discussions during the meeting focused on the current status and future directions of various ACME draft documents. The dialogue was rich with technical insights and highlighted the collaborative efforts of the community to enhance the ACME protocol. Key draft documents such as draft-ietf-acme-dtnnodeid and draft-ietf-acme-ari were discussed in detail, with the potential to significantly impact the way certificates are managed and renewed.

Meeting materials are available via the direct link IETF 119 Materials.

## 4.2 Meeting Discussions

### 4.2.1 Document Status Update

The chairs provided an update on the status of various documents, noting the expiration of ACME-Client and the readiness of draft-ietf-acme-dtnnodeid for Working Group Last Call (WGLC).

### 4.2.2 DTN NodeId Validation

Brian Sipos presented the latest updates to the draft-ietf-acme-dtnnodeid, indicating its readiness for WGLC, which suggests a significant step towards standardization.

### 4.2.3 ACME Renewal Information (ARI)

Aaron Gable discussed the draft-ietf-acme-ari and the open questions regarding the handling of certificate replacement requests, which could lead to a more streamlined renewal process upon resolution.

### 4.2.4 Device Attestation

Brandon Weeks highlighted the progress on draft-acme-device-attest, including an upcoming open-source implementation by Google, indicating a growing interest in device attestation within the ACME protocol.

### 4.2.5 ACME-Onion

Q Misell called for additional reviews of draft-ietf-acme-onion following a quiet WGLC, emphasizing the importance of community feedback for the advancement of the document.

### 4.2.6 ACME-based Provisioning of IoT Devices

Mike Sweet raised the question of whether draft-sweet-iot-acme's focus on IoT device provisioning aligns with ACME's scope, sparking a discussion on the appropriate venue for this work.

### 4.2.7 ACME Auto Discovery

Mike Ounsworth reported on the progress made by the design team formed in Prague, resulting in two drafts ready for community consideration, with draft-vanbrouwershaven-acme-auto-discovery being particularly ripe for an adoption call.

### 4.2.8 Any Other Business

Aaron Gable encouraged the review of the updated dns-account-01 draft, highlighting its significance in the context of DNS-based ACME account management.

# 5 Adaptive DNS Discovery (ADD)

## 5.1 Attendees Overview

### 5.1.1 Participation

The meeting was attended by representatives from prominent companies and institutions such as Salesforce, Meta Platforms, Inc., Comcast-NBCUniversal, Cisco, Apple, Microsoft, and Verisign, with a total attendance of 72 participants.

The discussions were rich and covered a range of topics from encrypted DNS server redirection to architectural directions for the group. The dialogue was particularly focused on the challenges of deploying encrypted DNS on Customer Premises Equipment (CPEs) and the associated certificate management issues. The conversation highlighted the need for a solution that could handle opportunistic encryption and authenticated discovery in a scalable and efficient manner. For more details on the discussions, refer to the draft-ietf-add-ddr.

Meeting materials are available at the ADD session materials page.

## 5.2 Meeting Discussions

### 5.2.1 Encrypted DNS Server Redirection

Tommy Jensen presented on the topic of encrypted DNS server redirection, emphasizing the need for a redirection mechanism that maintains protocol parameters continuity for the client. The discussion underscored the importance of performance optimizations and the challenges when transitioning from anycast to unicast services.

### 5.2.2 Architectural Directions

Tiru Redda revisited the morning's presentation, sparking a debate on whether the scope of the problem should be broadened beyond the working group. The consensus was that the problem statement needed to be refined and documented, potentially leading to the formation of a new working group. The session concluded with a call for volunteers to draft a problem statement that captures the essence of the challenges faced.

### 5.2.3 Next Steps

The group agreed on the importance of formulating a clear problem statement that includes the limitations of current solutions and why they are not viable. The chairs will initiate a discussion on the mailing list to encourage the creation of a problem statement before the next IETF meeting, reflecting the collective understanding and direction of the working group.

# 6 IETF-Wide "Dispatch" Session

## 6.1 Attendees Overview

### 6.1.1 Prominent Attendees and Total Attendance

The meeting was attended by representatives from prominent companies and institutions such as Google, Cisco Systems, Microsoft, and Huawei, with a total attendance of 370 participants.

### 6.1.2 Summary of Discussions

The discussions at the meeting covered a wide range of topics, from advancements in SSH protocols to new proposals for handling misdirected emails. The dialogue was rich with technical insights and strategic considerations, particularly with regard to the adoption of new standards and the evolution of existing ones. Notable discussions included the proposal for SSH3, which garnered interest for collaboration and raised questions about its implementation over HTTP. The Transport Layer Security (TLS) Authentication with Verifiable Credential (VC) presentation sparked a debate on the readiness of such a specification for IETF consideration, suggesting that more discussions with the TLS working group or a potential handover to W3C might be appropriate.

Meeting materials are available directly via the [IETF 119 Materials](#) page.

## 6.2 Meeting Discussions

### 6.2.1 Towards SSH3

The proposal for SSH3 was met with enthusiasm and suggestions for a Birds of a Feather (BoF) session to further explore the concept. There was a consensus that the work fits within the IETF's remit but should be reviewed in light of similar ongoing efforts.

### 6.2.2 TLS Authentication with VC

The TLS Authentication with VC presentation led to a discussion about whether the work should be pursued within the IETF or be better suited for W3C. The need for more discussions within the TLS working group was highlighted to determine the appropriate course of action.

### 6.2.3 Happy Eyeballs Version 3

The Happy Eyeballs Version 3 presentation suggested the formation of a new working group, with expertise from both the Transport and Security areas. The potential re-chartering of the TAPS working group was also considered.

### 6.2.4 Wrong-Recipient URL for Handling Misdirected Emails

The proposal for handling misdirected emails through a Wrong-Recipient URL was directed to the MAILMAINT working group for further discussion and consideration.

### 6.2.5 Human Rights Privacy through Deterministic Hashed Based Elision

The discussion on Human Rights Privacy through Deterministic Hashed Based Elision concluded that the work might be more appropriate for the IRTF. It was recommended that the work be broken down into more specific pieces for further feedback.

### 6.2.6 WebSocket Extension to Disable Masking

The WebSocket Extension to disable masking was advised to gather more feedback from the HTTPBIS working group. The necessity and motivations for this work require further elaboration.

### 6.2.7 Extended YANG Data Model for DOTS

An Extended YANG Data Model for DOTS was proposed to be discussed in a BoF session, including related topics, to gather more feedback throughout the week.

### 6.2.8 Unicode Character Repertoire Subsets

The discussion on Unicode Character Repertoire Subsets leaned towards using PRECIS profiles and potentially documenting the work as an ART AD-sponsored document.

### 6.2.9 Modern Network Unicode

The Modern Network Unicode presentation led to a decision to continue discussions on the ART mailing list, with caution advised regarding overlaps with existing work.

# 7 Advanced Network Management and Automation (ANIMA)

## 7.1 Attendees Overview

### 7.1.1 Prominent Companies and Total Attendance

The ANIMA working group meeting was attended by representatives from prominent companies and institutions such as China Unicom, Siemens, Meetecho, National Institute of Information and Communications Technology, and Cisco, with a total of 36 attendees.

The main points and dialogues of the meeting revolved around the progress and updates on various Internet Engineering Task Force (IETF) draft documents. The discussions were rich in technical content and focused on the evolution of protocols and standards within the ANIMA scope. For instance, the updates on BRSKI-AE and BRSKI-PRM draft documents were particularly noteworthy, as they highlighted the ongoing efforts to refine bootstrap and resource management protocols. Attendees engaged in a constructive exchange of ideas, emphasizing the need for interoperability and security in automated network management. The relevance of these discussions is underscored by the embedded hyperlinks to the IETF draft documents, such as draft-ietf-anima-brski-ae and draft-ietf-anima-brski-prm.

Meeting materials and notes can be accessed directly via IETF ANIMA WG meeting notes.

## 7.2 Meeting Discussions

### 7.2.1 BRSKI-AE Update

The BRSKI-AE draft has been updated to version 10, as presented by David von Oheimb. The discussion highlighted the importance of this draft in enhancing the security of automated enrollment processes.

### 7.2.2 BRSKI-PRM Update

Steffen Fries presented the latest version 12 of the BRSKI-PRM draft. The conversation focused on new use-cases such as transport over USB stick, and the potential impact on the document.

### 7.2.3 RFC8366bis Update

Michael Richardson and Toerless Eckert discussed the challenges with the YANG toolchain for validating YANG to CBOR/JSON conversions. The possibility of switching to CDDL specification was debated, with an emphasis on maintaining backward compatibility.

### 7.2.4 Lightweight GRASP

Sheng Jiang introduced the concept of Lightweight GRASP, which sparked a debate on the necessity and potential benefits of this approach over existing protocols.

### 7.2.5 Constrained ACP

Toerless Eckert discussed the idea of a constrained Autonomic Control Plane (ACP), which would cater to environments with limited resources.

### 7.2.6 Conclusion and Next Steps

The meeting concluded with a consensus on the need to maintain momentum on the current drafts while considering the introduction of new work items like Lightweight GRASP and constrained ACP. The working group acknowledged the potential shifts in technical direction and strategy, and outlined the next steps to further contribute to the field of network management and automation.

# 8 Audio/Video Transport Core Maintenance (AVTCORE)

## 8.1 Attendees Overview

### 8.1.1 Prominent Companies and Total Attendance

The AVTCORE working group meeting was attended by representatives from prominent companies and institutions, including Microsoft, Google, Cisco, and Nokia, with a total of 37 attendees.

### 8.1.2 Main Points and Dialogues

The discussions at the AVTCORE working group meeting centered around several key draft documents, including advancements in RTP payload formats and transport mechanisms. Notably, the [draft-ietf-avtcore-rtp-over-quic](draft-ietf-avtcore-rtp-over-quic) was a focal point, addressing the potential of RTP over QUIC protocol. The meeting also highlighted the [draft-ietf-avtcore-rtp-v3c](draft-ietf-avtcore-rtp-v3c) for volumetric video coding and the [draft-ietf-avtcore-rtp-sframe](draft-ietf-avtcore-rtp-sframe) for secure frame encryption. These discussions contextualized the technical direction of the working group and set the stage for future actions.

Meeting materials are available through the direct link: [AVTCORE Meeting Slides](AVTCORE Meeting Slides).

## 8.2 Meeting Discussions

### 8.2.1 Galois Counter Mode with Secure Short Tags (GCM-SST)

The group reviewed the [draft-mattsson-cfrg-aes-gcm-ss](draft-mattsson-cfrg-aes-gcm-ss), which addresses the security and performance of AES-GCM with short tags. The discussion concluded with a consensus on the interest in this work, pending further review by the CFRG.

### 8.2.2 RTP Payload Format for Visual Volumetric Video-based Coding (V3C)

Updates on the [draft-ietf-avtcore-rtp-v3c](draft-ietf-avtcore-rtp-v3c) were presented, including resolutions to several issues. The group considered simplifications based on SRST usage and the relevance of the Dependency Descriptor.

### 8.2.3 RTP over QUIC

The [draft-ietf-avtcore-rtp-over-quic](draft-ietf-avtcore-rtp-over-quic) sparked debates on experimental status, coalescing RTP packets, and the future of multiplexing protocols. The group agreed that while there are open questions, it is time to move towards a working group last call.

### 8.2.4 HEVC Profile for WebRTC

The [draft-ietf-avtcore-hevc-webrtc](draft-ietf-avtcore-hevc-webrtc) raised questions about codec negotiation in WebRTC, specifically regarding send/recv m-lines and browser support. The group encouraged further discussion on GitHub.

### 8.2.5 RTP Payload Format for SFrame

Concerns were raised about the need for implementation experience before progressing the [draft-ietf-avtcore-rtp-sframe](draft-ietf-avtcore-rtp-sframe), given its security implications.

### 8.2.6 RTP Payload for Haptics

The group discussed the adoption of the [draft-hsyang-avtcore-rtp-haptics](draft-hsyang-avtcore-rtp-haptics) and agreed to proceed with calling for adoption.

### 8.2.7 Next Steps

The meeting concluded with a plan for document shepherding and working group last calls on key drafts, reflecting the group's commitment to advancing the field of audio and video transport.

# 9 Working Group Update [WG]

## 9.1 Attendees Overview

### 9.1.1 Attendance Summary

The meeting was attended by representatives from prominent companies and institutions such as Cisco Systems, Nokia, Arista, Huawei, Juniper Networks, and APNIC, with a total of 71 attendees.

### 9.1.2 Main Points and Discussions

The discussions were rich and covered a range of topics from updates on working group activities to detailed technical debates on specific Internet Engineering Task Force (IETF) draft documents. The dialogue contextualized the importance of each draft in the broader scope of network engineering and interoperability, with references to drafts such as draft-ietf-[wg]-mpls-1stnibble-04 and draft-ietf-[wg]-evpn-ip-aliasing-01. The meeting materials can be accessed directly via meeting materials.

## 9.2 Meeting Discussions

### 9.2.1 Working Group Update

The chairs provided an update on the working group's progress, highlighting new RFCs, documents under review, and those ready for Working Group Last Call (WGLC). The discussions suggested a need for a stronger normative language in some documents and a call for more feedback before proceeding to WGLC.

### 9.2.2 MPLS First Nibble

Greg Mirsky presented updates to the MPLS first nibble draft, emphasizing the need for clear terminology and the deprecation of non-IP payloads without a control word (CW) to improve load balancing.

### 9.2.3 EVPN IP Aliasing

Jorge Rabadan discussed the EVPN IP aliasing draft, which addresses concerns for use case 3 and seeks more feedback for potential WGLC.

### 9.2.4 EVPN Inter-Domain Option B

Jorge Rabadan requested working group adoption for a draft that provides solutions for EVPN inter-domain option B without changing the control plane.

### 9.2.5 EVPN Multicast EEG

Jorge Rabadan solicited more feedback and working group adoption for a draft addressing intra and inter subnet multicast forwarding across EVPN domains.

### 9.2.6 BGP MVPN Source Active Route

Jeffery presented a solution for S,G only mode in MVPN and raised the question of whether to progress this document separately or incorporate it into RFC6514bis.

### 9.2.7 RFC6514bis Consideration

Jeffery shared thoughts on consolidating RFC6514 and its extensions into a single document, highlighting the need for a comprehensive review and potential simplification.

### 9.2.8 EVPN First-Hop Security

Krishnaswamy presented updates to the EVPN first-hop security draft, which is implemented in Cisco Nexus devices, and called for working group adoption.

### 9.2.9 IPSec over SRv6

Linda discussed a draft for IPSec over SRv6, proposing a new tunnel type and TLVs, and suggested integrating this work with the secure EVPN draft.

### 9.2.10 EVPN Group Policy

Ali Sajassi presented a merged draft covering both data plane and control plane aspects of EVPN group policy, implemented by multiple vendors, and sought working group adoption.

### 9.2.11 Centralized Anycast Gateway

Krishna, on behalf of Neeraj, discussed the centralized anycast gateway draft, which now references 7432bis and includes new requirements such as ARP/ND snooping on access L2 PEs, requesting working group adoption.

### 9.2.12 EVPN L3 Optimized IRB

Chuanfa presented the EVPN L3 optimized IRB draft, explaining two options for implementation and emphasizing the preferred option to avoid packet loss. The draft calls for more input and a working group call.

# 10 Bit Indexed Explicit Replication (BIER)

## 10.1 Attendees Overview

### 10.1.1 Participation Summary

The meeting was attended by representatives from prominent companies and institutions such as Deutsche Telekom, Juniper, Cisco, Nokia, ZTE Corporation, Huawei, and China Mobile, with a total attendance of 32 participants.

The session focused on the latest developments and draft discussions within the BIER working group. Key topics included advancements in BGP-LS extensions, source protection mechanisms, and implementation experiences with P4 Tofino. The discussions were enriched by the presence of industry experts, which provided a practical perspective on the theoretical constructs. Draft documents such as [draft-ietf-bier-bgp-ls-bier-ext-16](draft-ietf-bier-bgp-ls-bier-ext-16) and [draft-ietf-bier-source-protection-05](draft-ietf-bier-source-protection-05) were central to the dialogue, indicating the group's ongoing efforts to refine BIER specifications.

Meeting materials are available through the direct link: [BIER Session Slides](BIER Session Slides).

## 10.2 Meeting Discussions

### 10.2.1 BGP-LS Extensions for BIER

Ran Chen presented updates on the BGP-LS extensions for BIER, emphasizing the readiness of the document for further progression. The discussion highlighted the longevity of the work and the need for cross-working group review, particularly with the LSR group.

### 10.2.2 Source Protection for BIER

Sandy Zhang discussed the informational draft on source protection, which aligns with the three modes from RFC9026. The use of MVPN as a BIER overlay was a focal point, with a call for Working Group Last Call (WGLC). The conversation also touched on the need for synchronization with OAM BIER documents and the potential impact of the DF election draft.

### 10.2.3 P4 Tofino Implementation Experiences

Steffen Lindner shared experiences with P4 Tofino implementation, showcasing advanced stateless multicast source routing. The working group expressed interest in documenting this work in a draft, recognizing the potential contributions to the field.

### 10.2.4 EANTC Achievements

Hooman Bidgoli reported on the successful EANTC testing of BIER signaling with ISIS and BIER packet forwarding. The tests involved major industry players and set the stage for future work on NG-MVPN overlays and a live demo at the Paris MPLS Congress 2024.

### 10.2.5 Anycast Label for BIER

Siyu Chen introduced a draft on anycast label for BIER, aiming to minimize the impact on IGP state. The discussion revolved around understanding the specific problem being addressed and considering SDWAN use cases.

### 10.2.6 BIER In-situ OAM

Xiao Min presented options for encapsulating In-situ OAM (IOAM) over BIER. The proposal to treat IOAM headers as BIER extension headers sparked a debate on the need for WG adoption of related drafts and the value of documenting WG history through use-case drafts and problem statements.

# 11 Benchmarking Methodology Working Group (BMWG)

## 11.1 Attendee Overview

### 11.1.1 Participation Summary

The BMWG session at IETF 119 in Brisbane saw participation from key industry players including Huawei Technologies, Cisco, Telefonica Innovacion Digital, and Apple Inc., with a total of 19 attendees.

The discussions focused on the latest developments in benchmarking methodologies, with particular attention to containerized infrastructures, network tester management, and segment routing. The integration of YANG models for network tester configuration and the implications of multiple IP address usage in benchmarking tests were also key points of interest. The session provided valuable insights into the evolving landscape of network performance benchmarking, as represented by the diverse institutional affiliations of the participants.

Meeting materials are available via the direct link: IETF 119 BMWG Slides.

## 11.2 Meeting Discussions

### 11.2.1 Multiple Loss Ratio Search and Test Results

Maciek Konstantynowicz and Vratko Polak presented the draft-ietf-bmwg-mlrsearch, which proposes enhancements to the RFC 2544 search methodology, tailored for software platforms. The readiness of the draft for Working Group Last Call was affirmed, pending further review.

### 11.2.2 YANG Data Model for Network Tester Management

Vladimir Vassilev discussed the draft-ietf-bmwg-network-tester-cfg, highlighting the Hackathon results and the flexibility of the YANG model to accommodate future augmentations. The model's compatibility with various server types and the potential for detailed frame generation were noted.

### 11.2.3 Benchmarking in Containerized Infrastructures

Minh-Ngoc Tran presented the draft-ietf-bmwg-containerized-infra, which addresses the benchmarking of network performance in containerized environments. The draft includes considerations for kernel and user space acceleration, eBPF, and Smart-NIC acceleration.

### 11.2.4 Multiple IP Addresses in Benchmarking Tests

Gabor Lencse introduced the draft-lencse-bmwg-multiple-ip-addresses, emphasizing the significance of address and port randomization based on empirical measurements.

### 11.2.5 Segment Routing Benchmarking Methodology

Luis M. Contreras presented the merged draft-vfv-bmwg-sr-bench-meth, which consolidates previous individual drafts. The readiness of the draft for an adoption call was supported.

### 11.2.6 Power Benchmarking in Networking Devices

Qin Wu discussed the draft-cprjgf-bmwg-powerbench, which outlines a methodology for characterizing and benchmarking power consumption in networking devices.

### 11.2.7 SRv6 Service Benchmarking Guideline

Xuesong Geng introduced the draft-geng-bmwg-srv6-service-guideline, a guideline document for benchmarking SRv6 services. Feedback and comments from the working group were solicited.

The discussions and presentations at the BMWG session reflected a strategic shift towards more granular and sophisticated benchmarking methodologies, with an emphasis on accommodating the complexities of modern network infrastructures. The next steps involve further review and refinement of the drafts, with the anticipation of advancing the field of network performance benchmarking.

# 12 BPF/EBPF Working Group (BPF/EBPF)

The BPF/EBPF Working Group meeting brought together key stakeholders from the technology community to discuss advancements and challenges in the development of the BPF Instruction Set Architecture and its applications. The session featured presentations and discussions on various topics including the BPF ISA specification, the callx instruction, and the implementation and performance evaluation of PDM using eBPF.

## 12.1 Attendees Overview

### 12.1.1 Participation Summary

The meeting was attended by 37 participants representing prominent companies and institutions such as Cisco, Meta, Fastly, Dell Technologies, Ericsson, Huawei, and several universities. The diverse attendance underscores the wide interest and investment in the development of BPF/EBPF technologies.

### 12.1.2 Main Points and Contextualization

The discussions were centered around the latest updates to the BPF ISA specification document, with a call to action for participants to review the draft before the Working Group Last Call (WGLC) closes. The importance of a clear policy for deprecating instructions was highlighted, as well as the need for consensus on terminology used for pointers and addresses within the specification. The [draft-ietf-bpf-isa](draft-ietf-bpf-isa) document was a focal point, with the group deliberating on the implications of the callx instruction and its documentation within the IETF context. The session concluded with a presentation on the performance evaluation of PDM using eBPF, which sparked discussions on the need for improved documentation and functionality for packet manipulation.

Meeting materials are available at the following link: [IETF 119 Materials](IETF 119 Materials).

## 12.2 Meeting Discussions

### 12.2.1 BPF ISA Specification Document Update

Dave Thaler presented the recent changes to the BPF ISA specification document, emphasizing the actions required from the participants to review the draft. The discussion also touched upon the conformance suite and the eBPF verifier, PREVAIL. A policy change was suggested to differentiate between deprecated and historical instructions, with a focus on providing clarity for the designated expert on acceptable specifications.

### 12.2.2 BPF callx Instruction Discussion

The callx instruction was a topic of debate, with Dave Thaler setting up the problem through an example. The group discussed the implications of documenting the callx opcode, which is currently used by compilers like clang and gcc, within the IETF context. The conversation highlighted the complexities of instruction usage and the potential need to mark certain opcodes as dangerous.

### 12.2.3 Implementation and Performance Evaluation of PDM using BPF

The final presentation showcased the performance of eBPF in the context of PDM implementation. The discussion that followed pointed out the adequacy of documentation for helper functions and the desire for enhanced packet manipulation capabilities, particularly concerning packet splitting and header insertion with respect to MTU handling.

# 13 Calendar Extensions Working Group (CALEXT)

## 13.1 Attendee Overview

### 13.1.1 Participation Summary

The CALEXT session at IETF119 in Brisbane was attended by representatives from key organizations such as Fastmail, Ericsson, Bedework.com, ACSC, ICANN, Futurewei Technologies, Transmute, audriga GmbH, and APNIC, totaling 14 attendees.

### 13.1.2 Meeting Context

The discussions were centered around the progress of various drafts, including those in the AUTH48 state, WG Last Call, and expired drafts. The session provided an opportunity for the working group to address action items, clarify next steps, and discuss the future of certain drafts. Notable mentions include the [draft-ietf-calext-jscontact](draft-ietf-calext-jscontact) and its related documents, which are nearing completion, and the [draft-ietf-calext-icalendar-series](draft-ietf-calext-icalendar-series), which is still seeking active engagement for further progress.

Meeting materials can be found at [IETF 119 Materials](IETF 119 Materials).

## 13.2 Meeting Discussions

### 13.2.1 Document Progress

The working group acknowledged the near completion of the jscontact suite of documents, with a call to action for Robert to finalize AUTH48 formatting edits. Discussions on the ical tasks and subscription upgrade drafts led to action items for Ken and Mike to resolve outstanding issues and prepare for submission to the IESG.

### 13.2.2 Expired Drafts

The expired drafts, including jscalendar and vpoll, were discussed with a focus on the importance of interoperability testing and the potential revival of interest in these drafts. The group agreed on actions to make the jscalendar-icalendar tests public and to refresh the icalendar-series draft.

### 13.2.3 Future Directions

The session concluded with a debate on the necessity of a future meeting in Vancouver, weighing the work that justifies taking up a slot against the current progress of drafts. The group decided to hold off on planning a meeting unless significant interest and engagement are observed in the interim.

# 14 Computing-Aware Traffic Steering Working Group (CATS WG)

## 14.1 Attendees Overview

### 14.1.1 Prominent Companies and Institutions

The meeting was attended by representatives from prominent companies and institutions, including China Mobile, Huawei, Orange, Nokia, Qualcomm, and many others, with a total attendance of 67 participants.

### 14.1.2 Meeting Summary

The CATS WG meeting at IETF-119 focused on the progression of use case drafts, framework drafts, and discussions on computing service definitions. The dialogue centered on the relationship between CATS and Service Function Chaining (SFC), with an emphasis on clarifying the definition of a computing service within the CATS architecture. The meeting also included presentations on various drafts, which provided updates on computing information description, metrics, and traffic steering for midhaul networks. The discussions highlighted the need for a clear understanding of the metrics required for traffic steering and the importance of compatibility between computing and network metrics. The meeting concluded with flash presentations on potential solution technologies and a consensus on the need for further discussion on metrics distribution methods.

Meeting materials are available at IETF-119 CATS WG materials.

## 14.2 Meeting Discussions

### 14.2.1 Intro, WG Status

The chairs introduced the progress of the CATS WG, including the acceptance of use case and framework drafts. The agenda was presented without comments.

### 14.2.2 CATS Problem Statement, Use Cases, and Requirements

Kehan Yao presented updates to the use cases and requirements draft, which included merging content from Alibaba's draft and discussions on the relationship between CATS and SFC. The need to update the service definition in WG drafts was emphasized. The draft can be found at draft-ietf-cats-usecases-requirements.

### 14.2.3 Framework for Computing-Aware Traffic Steering

Cheng Li discussed the framework draft, addressing adoption comments and the need to clarify the service instance within the architecture. The draft is available at draft-ldbc-cats-framework.

### 14.2.4 Compute Modeling and Metrics

Presentations by Zongpeng Du and Luis M. Contreras focused on computing information description and joint exposure of network and compute metrics for service deployment. The discussions highlighted the need for simple and compatible metrics for traffic steering. Relevant drafts can be found at draft-du-cats-computing-modeling-description and draft-rcr-opsawg-operational-compute-metrics.

### 14.2.5 Compute-Aware Traffic Steering for Midhaul Networks

Luis M. Contreras presented a draft on traffic steering for midhaul networks, addressing the interaction between service management and orchestration (SMO) and transport management entities. The draft is available at draft-lcmw-cats-midhaul.

### 14.2.6 Analysis of Methods for Distributing the Computing Metric

Hang Shi's presentation analyzed methods for distributing computing metrics within the CATS architecture. The discussion suggested a feedback loop between this analysis and the metrics draft. The draft can be found at draft-shi-cats-analysis-of-metric-distribution.

### 14.2.7 Flash Presentations of Possible Solution Technologies

Flash presentations were given on various solution technologies, including segment routing and IP address anchoring for CATS. The presentations were well-received, and a show of hands indicated support for driving the metrics discussion through scheduled meetings.

The meeting concluded with an open discussion and next steps, with no further points raised. The potential impact of the discussions suggests a shift towards a more defined and structured approach to computing-aware traffic steering, with an emphasis on the development of compatible metrics and the distribution of computing metrics within the CATS architecture.

# 15 Concise Binary Object Representation (CBOR) Working Group (WG) [CBOR]

The CBOR Working Group (WG) meeting at IETF 119 saw a total of 33 attendees, including representatives from prominent organizations such as APNIC, CZ.NIC, Futurewei Technologies, RISE Research Institutes of Sweden, Vigil Security, LLC, Ericsson, Uni Bremen TZI, Security Theory LLC, Comcast, High North Inc, Huawei, Sandelman Software Works, TU Dresden, mesur.io, NIST, Arm, JPRS, LabN Consulting, Sateliot, Transmute, AIST Japan, IETF LLC, DataTrails, and Aoyama Gakuin University.

The discussions focused on the latest advancements in the use of CBOR in specifications, with a particular emphasis on the development and implementation of new draft documents. The meeting provided a platform for the presentation of new ideas, the refinement of existing proposals, and the setting of goals for future work. The attendees engaged in technical debates that highlighted the potential shifts in strategy and technical direction, as well as the next steps that would contribute to the field of CBOR applications.

Meeting materials are available at IETF 119 CBOR WG Materials.

## 15.1 Meeting Discussions

### 15.1.1 Using CBOR in Specifications

A tutorial block titled "Using CBOR in Specifications" was presented, covering topics such as the Extended Diagnostic Notation (EDN) and Concise Data Definition Language (CDDL). The tutorial included discussions on drafts such as draft-ietf-cbor-edn-literals, draft-bormann-cbor-e-ref, draft-bormann-cbor-draft-numbers, draft-ietf-cbor-cddl-modules, and draft-ietf-cbor-cddl-more-control. These discussions provided insights into the use of placeholders in examples, the establishment of conventions for draft numbers, and the introduction of new control operators in CDDL.

### 15.1.2 YANG Stand-in

The draft-bormann-cbor-yang-standin was discussed with the goal of engaging a wider audience, particularly from the network management and operations sectors. The draft aims to address the efficient representation of YANG data types, such as date/time and IP address/prefix, using CBOR tags. The discussion highlighted the need for a write-up that is understandable in the YANG context and the challenges associated with announcing capabilities.

### 15.1.3 Packed CBOR

The draft-ietf-cbor-packed was presented with the goal of making its state known more widely and deciding on the next steps. The draft addresses redundancy at the data model level by applying compression, and it is currently stable with ongoing work on table building and error representation.

### 15.1.4 Packed CBOR by Reference

Finally, the draft-amsuess-cbor-packed-by-reference was proposed for adoption with the goal of optimizing the setup of reference tables for applications like CoRAL. The draft describes a general method for avoiding large or inconvenient table setups by instructing a shift in the entries of the used reference tables.

The meeting concluded with an agreement to continue interim meetings with the usual cadence, alternating with those of the CoRE WG.

# 16 Common Control and Measurement Plane (CCAMP)

## 16.1 Attendee Overview

### 16.1.1 Participation Summary

The CCAMP working group meeting at IETF 119 saw a robust attendance from a variety of prominent companies and institutions. Notable attendees included representatives from Huawei Technologies, Nokia, Cisco, Ericsson, Telefonica Innovacion Digital, and ATT, among others. The meeting was attended by 35 participants, indicating a strong interest in the ongoing work of the CCAMP.

### 16.1.2 Meeting Summary

The discussions at the CCAMP working group meeting were centered around the advancement of several Internet Engineering Task Force (IETF) draft documents. These drafts focused on topics such as Optical Impairment-aware Topology, WDM Tunnels, Path Computation, Performance Monitoring, and the applicability of Generalized Multi-Protocol Label Switching (GMPLS) for fine-grain Optical Transport Networks. The meeting featured a mix of presentations and dialogues that highlighted the technical intricacies and strategic implications of the drafts. For instance, the [draft-ietf-ccamp-optical-impairment-topology-yang](#) was discussed in the context of its potential to enhance network topology awareness and efficiency.

Meeting materials and minutes can be accessed through the direct link [here](#).

## 16.2 Meeting Discussions

### 16.2.1 Optical Impairment-aware Topology

The presentation on the YANG Data Model for Optical Impairment-aware Topology sparked a conversation about the need for a unified approach to modeling optical impairments. The discussion underscored the importance of this work in enabling more accurate and efficient optical network planning and operation.

### 16.2.2 WDM Tunnels and Path Computation

The group reviewed the YANG Data Model for WDM Tunnels and Path Computation, emphasizing the significance of these models in supporting wavelength division multiplexing (WDM) technologies. The consensus was that these models are critical for the future of high-capacity optical networks.

### 16.2.3 Transport NBI Design Team Activities

Updates from the Transport NBI Design Team highlighted the collaborative efforts to align terminology and models across different standardization bodies. The dialogue suggested a strategic shift towards greater interoperability and coherence in network model design.

### 16.2.4 Pluggable Progress and Documents

The discussions on pluggable interfaces and their progress brought to light the evolving requirements and use cases in the industry. The group acknowledged the potential impact of these developments on the standardization of data models for network equipment.

### 16.2.5 Performance Monitoring

The YANG Data Model for Performance Monitoring was another focal point, with the group recognizing the model's potential to enhance network diagnostics and maintenance capabilities.

### 16.2.6 Next Steps

The meeting concluded with an agreement on the importance of these discussions for the future of network control and measurement. The next steps involve further refinement of the drafts, with an emphasis on addressing the feedback received and continuing the collaborative efforts to advance the CCAMP's objectives.

# 17 Congestion Control Working Group (CCWG)

The Congestion Control Working Group (CCWG) convened to discuss advancements and updates in congestion control mechanisms. The meeting was attended by representatives from prominent companies and institutions, with a total attendance of 78 participants.

## 17.1 Attendees Overview

### 17.1.1 Representation and Attendance

The meeting saw a diverse representation from the industry, including key figures from Meta, Google, Cisco, and Apple, among others. The total number of attendees was 78, indicating a strong interest in the topic of congestion control within the networking community.

### 17.1.2 Main Points and Contextualization

The discussions focused on various congestion control proposals, with an emphasis on their potential impact on internet traffic management. The group reviewed updates to existing algorithms and considered new approaches that could better address the evolving needs of network traffic. Key documents such as [draft-ietf-ccwg-rfc5033bis](draft-ietf-ccwg-rfc5033bis) were discussed, highlighting the importance of continuous evaluation and improvement of congestion control mechanisms.

Meeting materials are available through the direct link: [IETF 119 Materials](IETF 119 Materials).

## 17.2 Meeting Discussions

### 17.2.1 5033bis: Specifying New Congestion Control Algorithms

Martin Duke presented an overview of the 5033bis draft, which aims to provide guidelines for the specification of new congestion control algorithms. The discussion underscored the need for more reviews and the importance of considering multiple dimensions of fairness in congestion control.

### 17.2.2 Increase of the Congestion Window when the Sender Is Rate-Limited

Gorry Fairhurst introduced the draft-welzl-ccwg-ratelimited-increase, which addresses the increase of the congestion window when the sender is rate-limited. The conversation highlighted the need to distinguish between principles and implementation, with a focus on the broader goal of not augmenting capacity when not fully utilized.

### 17.2.3 8298bis: Updated SCReAM congestion control

Ingemar Johansson discussed the updates to the SCReAM congestion control algorithm, emphasizing the simplification and stabilization of video rate computation. The group considered the necessity of renaming certain variables to avoid confusion.

### 17.2.4 HPCC++

Rui Miao presented the HPCC++ draft, which proposes a high-precision congestion control algorithm. The dialogue revolved around the encoding of congestion notification information and its impact on performance.

### 17.2.5 BBRv3 Update and Deployment Status

Neal Cardwell provided an update on BBRv3, including its deployment status and performance improvements. The group expressed a strong interest in adopting BBRv3 as a working group item, recognizing its significant deployment and potential for further optimization.

The discussions suggested a shift towards a more unified approach to congestion control, with an emphasis on principles that can be applied across different algorithms. The next steps for the working group include refining the charter and continuing to work on the drafts discussed, with the anticipation that these efforts will contribute significantly to the field of congestion control.

# 18 Content Delivery Network Interconnection Working Group (CDNI)

## 18.1 Attendee Overview

### 18.1.1 Attendance Summary

The CDNI working group meeting saw participation from a diverse range of companies and institutions, including Vecima, Comcast, Verizon, FCC, Qwilt, Meetecho, SVTA, Interdigital Europe, Soongsil University, NetPico Labs Pvt. Ltd., ISOC-PH, Ciena, Web Civics, AIST Japan, Amazon Web Services, and Ericsson. The total attendance was recorded at 23 individuals.

### 18.1.2 Meeting Materials

Meeting materials are available at [IETF 119 Materials](#).

## 18.2 Meeting Discussions

### 18.2.1 CDNI Control Interface / Triggers 2nd Edition

Jay Robertson led a discussion on the second edition of the CDNI Control Interface / Triggers, focusing on the [draft-ietf-cdni-ci-triggers-rfc8007bis-11](#). The conversation highlighted the need for clarity around timezone handling and daylight savings time, with a consensus on the importance of providing implementers with clear guidelines.

### 18.2.2 Capacity Capability Advertisement Extension

Ben Rosenblum presented the [draft-ietf-cdni-capacity-insights-extensions-04](#), which proposes an extension for capacity capability advertisement in CDNI. The presentation was concise and focused on the technical aspects of the extension.

### 18.2.3 Logging Extensions

The group reviewed the logging extensions as outlined in [draft-rosenblum-cdni-logging-extensions-01](#), with Ben Rosenblum leading the discussion. The working group considered an adoption request for this draft, emphasizing the need for further review.

### 18.2.4 CI/T v2 Triggers Draft Additions

Alan Arolovitch proposed additions to the CI/T v2 triggers draft beyond the current version 11. The discussion underscored the importance of reaching a consensus on the proposed changes and the potential impact on the timeline for the working group's deliverables.

### 18.2.5 Named Footprints

The [draft-arolovitch-cdni-named-footprints-01](#) was briefly presented by Alan Arolovitch, introducing the concept of named footprints in CDNI.

### 18.2.6 Metadata Drafts

Glen Goldstein and his team covered a series of metadata-related drafts, requesting adoption for several and moving others towards working group last call (WGLC). The discussions highlighted the need for clear charter scope and the importance of community review and implementation support.

The meeting concluded with a wrap-up by the chairs, emphasizing the need for shepherds and reviewers to facilitate progress on the drafts and ensure adherence to the working group's charter.

# 19 Crypto Forum Research Group (CFRG)

The Crypto Forum Research Group (CFRG) serves as a bridge between theory and practice, bringing new cryptographic techniques to the Internet community and promoting an understanding of the use and applicability of these mechanisms.

## 19.1 Attendees Overview

### 19.1.1 Attendance Summary

The meeting saw participation from a diverse set of individuals, representing prominent companies and institutions such as Google, Apple, Cloudflare, and Ericsson, with a total attendance of 104.

### 19.1.2 Meeting Context

The discussions at the meeting were rich and varied, covering topics from hedged signatures to federated machine learning. The presentations sparked debates on the adoption of new cryptographic mechanisms and their implications for industry standards. Notably, the dialogue on hybrid Post-Quantum Cryptography (PQC) KEMs and the need for RSA hybrids in industry highlighted the group's commitment to addressing real-world cryptographic challenges. References to IETF draft documents were made throughout the discussions, such as the [draft-ietf-cfrg-hedged](draft-ietf-cfrg-hedged) and [draft-ietf-cfrg-ml-kem-hpke](draft-ietf-cfrg-ml-kem-hpke).

Meeting materials can be accessed directly via the [IETF 119 Materials](IETF 119 Materials) page.

## 19.2 Meeting Discussions

### 19.2.1 Hedged Signatures

John Mattson presented on "Hedged Signatures," discussing recent changes, open issues, and next steps. The presentation prompted a discussion on the motivation behind certain issues and the potential for collaboration on providing proofs.

### 19.2.2 GCM-SST

John Mattson also spoke about "GCM-SST," a variant of Galois/Counter Mode with secure short tags. The presentation highlighted the industry interest and led to a conversation about the possibility of CFRG adoption.

### 19.2.3 ML-KEM for HPKE

Deirdre Connolly's presentation on "ML-KEM for HPKE" delved into the goals, requirements, and potential re-encapsulation attacks. The talk sparked a debate on whether to adapt HPKE specifically for ML-KEM or to modify it more broadly to accommodate other KEMs.

### 19.2.4 Next Steps for draft-mouris-cfrg-mastic

Chris Patton discussed the "Next steps for draft-mouris-cfrg-mastic," which is related to VDAF/PPM/heavy-hitters. The authors are considering a next draft for Research Group adoption and raised questions about naming conventions and handling the development and review of new VDAFs.

### 19.2.5 PINE, a VDAF for Federated Machine Learning

Junye Chen introduced "PINE, a VDAF for federated machine learning," and discussed the draft's progress and the possibility of Research Group adoption. The presentation highlighted the need for further discussion on workload distribution among participants in federated systems.

### 19.2.6 The Asynchronous Remote Key Generation (ARKG) Algorithm

John Bradley presented "The Asynchronous Remote Key Generation (ARKG) algorithm," asking for review and feedback on the draft. The talk covered the motivation and methods for generating key pairs from seeds.

### 19.2.7 Why P-256 and RSA Hybrids are Needed in Industry

Mike Ounsworth's presentation on the necessity of P-256 and RSA hybrids in industry underscored the challenges faced by private and enterprise PKIs. The discussion emphasized the need for a careful balance between supporting legacy systems and advancing towards hybrid KEM use.

The meeting concluded with no additional business, reflecting a productive session with potential impacts on the direction of cryptographic research and its application in the industry.

# 20 Constrained RESTful Environments (CoRE) Working Group

## 20.1 Attendees Overview

### 20.1.1 Attendance Summary

The meeting was attended by representatives from prominent companies and institutions such as Ericsson, Huawei Technologies, and RISE Research Institutes of Sweden, with a total attendance of 24 participants.

### 20.1.2 Meeting Summary

The CoRE Working Group meeting focused on the advancement of several Internet Engineering Task Force (IETF) draft documents related to constrained environments. Discussions centered on the development of protocols and extensions that facilitate efficient and secure communication in resource-constrained settings. Key topics included updates to the CORECONF cluster, the introduction of Constrained Resource Identifiers (CRIs), and the exploration of DNS over CoAP (DoC) and CoRE DNR. The meeting also delved into the implications of CoAP Protocol Indication and Proxy Operations for CoAP Group Communication. The integration of OSCORE for secure communication was a recurring theme, with presentations on OSCORE-capable proxies and updates to OSCORE key management. The meeting materials are directly available via the CoRE WG session page.

## 20.2 Meeting Discussions

### 20.2.1 CORECONF Cluster

Carsten Bormann presented updates on the CORECONF cluster, which includes drafts on YANG-CBOR mapping, SID, and COMI. The discussion highlighted the need for tool support and addressed comments from the last call. The drafts aim to optimize data representation and management in constrained environments. The next steps involve addressing remaining comments and adding further examples.

### 20.2.2 Constrained Resource Identifiers

The draft on Constrained Resource Identifiers (CRIs) was discussed, focusing on the exchange format for parsed URIs and the introduction of new CoAP options for proxy operations. The plan is to finalize the draft by adding test vectors and clarifying the determinism objective.

### 20.2.3 DNS over CoAP and CoRE DNR

Martine Lenders presented the DNS over CoAP (DoC) and CoRE DNR drafts, which aim to protect DNS requests in constrained environments. The group discussed the readiness for a working group last call and the potential integration with EDHOC and OSCORE for secure discovery.

### 20.2.4 CoAP Protocol Indication

Christian Amsüss discussed the CoAP Protocol Indication draft, which enables the discovery of alternative transports for CoAP. Open questions regarding the scope of "has-proxy" and the handling of self-description were raised, with a call for reviews and interoperability tests.

### 20.2.5 Proxy Operations for CoAP Group Communication

Esko Dijk presented the newly adopted draft on proxy operations for CoAP group communication, which addresses forward and reverse proxies, and cross-proxying between HTTP and CoAP. The discussion included updates on terminology, option handling, and the renaming of the Response-Forwarding option to "Reply-To."

### 20.2.6 OSCORE-capable Proxies

Rikard Höglund introduced the draft on OSCORE-capable proxies, which allows for OSCORE-protected communication between proxies or between client/server and a proxy. The draft specifies rules for encrypting CoAP options and supports nested OSCORE protection.

### 20.2.7 OSCORE Key Update and ID Update

The OSCORE Key Update (KUDOS) and ID Update drafts were split, with the former focusing on key update procedures and the latter on OSCORE ID management. The group discussed the use of non-random nonces and the independence of KUDOS execution flows from the request-response flow.

### 20.2.8 CoAP: Non-traditional Response Forms

Christian Amsüss presented a draft that defines common terms and concepts for non-traditional responses in CoAP. The draft provides guidelines for secure handling with OSCORE and seeks reviews from experts.

### 20.2.9 CoRE Resource Directory Extensions

The draft on CoRE Resource Directory (RD) Extensions was discussed, with a focus on integrating the RD with the ACE framework and EDHOC. The group considered enhancements to ACE and evaluated the suitability of ACE and EDHOC for use with RD and constrained endpoints.

### 20.2.10 CoAP over BP and in Space

Carles Gomez introduced drafts on CoAP over Bundle Protocol (BP) and CoAP in space. The drafts propose architectures and messaging models for CoAP in deep space communication, considering long delays and intermittent connectivity. The group discussed the need for congestion control and the definition of a new URI scheme for CoAP+BP.

# 21 CBOR Object Signing and Encryption (COSE)

## 21.1 Attendees Overview

### 21.1.1 Attendance Summary

The meeting was attended by representatives from prominent companies and institutions such as Google, Siemens, Ericsson, Intel, Cisco, and many others, totaling 56 attendees.

The discussions during the meeting revolved around several key drafts and proposals, with a focus on cryptographic standards and protocols. Notable discussions included advancements in hybrid public key encryption (draft-ietf-cose-hpke), key thumbprints (draft-ietf-cose-key-thumbprint), and the integration of post-quantum cryptography into existing frameworks. The dialogue was rich with technical insights and underscored the community's commitment to evolving security standards in light of emerging challenges.

Meeting materials are available at this link.

## 21.2 Meeting Discussions

### 21.2.1 draft-ietf-cose-hpke

Hannes Tschofenig presented improvements in terminology and context setting for hybrid public key encryption, addressing potential attack vectors. The draft aims for stability by the next IETF meeting.

### 21.2.2 draft-ietf-cose-key-thumbprint

Feedback on the key thumbprint draft was discussed, with a call for additional input before finalization.

### 21.2.3 draft-tschofenig-cose-cwt-chain

The draft, which allows for the presentation of a chain of CWTs, was extracted from the SUIT document to enable broader application. A call for adoption was supported by attendees.

### 21.2.4 draft-demarco-cose-header-federation-trust-chain

John Bradley discussed the federation meta-data trust chain and potential collaboration on a common document with other drafts.

### 21.2.5 draft-ietf-cose-dilithium and draft-ietf-cose-sphincs-plus

Mike Prorock updated on post-quantum signature algorithms, with discussions on when to register these algorithms and the importance of interoperability.

### 21.2.6 draft-ietf-cose-merkle-tree-proofs

Orie Steele presented a generic scheme for creating receipts for logs, with support from the SCITT group.

### 21.2.7 draft-ietf-cose-cbor-encoded-cert

John Mattsson aimed for the document to reach WGLC before the next meeting, with discussions on performance improvements and potential changes to the signature algorithm field.

### 21.2.8 draft-ietf-cose-bls-key-representations

Tobias Looker's presentation led to discussions on the utility of compressed curve representations.

### 21.2.9 draft-reddy-cose-jose-pqc-kem

Tiru Reddy presented on KEM constructions, with feedback suggesting a preference for not having multiple similar algorithms in the COSE registry.

### 21.2.10 AOB

The meeting concluded with votes on moving the algorithm in c509 and whether to mark old numbers as reserved, both receiving affirmative responses.

# 22 New DNS Delegation BoF (NDD BoF)

## 22.1 Attendees Overview

### 22.1.1 Prominent Attendees and Total Attendance

The New DNS Delegation BoF session was attended by representatives from prominent companies and institutions such as ICANN, Cox, SIDN, IBM, NCSC, USC/ISI, Stanford University, Internet Systems Consortium, Google, Meta Platforms, Inc., and many others. The total attendance was 153 individuals.

### 22.1.2 Summary of Discussions

The session focused on exploring better methods for DNS delegation, with discussions contextualized around the limitations of current practices and the potential for new standards to improve security and efficiency. Key presentations highlighted the need for innovation in DNS delegation, with contributions from experts across the industry. The dialogue centered on the [draft-ietf-dd-deleg](draft-ietf-dd-deleg) document, which proposes a new approach to DNS delegation that could significantly impact the future of internet infrastructure.

Meeting materials are available directly via the [IETF 119 materials page](IETF 119 materials page).

## 22.2 Meeting Discussions

### 22.2.1 Presentations

Presentations were given on various aspects of DNS delegation, including its limitations, goals, and requirements. Speakers from different organizations shared their perspectives on the importance of developing the DELEG method for DNS delegation, discussing its potential applications in encrypted DNS and its relationship with the DBOUND project.

### 22.2.2 Debates and Outcomes

The session featured an open discussion where participants debated the technical direction and strategy for DNS delegation. The debates were constructive, with attendees expressing a range of views on the proposed changes. The outcomes of the discussions suggested a shift towards a more secure and efficient DNS delegation process, with a consensus on the need for further development of the DELEG method.

### 22.2.3 Prospective Actions

The next steps include refining the [draft-ietf-dd-deleg](draft-ietf-dd-deleg) document, gathering broader community feedback, and working towards a potential new standard for DNS delegation. The anticipated contribution of these actions to the field is significant, as they aim to address longstanding issues with DNS delegation and pave the way for a more robust internet infrastructure.

# 23 Deterministic Networking (DetNet) Working Group

## 23.1 Attendees Overview

### 23.1.1 Prominent Participants and Attendance

The DetNet Working Group meeting was attended by representatives from prominent companies and institutions such as Ericsson, Huawei, ZTE, China Mobile, Cisco, and the U.S. Department of Defense, among others. The total attendance for the meeting was 87 participants.

The discussions during the DetNet Working Group meeting were centered around the advancement of deterministic networking technologies and their applications. The meeting featured presentations on various topics, including RAW Architecture, Reliable Wireless Industrial Services, and Latency Analysis of Mobile Transmission. The group engaged in debates on the technical direction and strategies for DetNet, with a focus on ensuring reliable and timely data transport over both wired and wireless networks. The meeting also highlighted the importance of interoperability and the need for a common understanding of quality of service (QoS) mechanisms, as evidenced by the discussions on IPv6 extension headers and echo request/reply for capability discovery. The presentations and discussions underscored the group's commitment to refining the DetNet specifications and addressing the challenges of deploying deterministic services in diverse environments.

Meeting materials are available directly via this link.

## 23.2 Meeting Discussions

### 23.2.1 RAW Architecture

Janos Farkas presented the latest draft of the RAW Architecture, which aims to provide reliable and available wireless connectivity for industrial services. The discussion highlighted the need for consistency in definitions and the potential removal of technology-specific details to maintain a focus on requirements.

### 23.2.2 Reliable Wireless Industrial Services

Carlos Jesus Bernardos Cano discussed the requirements for reliable wireless industrial services, emphasizing the need for technology-agnostic solutions and soliciting feedback on the draft.

### 23.2.3 Latency Analysis of Mobile Transmission

Balazs Varga's presentation on latency analysis of mobile transmission sparked a conversation about the applicability of DetNet in 5G systems and the potential for DetNet as an overlay in mobile networks.

### 23.2.4 BFD Extension for DetNet RDI

Li Zhang introduced a draft for BFD Extension for DetNet Remote Defect Indication (RDI), which led to a debate on the appropriateness of BFD for notifications and the possibility of a more generic mechanism for multiple Service Level Objectives (SLOs).

### 23.2.5 Echo Request/Reply for DetNet Capability Discovery

Li Zhang also presented on Echo Request/Reply for DetNet Capability Discovery. The discussion raised questions about the consumers of the information and the suitability of ICMPv6 for the extension of discovery functionality.

### 23.2.6 QoS IPv6 Extension Headers

Toerless Eckert's presentation on QoS IPv6 extension headers prompted a debate on the necessity and design of such headers, with suggestions to seek feedback from the 6MAN working group before proceeding.

The DetNet Working Group's discussions concluded with a consensus on the importance of progressing with the drafts and refining the technical aspects of deterministic networking. The group agreed on the need for further discussions on the mailing list, particularly regarding the QoS IPv6 extension headers and the use of BFD for notifications. The outcomes of the meeting suggest a continued effort towards enhancing the DetNet specifications and a commitment to addressing the evolving requirements of deterministic networking applications.

# 24 Decentralized Internet Infrastructure Research Group (DINRG)

## 24.1 Attendee Overview

### 24.1.1 Participants

The meeting was attended by representatives from prominent companies and institutions such as Open-Commons, SIDN, HKUST, UCLA, Verisign, Internet Society, Cisco, NSA-CCSS, The Boeing Company, and many others, totaling 78 attendees.

### 24.1.2 Summary

The discussions at the DINRG meeting focused on the implications of centralization and decentralization in cyberspace, the delivery of social and municipal services, and the development of local-first application prototypes. The conversation was enriched by the diverse perspectives of the attendees, with particular attention given to the challenges and opportunities presented by decentralized systems. Key documents, such as the RFC 9518, were referenced to provide context and depth to the dialogue.

Meeting materials are available via the direct link: DINRG Meeting Materials.

## 24.2 Meeting Discussions

### 24.2.1 Cyberspace Regulation Case Study

Xing Li presented a case study on cyberspace regulation, prompting discussions on the interplay between business practices and regulatory frameworks. The debate highlighted the varying approaches to centralization in different jurisdictions, with a nod to China's unique regulatory landscape.

### 24.2.2 Delivering Social and Municipal Services

Wilfried Pinfold's presentation on delivering social and municipal services through decentralized systems sparked conversations about the challenges of interoperability and the need for secure identity management across various service platforms.

### 24.2.3 Workspace: A Local-First Application Prototype

The presentation by Xinyu Ma, Varun Patil, and Tianyuan Yu on Workspace, a local-first collaborative editing application, led to discussions on the technical challenges of ensuring eventual consistency and trust in decentralized networks.

### 24.2.4 RFC 9518 Discussion

A robust discussion ensued on the next steps following the publication of RFC 9518, which addresses centralization, decentralization, and internet standards. Participants debated the prioritization of decentralization efforts, the economic forces driving centralization, and the role of regulation and standardization in fostering a more decentralized internet infrastructure.

The meeting concluded with a consensus on the need for a clearer taxonomy of decentralization and a call for continued collaboration to address the complex interdependencies of technical, market, and regulatory factors in the evolution of the internet.

# 25 Dynamic Mobility Management Working Group (DMM)

The Dynamic Mobility Management Working Group (DMM) focuses on developing mobility management protocols and mechanisms that are dynamic, flexible, and efficient. The group aims to address the challenges posed by the evolving mobile Internet landscape.

## 25.1 Attendees Overview

### 25.1.1 Participation Summary

The meeting was attended by representatives from prominent companies and institutions such as Cisco, SoftBank, Futurewei, Telefonica Innovacion Digital, Verizon, China Mobile, Huawei, NEC, and BT. The total attendance was 29 individuals.

The discussions during the meeting revolved around various draft documents and proposals aimed at enhancing mobility management. The conversation included updates on existing drafts, presentations on new proposals, and debates on the technical direction and strategy for the group's work. Key topics included network slicing for 5G, the Mobile User Plane (MUP) architecture, and emergency calling over WiFi. Participants engaged in a constructive dialogue, with several drafts being considered for working group adoption and others receiving valuable feedback for further development.

Meeting materials are available directly via the [IETF Datatracker](#).

## 25.2 Meeting Discussions

### 25.2.1 draft-ietf-dmm-tn-aware-mobility

John Kaippallimalil presented updates on the "Mobility aware Transport Network Slicing for 5G" draft, highlighting the collaboration with the TEAS working group on slicing and the inclusion of control plane slices. The draft aims to provide guidance for 3GPP and the broader community on the potential use of transport network slicing in mobility management.

### 25.2.2 draft-mhkk-dmm-mup-architecture

Satoru Matsushima discussed the "MUP Architecture" draft, emphasizing its data plane independence and the definition of architecture principles. The draft outlines how MUP segments can be auto-discovered and how session information can be transformed into routing information.

### 25.2.3 draft-dcn-dmm-cats-mup

Minh Ngoc Tran introduced the "Computing Aware Traffic Steering Consideration for MUP" draft, which integrates the Computing Aware Traffic Steering (CATS) concept into the MUP architecture. The proposal aims to optimize service instance selection for anycast routing.

### 25.2.4 draft-zzhang-dmm-mup-evolution

Tianji Jiang provided a summary of the "Mobile User Plane Evolution" draft, addressing comments from the working group and suggesting that the draft could serve as a reference for future work in 3GPP, particularly for 6G.

### 25.2.5 draft-yan-dmm-man

Tianji Jiang also presented the "Mobility Capability Negotiation" draft, which underwent significant restructuring. The draft proposes a dichotomy of mobility management and capability negotiation protocols, with a call for working group adoption.

### 25.2.6 draft-gundavelli-dispatch-e911-wifi

Sri Gundavelli discussed the "Emergency Calling (911/112/*) over WiFi" draft, which addresses the challenge of making emergency calls over WiFi without public network access. The draft proposes a solution that leverages existing roaming architectures and ensures location identification for emergency services.

**25.2.7  Traffic Steering (No Draft)**

Marco Liebsch provided an update on the ongoing discussions about traffic steering, indicating that a draft is being compiled and will be shared for feedback. The topic explores routing aspects in the data network and its relevance to the DMM working group's charter.

The meeting concluded with an agreement to consider the traffic steering topic in line with the working group's charter and to encourage feedback on the upcoming draft.

# 26  DNS Operations (DNSOP)

## 26.1  Attendees Overview

### 26.1.1  Attendance Summary

The DNS Operations Working Group (DNSOP) meeting was attended by a diverse group of participants, including representatives from prominent companies such as Meta Platforms, Inc., IBM, and Verisign, as well as academic institutions like Keio University. The total attendance was recorded at 109 individuals.

### 26.1.2  Main Points and Contextualization

The DNSOP discussions were robust, with a focus on several key draft documents. Notable discussions included the [draft-ietf-dnsop-generalized-notify](), which raised concerns about potential DDoS attacks and the necessity of DNSSEC for synchronization of records. The [draft-ietf-dnsop-rfc8109bis]() sparked debate over the status of root zone operators, while the [draft-ietf-dnsop-rfc7958bis]() addressed the multiplicity of IANA publishing methods. The [draft-ietf-dnsop-compact-denial-of-existence]() and the [draft-ietf-dnsop-ns-revalidation]() were also discussed, with the latter emphasizing the need for implementation in resolvers.

Meeting materials and presentations can be accessed via the direct link [DNSOP Meeting Materials]().

## 26.2  Meeting Discussions

### 26.2.1  Generalized Notify

The [draft-ietf-dnsop-generalized-notify]() presentation led to a lively debate about the implications of syncing DNSSEC-related records and the potential for DDoS attacks. The consensus suggested a need for careful consideration of security implications.

### 26.2.2  RFC 8109bis and RFC 7958bis

Discussions on [draft-ietf-dnsop-rfc8109bis]() and [draft-ietf-dnsop-rfc7958bis]() highlighted the special status of root zone operators and the variety of IANA's publishing methods. The group considered significant changes to the documents to reflect these issues.

### 26.2.3  Compact Denial of Existence

The [draft-ietf-dnsop-compact-denial-of-existence]() was presented without major contention, suggesting general support for the approach to streamline denial of existence in DNS responses.

### 26.2.4  NS Revalidation

The [draft-ietf-dnsop-ns-revalidation]() sparked a discussion on the necessity of NS record revalidation in resolvers, with a call for its implementation to prevent potential stale data issues.

### 26.2.5  GREASE for DNS

The introduction of [draft-huque-dnsop-grease]() was met with enthusiasm, as it proposes a method to apply pressure on implementations to adhere to standards, similar to the concept of GREASE in TLS.

The discussions at DNSOP suggest a continued commitment to evolving DNS security and efficiency. The outcomes of these debates are expected to influence future technical directions and strategies within the field of DNS operations.

# 27 Delay/Disruption Tolerant Networking (DTN) WG

## 27.1 Attendees Overview

### 27.1.1 Prominent Companies and Total Attendance

The meeting was attended by representatives from prominent institutions such as Spacely Packets, LLC, JHUAPL, ISOC IPNSIG, SAIC/MSFC, SPATIAM CORPORATION, Hamburg University of Technology and ESA, Aalyria, Ericsson, Cisco, and many others, totaling 51 attendees.

### 27.1.2 Main Points and Dialogues

The discussions at the DTN WG meeting centered around the advancement of various Internet Engineering Task Force (IETF) draft documents and the exploration of new concepts such as Constrained Application Protocol (CoAP) over Bundle Protocol (BP) and Quality of Service (QoS) extensions for BP. The dialogue was rich with technical insights and contextualized within the broader scope of Delay/Disruption Tolerant Networking. Key draft documents such as draft-ietf-dtn-bpv7-admin-iana, draft-ietf-dtn-bpsec-cose, and draft-sipos-dtn-eid-pattern were discussed in detail.

Meeting materials are available directly via this link: IETF 119 Materials.

## 27.2 Meeting Discussions

### 27.2.1 BPv7 Admin. Record Types Registry

Brian Sipos presented the draft document on BPv7 Admin. Record Types Registry, emphasizing the need for updating the document which is currently in Working Group Last Call (WGLC) and has expired without comments.

### 27.2.2 BPSec COSE Context

The BPSec COSE Context presentation by Brian Sipos highlighted the adoption of COSE to avoid redundant work and noted that no major issues were raised during WGLC that would significantly alter the document.

### 27.2.3 BP EID Patterns

Brian Sipos also discussed the BP EID Patterns draft, proposing capabilities for DTN and IPN scheme patterns and emphasizing that EID patterns are not interchangeable with EIDs.

### 27.2.4 Constrained Application Protocol (CoAP) over BP

Carles Gomez's presentation on CoAP over BP outlined the main goal of specifying CoAP over BP and the need to keep both the CoRE and DTN working groups in the loop. The discussion touched on message formats, parameter settings, and the potential need for congestion control.

### 27.2.5 BIBE Updates

Scott Burleigh and Alberto Montilla provided updates on Bundle-in-Bundle Encapsulation (BIBE), discussing the need for control over parameters in BIBE source that condition forwarding and the addition of a processing flag for BIBE.

### 27.2.6 Quality of Service Extension for BP

Teresa Algarra introduced a proposal for QoS extension blocks for BP, focusing on Earth to Moon communications and the congestion that occurs with a single relay point. The proposal includes a User QoS Block and Network QoS Blocks that can be added and removed along the path.

### 27.2.7 3GPP SA1 Store-Forward Use Cases

Ed Birrane briefed the group on the 3GPP SA-1 Working Group's focus on store-and-forward satellite operations and the relevance of DTN service standards being produced by the IETF.

### 27.2.8 Open Mic

The open mic session included discussions on the need for maturity in specifying EIDs and the potential engagement with the DNS community for structured DTN-like names.

# 28 Enhanced Authentication Protocol Method Updates (EMU)

## 28.1 Attendees Overview

### 28.1.1 Prominent Attendees and Total Attendance

The meeting was attended by representatives from prominent companies and institutions such as AKAYLA, FreeRADIUS, Venafi, Radiator Software, NSA - CCSS, coreMem Limited, Siemens, DFN-Verein, Huawei, Ericsson, Carnegie Mellon University, Ping Identity, Cisco, Okta, Adana Alparslan Turkes Science and Technology University, Nokia, Comcast, Hewlett Packard Enterprise, and AIST, totaling 20 attendees.

The discussions were rich with technical insights and revolved around the latest draft proposals. The dialogue contextualized the need for updated EAP methods in light of evolving security requirements and technological advancements. Drafts such as draft-ietf-emu-aka-pfs and draft-ietf-emu-bootstrapped-tls were central to the conversation, indicating a shift towards more robust and flexible authentication mechanisms.

Meeting materials are available directly via this link.

## 28.2 Meeting Discussions

### 28.2.1 EAP-AKAFS

Discussion on draft-ietf-emu-aka-pfs indicated readiness for IESG processing with no technical issues reported.

### 28.2.2 Bootstrapped TLS

The draft-ietf-emu-bootstrapped-tls is proceeding with a downref for RFC 8773 due to unclear timelines for its elevation to a Proposed Standard within the TLS WG.

### 28.2.3 Charter Update

A proposed charter update was discussed, with the intention to incorporate new work on FIDO and EDHOC. The discussion highlighted the potential of EDHOC in IoT environments and the relevance of FIDO for user-friendly authentication. The next steps involve formalizing the charter update to include these deliverables.

### 28.2.4 EAP-ARPA

The draft-dekok-emu-eap-arpa presentation covered the proposal for an 'eap.arpa' domain to support EAP methods, with an ongoing WG adoption call signaling imminent Working Group Last Call (WGLC).

### 28.2.5 EAP EDHOC

draft-ingles-eap-edhoc is awaiting the revised charter for formal adoption, with noted industry interest and implementation efforts. A Hackathon in Paris was mentioned as an opportunity for further collaboration.

### 28.2.6 EAP FIDO

The draft-janfred-eap-fido presentation sparked a debate on the practicality of integrating FIDO with EAP, particularly regarding silent authentication and the use of web-tied credentials. The discussion concluded with an acknowledgment of the need for further exploration and liaison with the FIDO alliance.

### 28.2.7 EAP-AKA-PQC

The draft-ar-emu-pqc-eapaka presentation introduced the integration of post-quantum cryptography with EAP-AKA, with a call for formal analysis of the new hybrid design.

### 28.2.8 EAP Multiple Pre-Shared Keys (EAP-MPSK)

Interest in the draft-yan-emu-eap-multiple-psk was gauged, with the chair encouraging further discussion on the mailing list to assess community interest.

# 29 Global Network Advancement Program (GNAP)

## 29.1 Attendees Overview

### 29.1.1 Attendance Summary

The meeting was attended by representatives from prominent companies and institutions such as Cisco Systems, Intuit, the National Security Agency (NSA), and the National Institute of Standards and Technology (NIST), with a total of 14 attendees.

### 29.1.2 Meeting Context

The discussions were centered around the updates to the core protocol and the Resource Server (RS) protocol.

## 29.2 Meeting Discussions

### 29.2.1 Working Group Status

The chairs announced that the core protocol has been approved for publication, while the RS Protocol is currently in Working Group Last Call (WGLC). There was a discussion about the potential shutdown of the working group due to a lack of energy and the exploration of alternative venues for the RS draft.

### 29.2.2 Core Protocol and Resource Server Protocol Updates

The RS Draft did not receive any comments during the WGLC, raising questions about the community's interest and engagement. Justin Richer provided an update on the protocol, noting that while there are implementations of different pieces of the RS draft, the discovery parts are not widely implemented, and no known parties have implemented the entire suite.

### 29.2.3 Next Steps

The meeting concluded with an emphasis on the need to increase community involvement and to consider alternative strategies for advancing the RS protocol. The potential shutdown of the working group suggests a pivot in the approach to developing these protocols, with the next steps focused on finding a new platform to continue the work and to foster greater implementation and feedback.

# 30 Global Routing Operations (GROW)

## 30.1 Attendee Overview

### 30.1.1 Participation Summary

The GROW working group meeting was attended by representatives from major organizations such as Verisign, Google, Fastly, China Mobile, Deutsche Telekom, Amazon, and Huawei, among others, with a total attendance of 69 participants. This diverse assembly underscores the global interest and collaborative effort in routing operations and optimization.

The discussions were rich with technical insights and revolved around several Internet Engineering Task Force (IETF) draft documents. The group's deliberations aimed to address challenges and propose enhancements to the current state of global routing operations, with a focus on automation, security, and monitoring.

Meeting materials are available directly via the [IETF 119 Materials](#) page.

## 30.2 Meeting Discussions

### 30.2.1 Peering API Proposal

Jenny Ramseyer presented on the [draft-ramseyer-grow-peering-api](#), which introduces an automated peering management API. The proposal aims to reduce human errors and streamline peering operations. The discussion highlighted the need for security considerations and the potential fit of this draft within the GROW charter. A side meeting was suggested to further delve into the workflow implications and technical details.

### 30.2.2 BMP Enhancements

Paolo Lucente discussed several drafts related to the BGP Monitoring Protocol (BMP), including [draft-ietf-grow-bmp-tlv](#), [draft-ietf-grow-bmp-rel](#), and [draft-ietf-grow-bmp-tlv-ebit](#). These drafts propose enhancements to BMP for improved route monitoring and logging. Feedback from the attendees suggested the importance of visibility into routing policies and the balance between memory and CPU resources in implementations.

### 30.2.3 Operational Security Update

Tobias Fiebig presented the [draft-ietf-grow-bgpopsecupd](#), which addresses operational security in BGP. The conversation revolved around the document's maintenance and the challenge of keeping it up-to-date. There was a consensus on the need for a living document approach, possibly facilitated by a platform like GitHub, to allow for continuous updates.

### 30.2.4 BMP Statistics and Aggregation

Jinming Li and Yisong Liu introduced drafts on BMP statistics reports ([draft-liu-grow-bmp-stats-reports](#)) and route monitoring aggregation ([draft-liu-grow-bmp-rm-aggregated](#)). The discussions raised concerns about the practicality of implementing new counters and the potential loss of monitoring detail due to aggregation. The group debated the trade-offs between saving bandwidth and maintaining the fidelity of routing information.

### 30.2.5 BMP TCP Authentication Option

Jeffrey Haas discussed the [draft-hmntsharma-bmp-tcp-ao](#), which proposes the use of TCP Authentication Option (TCP-AO) for BMP sessions. The idea was met with interest, and an adoption request was anticipated.

The GROW meeting at IETF 119 concluded with a clear direction for future work, emphasizing the importance of security, automation, and efficient monitoring in global routing operations. The discussions and outcomes from this meeting are expected to significantly contribute to the advancement of routing technologies and practices.

# 31 HTTPAPI Working Group (HTTPAPI)

The HTTPAPI Working Group (WG) meeting brought together representatives from various prominent companies and institutions in the field of web technologies. A total of 22 attendees were present, including individuals from Microsoft, Akamai, PCCW Global, APNIC, Comcast, NTT, Cloudflare, Google, Fastly, Ericsson, AFRINIC, Deutsche Telekom, Vodafone, and Apple.

### 31.0.1 Attendees Overview

The meeting was attended by key stakeholders from the industry, with a notable presence from Darrel Miller (Microsoft), Rich Salz (Akamai), and Mark Nottingham (Cloudflare). The diverse attendance underscored the WG's commitment to broad-based collaboration in the development of HTTP APIs.

The discussions centered around the progress of various IETF draft documents and the strategic direction for the WG's work. The group reviewed the status of documents such as the newly published YAML media type RFC 9512 and the ongoing work on rate limit headers. The dialogue was enriched with references to IETF draft documents, which can be accessed via hyperlinks like [draft-ietf-httpapi-yaml-mediatype](draft-ietf-httpapi-yaml-mediatype).

Meeting materials can be found directly at [HTTPAPI WG Materials](HTTPAPI WG Materials).

## 31.1 Meeting Discussions

### 31.1.1 Document Progress

The YAML media type has achieved the status of RFC 9512, marking a significant milestone for the WG.

### 31.1.2 Authentication Links

A review from the Security Directorate (SECDIR) has been requested for the Authentication Links draft to ensure robust security practices.

### 31.1.3 Link Hints

The Link Hints draft requires further exploration, particularly in the context of serialization into HTML and compatibility with HTTP structured fields.

### 31.1.4 MediaTypes

The MediaTypes draft has been updated with a Security Considerations document. There is a proposal to include application/openapi, with ongoing contributions from the community.

### 31.1.5 Idempotency

The Idempotency draft encountered some issues during the Working Group Last Call (WGLC) and is undergoing further discussion and revision. The WG encourages review of the ongoing discussions at [GitHub - Idempotency Issues](GitHub - Idempotency Issues).

### 31.1.6 ByteRange

The ByteRange draft has seen a new version with significant updates, including new syntax for content offset and the ability to overwrite regions. The WG is considering limiting the scope to publishing the media type and deferring other use cases.

### 31.1.7 Deprecation

There has been progress on the Deprecation draft, with a new version addressing previous concerns about time format. The WG chairs are set to review and potentially initiate a WGLC.

The meeting concluded with a recognition of the productive discussions and the anticipation of further advancements in the HTTPAPI space.

# 32 HTTPBIS

## 32.1 Attendees Overview

### 32.1.1 Attendance Summary

The meeting was attended by representatives from prominent companies such as Apple, Google, and Microsoft, as well as institutions like the University of Auckland. The total attendance was recorded at 58 individuals.

## 32.2 Main Points and Discussions

The Working Group (WG) meeting featured a robust discussion on several key topics, including the advancement of draft proposals and the exploration of new ideas. The dialogue was rich with technical insights, contextualizing the importance of each draft within the broader scope of internet standards and their implementation.

Meeting materials are available directly via the [WG materials page](#).

## 32.3 Meeting Discussions

### 32.3.1 Cookies Draft

The discussion on the Cookies draft highlighted a reversion in the specification due to unforeseen dependencies by websites on a particular bug. The group is nearing a consensus for a Working Group Last Call, with a few minor issues remaining.

### 32.3.2 Unprompted Authentication

The Unprompted Authentication draft has seen progress with multiple open-source implementations and a successful interoperability test. A security analysis has been conducted, and the group is considering renaming the authentication scheme to better reflect its benefits to users.

### 32.3.3 Resumable Uploads

The Resumable Uploads presentation sparked a conversation on defining a media type for PATCH requests and how clients can discover server limits. The group discussed the semantics of interrupted PATCH requests and handling content encoding during resumptions.

### 32.3.4 Compression Dictionary Transport

The Compression Dictionary Transport draft raised questions about the practicality of pattern matching on the client side and potential performance concerns. The group acknowledged the need for further discussion on this topic, particularly regarding the implications for client-side processing.

The meeting concluded with a sense of forward momentum, as the WG members recognized the potential impact of their discussions on the future of internet protocols. The next steps involve refining the drafts based on feedback and moving towards formalizing the proposals, reflecting their anticipated contribution to the field.

# 33 Internet Congestion Control Research Group (ICCRG)

## 33.1 Attendees Overview

### 33.1.1 Participation Summary

The meeting was attended by representatives from prominent companies and institutions such as Netflix, Apple Inc., Huawei, Cisco, and Mozilla, with a total attendance of 47 participants.

The discussions at the ICCRG meeting were insightful and covered a range of topics relevant to the evolution of internet congestion control. The presentations and debates revolved around the latest research and proposals, including updates on congestion control algorithms and evaluation suites. The dialogue was enriched by contributions from various stakeholders, providing a comprehensive view of the current state and future directions of congestion control mechanisms. Key documents, such as IETF draft proposals, were referenced throughout the discussions, providing a foundation for the technical discourse. For instance, the draft-ietf-iccrg-tcpeval document was a focal point for evaluating TCP performance.

Meeting materials are available through the direct link: ICCRG Meeting Materials.

## 33.2 Meeting Discussions

### 33.2.1 Revisiting Common TCP Evaluation Suite

David Hayes presented the "Revisiting Common TCP Evaluation Suite," which sparked a dialogue on the need for a more generic and inclusive approach to congestion control evaluation. The discussion highlighted the importance of considering a variety of traffic patterns, including video-like and inelastic traffic, to ensure a comprehensive assessment of new congestion control algorithms.

### 33.2.2 Implicit Congestion Notification

Zili Meng's presentation on "Implicit Congestion Notification" led to an exchange on the practical implications of delaying acknowledgments and its compatibility with different congestion control protocols. The potential integration with ECN and the challenges posed by specific Wi-Fi congestion behaviors were also explored.

### 33.2.3 rLEDBAT Update

The update on rLEDBAT by Vidhi Goel prompted a review of the draft's readiness for progression, with a consensus on the need for a final review before proceeding to the last call.

The ICCRG meeting concluded with a clear direction for future work, emphasizing the importance of refining evaluation tools and methodologies to keep pace with the evolving landscape of congestion control. The next steps involve community engagement to further develop these tools and a commitment to reviewing and advancing the rLEDBAT draft.

# 34 Information-Centric Networking Research Group (ICNRG)

## 34.1 Attendees Overview

### 34.1.1 Prominent Attendees and Total Attendance

The ICNRG meeting was attended by representatives from prominent institutions such as MIT Media Lab, HKUST, SRI/PARC, TUM, NICT, and UCLA, with a total of 27 attendees.

### 34.1.2 Main Points and Contextualization

The meeting discussions revolved around the evolution of transactions in the context of Information-Centric Networking (ICN). The group examined various approaches to achieving transactions with ACID properties, contrasting traditional client-server systems with ICN's network layer and shared data structures. Key presentations included topics on Secure Web Objects, Transaction Manifests, and the Vanadium framework for secure, distributed applications. The implications of these discussions were significant, suggesting a shift towards more secure and efficient transactional models in ICN, with references to IETF draft documents such as draft-ietf-icnrg-secure-web-objects.

Meeting materials are available directly via the IETF materials repository.

## 34.2 Meeting Discussions

### 34.2.1 Secure Web Objects and Transactions

Dirk Kutscher's presentation highlighted the transition from the original web design to today's data-oriented web, emphasizing the need for ICN components to support real-time media and transactions. The discussion touched upon the use of Reflexive Forwarding for transactions, raising questions about authentication and authorization within ICN.

### 34.2.2 Transaction Manifests

Marc Mosko introduced the concept of Transaction Manifests (TMs) as a means to map distributed ledger-style transactions to ICN. The presentation sparked a debate on the challenges of multi-bookkeeper systems, including livelocks and deadlocks, and the necessity of total order for resolution.

### 34.2.3 Vanadium: Secure, Distributed Applications

The Vanadium framework was discussed as a potential model for ICN, with its secure, distributed RPC system and dynamic access control mechanisms. The comparison with NDN trust schema and the adaptability of Vanadium's security features to ICN were key points of interest.

### 34.2.4 Global vs. Scoped Namespaces

The conversation on namespaces led by Marc Mosko delved into the trust models of IPFS and CCNX, discussing the implications of scoped versus global namespaces. The dialogue with Dave Oran explored the concept of growing namespaces from the bottom up and the computational challenges of comparing names within relative naming systems.

### 34.2.5 Conclusions and Next Steps

The meeting concluded with a look towards future ICNRG meetings and a call for ICN submissions to ICNP. The discussions underscored the importance of continued research into secure and efficient transactional models within ICN, with a focus on the practical implementation and analysis of the proposed concepts.

# 35 Inter-Domain Routing Working Group (IDR)

## 35.1 Attendees Overview

### 35.1.1 Prominent Companies and Total Attendance

The IDR meeting at IETF 119 saw a total of 136 attendees, including representatives from prominent companies and institutions such as Verisign Inc, Cisco, Juniper Networks, Huawei, Nokia, and China Telecom.

The discussions during the meeting were rich and varied, covering a range of topics from BGP extensions for Segment Routing to Flow Specification rules for Advanced Packet Network (APN). The meeting materials can be found [here](here).

## 35.2 Meeting Discussions

### 35.2.1 Generic Metric for the AIGP Attribute

Srihari Sangli presented the [draft-ssangli-idr-bgp-generic-metric-aigp-07](draft-ssangli-idr-bgp-generic-metric-aigp-07), which proposes a generic metric for the AIGP attribute in BGP. The discussion highlighted the need for a brief question due to time constraints, with no questions raised in the room.

### 35.2.2 MP-BGP Extension for IPv4/IPv6 Mapping Advertisement

The [draft-ietf-idr-mpbgp-extension-4map6-01](draft-ietf-idr-mpbgp-extension-4map6-01) by Xing Li and Chongfeng Xie was discussed, with questions about the draft's relevance to IDR versus BESS. The consensus was to continue the discussion on the mailing list.

### 35.2.3 Advertising SID Algorithm Information in BGP

Yao Liu's presentation on [draft-peng-idr-segment-routing-te-policy-attr-08](draft-peng-idr-segment-routing-te-policy-attr-08) led to a query about implementation plans for Segment Types extensions, with the presenter indicating ongoing implementations in the SR Policy WG documents.

### 35.2.4 Segment Routing BGP Egress Peer Engineering over Layer 2 Bundle

Mengxiao Chen's [draft-lin-idr-sr-epe-over-l2bundle-04](draft-lin-idr-sr-epe-over-l2bundle-04) sparked a debate on the addressability of L2 bundle members and the implications for forwarding behavior. The discussion underscored the need for clarity on the use of Adj-SIDs and PeerAdj SIDs in the context of BGP Egress Peer Engineering (EPE).

### 35.2.5 BGP Extensions of SR Policy for Headend Behavior

The [draft-lin-idr-sr-policy-headend-behavior-03](draft-lin-idr-sr-policy-headend-behavior-03) was presented without comments, suggesting a straightforward reception of the proposed extensions.

### 35.2.6 SR Policies Extensions for NRP in BGP-LS

Ran Chen's [draft-chen-idr-bgp-ls-sr-policy-nrp-05](draft-chen-idr-bgp-ls-sr-policy-nrp-05) led to discussions on the applicability of SR Policy with Network Resource Partitioning (NRP) and the need for coordination with the TEAS WG.

### 35.2.7 BGP Link State Extensions for Scalable NRP

Jie Dong's [draft-dong-idr-bgp-ls-scalable-nrp-00](draft-dong-idr-bgp-ls-scalable-nrp-00) presentation highlighted the challenges of disseminating NRP information and the relationship with Flex-Algo for path computation.

### 35.2.8 Dissemination of BGP Flow Specification Rules for APN

Shuping Peng's [draft-peng-idr-apn-bgp-flowspec-00](draft-peng-idr-apn-bgp-flowspec-00) raised concerns about the timing of protocol extensions for APN, given the ongoing discussions about APN's definition within the IETF.

### 35.2.9 FC-BGP Protocol Specification

Zhuotao Liu's draft-sidrops-wang-fcbgp-protocol-00 led to a debate on the security equivalence with BGPSec, the impact on routing infrastructure, and the need for strategic deployment.

### 35.2.10 Destination-IP-Community Filter for BGP Flow Specification

Tianhao Wu's draft-wu-idr-flowspec-dip-community-filter-00 presentation prompted a discussion on the feasibility of filtering based on BGP community in the forwarding plane and the potential scalability issues.

### 35.2.11 BGP Flow Specification for Source Address Validation

Lastly, Nan Geng's draft-geng-idr-flowspec-sav-03 presentation on source address validation using BGP Flow Specification sparked a conversation on the generation of rules and the integration with VRF and Flowspec modes.

The meeting concluded with an emphasis on the importance of these discussions for the evolution of BGP and the anticipation of further contributions to the field.

# 36 Internet Area Working Group (IntArea)

## 36.1 Attendee Overview

### 36.1.1 Attendance Summary

The Internet Area Working Group (IntArea) meeting at IETF 119 saw a robust attendance of 75 participants, including representatives from prominent companies and institutions such as Cisco, Ericsson, Google, Apple, Huawei, and several universities and research organizations.

The discussions were rich and covered a range of topics from proxy configurations in provisioning domains to probing interfaces and extending ICMP for system identification. The meeting materials can be accessed via this [link](link).

## 36.2 Meeting Discussions

### 36.2.1 Communicating Proxy Configurations in Provisioning Domains

Tommy Pauly presented the [draft-pauly-intarea-proxy-config-pvd-02](draft-pauly-intarea-proxy-config-pvd-02), which discusses methods for communicating proxy configurations within provisioning domains. The presentation sparked a vote for adoption, resulting in a strong consensus to move forward.

### 36.2.2 Probe: A Utility for Probing Interfaces

Bill Fenner introduced [draft-fenner-intarea-probe-clarification-00](draft-fenner-intarea-probe-clarification-00), aimed at clarifying the use of the Probe utility for probing interfaces. The audience engaged with questions about the draft's readability and the identification of inconsistencies, indicating a healthy interest and potential for further refinement.

### 36.2.3 Extending ICMP for System Identification

The session continued with Bill Fenner's brief on [draft-fenner-intarea-extended-icmp-hostid-00](draft-fenner-intarea-extended-icmp-hostid-00), which proposes an extension to ICMP for better system identification. The presentation was concise, setting the stage for future discussions on its implications and adoption.

### 36.2.4 A SAVI Solution for WLAN

Lin He discussed [draft-bi-intarea-savi-wlan-02](draft-bi-intarea-savi-wlan-02), focusing on a Source Address Validation Improvement (SAVI) solution for WLAN. The conversation highlighted alternatives and existing solutions, such as RFC6620, and addressed concerns regarding MAC address translation and DHCPv6 prefixes. The feedback received was constructive, suggesting a path for draft refinement.

The meeting concluded with a forward-looking perspective, emphasizing the importance of addressing current gaps in wireless LAN security and configuration, and the potential for IntArea's work to contribute significantly to the evolution of internet protocols and practices.

# 37 Internet of Things Operations (IOTOPS)

## 37.1 Attendee Overview

### 37.1.1 Attendance Summary

The IOTOPS session at IETF 119 in Brisbane was attended by a diverse group of 38 participants, representing prominent companies and institutions such as Isode Limited, RISE Research Institutes of Sweden, High North Inc, Itron, ALAXALA Networks, Corp., Arm, Ericsson, and Google. The attendance reflected a wide interest in the operational aspects of IoT from various sectors of the industry.

The discussions during the session were focused on the operational challenges and the security aspects of IoT. Notably, the session included presentations on the terminology for constrained-node networks, a comparison of CoAP security protocols, and an update on the IoT operational security summary. These discussions were enriched by contributions from attendees, which provided insights into the practical implications of the topics at hand. For instance, the terminology presentation by Carsten Bormann (draft-bormann-iotops-ietf-lwig-7228bis-00) aimed to update the classification of constrained-node networks, reflecting the evolution of technology since the original RFC 7228.

Meeting materials are available via the direct link: IOTOPS Session Notes.

## 37.2 Meeting Discussions

### 37.2.1 Terminology for Constrained-Node Networks

Carsten Bormann presented the revised terminology for constrained-node networks, which is crucial for understanding and operating IoT systems. The updated document aims to reflect the technological advancements and provide a common language for IoT operations. The group discussed the potential need for a charter update to accommodate the terminology draft, with a consensus that the terms are indeed part of IoT operations.

### 37.2.2 Comparison of CoAP Security Protocols

John Mattsson presented a comparison of CoAP security protocols (draft-ietf-iotops-security-protocol-comparison-04), summarizing the recent changes and suggesting that the document is stable enough for publication. The group agreed to initiate a Working Group Last Call (WGLC) while parking the document for normative references.

### 37.2.3 IOTOPS Security Summary Update

Brendan Moran provided an update on the IoT operational security summary (draft-ietf-iotops-security-summary-01), which references baseline security documents and technologies relevant to IoT. The discussion highlighted the importance of considering regional regulatory requirements, such as the EU Cyber Resilience Act, and the potential need for additional authors to address these aspects.

### 37.2.4 IoT Operational Issues

Karsten Walther shared practical operational issues encountered in IoT deployments, emphasizing the need for standardization to address common problems. The presentation sparked a discussion on the role of browsers in IoT and the challenges of network configuration in virtualized environments. The group considered the value of documenting these experiences in an Internet-Draft to facilitate broader discussions and potential solutions within the IETF community.

The session concluded with a commitment to further explore the presented issues, with the potential for new work items to emerge from these discussions. The next steps include initiating a WGLC for the security protocol comparison, incorporating regulatory considerations into the security summary, and potentially drafting a document to capture the operational issues presented.

# 38 Internet Protocol Performance Metrics (IPPM)

## 38.1 Attendees Overview

### 38.1.1 Prominent Companies and Total Attendance

The meeting was attended by representatives from notable companies and institutions such as Apple, Google, Huawei, Cisco, and Ericsson, with a total of 56 attendees.

### 38.1.2 Summary and Context

The discussions focused on a range of topics including IPv6 Performance and Diagnostic Metrics, Quality of Outcome, Alternate Marking Deployment Framework, and Integrity of In-situ OAM Data Fields. The debates were rich with technical insights and contextualized by the need for improved network performance measurements. Several draft-ietf-ippm-encrypted-pdmv2 documents were referenced to support the discussions.

Meeting materials can be found at the following link.

## 38.2 Meeting Discussions

### 38.2.1 IPv6 Performance and Diagnostic Metrics v2 (PDMv2) Destination Option

Michael Ackermann presented an update on the draft-elkins-ippm-encrypted-pdmv2, highlighting its current status and potential implications for network diagnostics.

### 38.2.2 Quality of Outcome

Bjørn Ivar Teigen reviewed the draft-olden-ippm-qoo, proposing additional metrics for packet loss to better represent application requirements. The discussion raised questions about the linearity of the "perfect" to "unusable" scale and how different applications adapt to network conditions.

### 38.2.3 Alternate Marking Deployment Framework

Giuseppe Fioccola presented the draft-fz-ippm-alt-mark-deployment, which did not elicit comments from the attendees, suggesting a consensus on its content.

### 38.2.4 Integrity of In-situ OAM Data Fields

Justin Iurman's presentation on the draft-ietf-ippm-ioam-data-integrity sparked a debate on the necessity of different IOAM trace options and the potential impact of IPSEC on measurement accuracy.

### 38.2.5 Next Steps

The meeting concluded with a consensus to focus on Option 2 for IOAM integrity and to consider the integration of a new STAMP TLV for reflecting IPv6 headers. The discussions suggested a shift towards more precise and less intrusive measurement techniques, with a call for further feedback and implementation experiences.

# 39 Internet Research Task Force (IRTF)

## 39.1 Attendee Overview

### 39.1.1 Attendance Summary

The meeting was attended by representatives from prominent companies and institutions such as Google, Huawei, Nokia, and MITRE, with a total attendance of 85 participants.

The discussions at the IRTF Open Meeting centered around the latest developments in applied networking research. The highlight of the meeting was the presentation of the Applied Networking Research Prize (ANRP), which serves to acknowledge significant contributions to the field. The discourse provided insights into the current trends and future directions of internet standards and technologies.

Meeting materials and recordings can be found at IRTF ANRP.

## 39.2 Meeting Discussions

### 39.2.1 Applied Networking Research Prize (ANRP) Presentation

The ANRP presentation by Dongqi Han focused on anomaly detection in network systems. The research presented is expected to have a significant impact on the field of network security and anomaly detection methodologies. The discussion that followed the presentation highlighted the importance of accurate data labeling and the challenges associated with training anomaly detection systems. The potential for collaboration with the IETF and IRTF communities was also explored, particularly in terms of facilitating research through the provision of test data.

### 39.2.2 Main Points and Dialogues

The dialogue following the presentation delved into the practical aspects of implementing anomaly detection systems, such as the level of expertise required for data labeling and the robustness of the systems against inaccuracies in training data. The conversation underscored the need for collaboration between researchers and the broader internet standards community to enhance the effectiveness of such systems.

### 39.2.3 Conclusions and Next Steps

The meeting concluded with an emphasis on the importance of continued research in the field of applied networking and the potential benefits of increased interaction between the research community and internet standards bodies. The call for papers for the upcoming ANRW 2024 was highlighted as an opportunity for researchers to contribute and engage with the community.

# 40 Inventory and Visibility Working Group (IVY)

## 40.1 Attendees Overview

### 40.1.1 Participation Summary

The Inventory and Visibility Working Group (IVY) meeting saw participation from a diverse set of companies and institutions, including Huawei Technologies, Cisco Systems, Ericsson, Juniper Networks, and Deutsche Telekom, among others. A total of 48 attendees were present, indicating a strong interest in the topics discussed.

### 40.1.2 Meeting Context

The meeting discussions revolved around the development of YANG data models for network inventory management and related operations. The dialogue was rich with technical insights and highlighted the need for clear definitions of scope and terminology within the working group's charter. Attendees engaged in debates on the inclusion of physical versus virtual network elements, the delineation between inventory and capabilities, and the importance of planning and lifecycle management within the inventory models. Hyperlinks to relevant IETF draft documents were provided, allowing for a deeper understanding of the proposals and ongoing work.

Meeting materials can be accessed directly via the following link: [IVY Session Materials](IVY Session Materials).

## 40.2 Meeting Discussions

### 40.2.1 Introduction and Working Group Status

The chairs provided an introduction to the session and an update on the working group's status, setting the stage for the presentations and discussions that followed.

### 40.2.2 A Word from the New AD

Mahesh Jethanandani, the new Area Director, expressed gratitude for the support of the working group and discussed the liaison with the Broadband Forum (BBF) and the potential overlap between IVY and BBF work.

### 40.2.3 A YANG Data Model for Network Inventory

Chaode Yu presented the [draft-ietf-ivy-network-inventory-yang-01](draft-ietf-ivy-network-inventory-yang-01), which sparked discussions on the inclusion of component-specific parameters and the distinction between network-wide and domain-specific inventory models.

### 40.2.4 IVY Boundary

Nigel Davis's presentation on the IVY boundary led to a debate on what should be considered in scope for the base model and the importance of flexibility in defining the scope.

### 40.2.5 Inventory Terminology

Italo Busi's presentation on inventory terminology highlighted the need for clear definitions and the consideration of planning use cases within the inventory models.

### 40.2.6 A Network Inventory Location Model

Bo Wu discussed the [draft-wbbpb-ivy-network-inventory-location-00](draft-wbbpb-ivy-network-inventory-location-00), emphasizing the importance of location information in inventory models and the potential for future augmentation to address additional cases.

### 40.2.7 Asset Lifecycle Management and Operations

Diego Lopez presented problem statements and a YANG data model for asset lifecycle management and operations, touching upon the relationship between features, capabilities, and inventory.

### 40.2.8 A YANG Data Model for Energy Saving Management

Qin Wu's presentation on energy saving management led to a discussion on the relationship between this work and other power management initiatives, with the suggestion to defer certain topics to a newly forming working group focused on power management.

The meeting concluded with an open discussion, where attendees reflected on the presentations and considered the next steps for the working group. The discussions suggested a shift towards a more comprehensive and modular approach to inventory management, with a focus on defining clear boundaries and ensuring that the models developed can be augmented to address a wide range of use cases. The potential for rechartering to include additional work on extensions and augmentations was also acknowledged.

# 41 JSON Mail Access Protocol (JMAP)

## 41.1 Attendees Overview

### 41.1.1 Overview

The meeting was attended by representatives from prominent companies and institutions such as Fastmail, Meta Platforms, Inc., and the NSA Center for Cybersecurity Standards, with a total attendance of 26 participants.

The discussions revolved around the progress of various Internet Engineering Task Force (IETF) draft documents and the strategic direction for the working group's technical contributions. Key dialogues included the advancement of JMAP-related drafts, such as [draft-ietf-jmap-sieve](draft-ietf-jmap-sieve) and [draft-ietf-jmap-calendars](draft-ietf-jmap-calendars), as well as discussions on IMAP extensions and the potential rechartering of the working group. The meeting materials can be accessed directly via [meeting materials link](meeting materials link).

## 41.2 Meeting Discussions

### 41.2.1 JMAP Presentations and Debates

The JMAP session included presentations on drafts currently with the IESG, such as [draft-ietf-jmap-sieve](draft-ietf-jmap-sieve), which is awaiting updates based on IETF last-call feedback. Discussions also covered drafts in the working group's last call, like [draft-ietf-jmap-calendars](draft-ietf-jmap-calendars), with a proposal to rename "scheduleId" to "calendarUserAddress" to remove dependencies. The session concluded with a review of JMAP milestones and an agreement on the need for a draft on JMAP S/MIME extensions.

### 41.2.2 IMAP Extensions and Rechartering

The EXTRA session focused on IMAP extensions, including [draft-ietf-imap-jmapaccess](draft-ietf-imap-jmapaccess) and [draft-ietf-imap-messagelimit](draft-ietf-imap-messagelimit), with discussions on the necessity of capabilities and handling of message limits. The potential rechartering of the working group was also debated, with a call to review the proposed charter and consider the formation of a new group focused on mail maintenance.

### 41.2.3 Outcomes and Next Steps

The meeting underscored the importance of refining drafts based on community feedback and the need for clear guidelines on IMAP extensions. The discussions suggested a shift towards a more streamlined approach to JMAP and IMAP specifications, with an emphasis on interoperability and efficiency. The next steps involve updating drafts, finalizing reviews, and progressing documents to the next stages of approval, reflecting the working group's commitment to enhancing email protocols.

# 42 Lightweight Authenticated Key Exchange (LAKE)

The LAKE working group meeting at IETF 119 featured a diverse attendance, including representatives from prominent companies and institutions such as Ericsson, Huawei, Nokia, and the University of Luxembourg, with a total of 33 attendees.

The discussions focused on the latest advancements in the Lightweight Authenticated Key Exchange protocols, with particular attention to authorization, application profiles, and remote attestation. The group reviewed several [draft-ietf-lake-authz](draft-ietf-lake-authz) documents, highlighting the need for efficient and secure key exchange mechanisms in constrained environments. The meeting materials are available [here](here).

## 42.1 Meeting Discussions

### 42.1.1 Authorization

Geovane Fedrecheski presented updates on the authorization draft, emphasizing the use of OPAQUE_INFO for communication between devices and gateways. The discussion raised privacy concerns regarding the collection of nearby identifiers by devices, which the group agreed to explore further.

### 42.1.2 Application Profiles

Marco Tiloca discussed the application profiles draft, which aims to provide a canonical description of EDHOC application profiles using CBOR maps. The group debated whether to allow just-identifiers or all parameters in the profiles, with the consensus leaning towards a more granular approach.

### 42.1.3 Remote Attestation

Yuxuan Song introduced the remote attestation draft, focusing on the background check model where a gateway acts as the relying party. The group acknowledged the need to include the verifier role in the model and discussed the importance of the evidence being verified.

### 42.1.4 PSK-based Authentication

John Preuß Mattsson recapped the discussions on EDHOC with PSK-based authentication, outlining the requirements for resumption and the challenges associated with privacy when using external PSK identifiers. The group expressed interest in a new draft that would address these issues.

### 42.1.5 IoTDisco Presentation

Johann Großschädl presented IoTDisco, a framework for end-to-end security in constrained IoT environments. The group showed interest in the potential for post-quantum versions and discussed the comparison with existing protocols like EDHOC.

The LAKE working group's discussions suggested a shift towards more granular and flexible key exchange mechanisms, with a focus on privacy and efficiency. The next steps include further refinement of the drafts and consideration of the privacy implications of the protocols.

# 43 Link State Routing (LSR)

## 43.1 Attendees Overview

### 43.1.1 Prominent Participants and Attendance

The meeting was attended by representatives from major companies and institutions, including Cisco, Nokia, Juniper Networks, China Telecom, and Huawei Technologies, with a total attendance of 49 participants.

The discussions were rich with technical insights and revolved around the latest Internet Engineering Task Force (IETF) draft documents. The attendees engaged in debates over the technical direction and strategy of the LSR working group, with an emphasis on the evolution of routing protocols and their implementation. Key draft documents were referenced, providing a deeper context to the discussions. For example, the draft-ietf-lsr-multi-tlv was a focal point for deliberation on extending routing protocol capabilities.

Meeting materials can be accessed directly via the LSR session materials link.

## 43.2 Meeting Discussions

### 43.2.1 Multi-part TLVs in IS-IS

The presentation on Multi-part TLVs in IS-IS sparked a discussion on the readiness of the document for Last Call, with considerations on prioritizing it alongside the IS-IS and OSPF SR YANG Models. The consensus was to move forward with the document, reflecting a strategic shift towards enhancing protocol extensibility.

### 43.2.2 YANG Model for IS-IS PICS

The YANG Model for IS-IS PICS was proposed for adoption, indicating a move towards standardizing protocol implementation conformity statements through YANG models. The working group agreed to adopt the draft and use it as a learning opportunity for future model development.

### 43.2.3 Intra-domain SAVNET Method

The Intra-domain SAVNET method presentation led to a debate on the necessity of protocol extensions for intra-area prefix advertisement. The discussion concluded that existing mechanisms like tagging or local configuration filtering could suffice, suggesting a preference for leveraging existing protocol features over introducing new extensions.

### 43.2.4 Signaling Aggregate Header Size Limit via IGP

The proposal to signal the Aggregate Header Size Limit via IGP was met with skepticism, particularly regarding the appropriateness of using the IGP for such signaling. The conversation highlighted the complexity of the problem and the need for further discussion in the 6MAN working group, reflecting a cautious approach to extending IGP responsibilities.

### 43.2.5 Advertisement of Remote Interface Identifiers for Layer 2 Bundle Members

The discussion on advertising remote interface identifiers for Layer 2 bundle members highlighted the need for additional use cases beyond controller discovery. This suggests a strategic emphasis on broadening the applicability of protocol extensions to ensure their utility in diverse operational scenarios.

### 43.2.6 IS-IS and OSPF Extensions for TVR (Time-Variant Routing)

The presentation on IS-IS and OSPF extensions for TVR led to a critical examination of the necessity and implications of distributing schedule information via the IGP. The working group's feedback indicated a preference for alternative methods like YANG over IGP extensions for conveying such information.

### 43.2.7 IGP Color-Aware Shortcut

The IGP Color-Aware Shortcut proposal was discussed with reference to existing mechanisms like prefix tags and SR policies. The consensus was that the headend decision for shortcuts is a local matter and does not require additional specifications in the IGP, suggesting a strategic preference for simplicity and existing solutions.

### 43.2.8 Using Flex-Algo for Segment Routing (SR) based Network Resource Partition (NRP)

The conversation on using Flex-Algo for SR-based NRP highlighted the need for consensus on several issues before proceeding with the draft. The discussion on metric types and the use of bundles for NRP separation indicated a need for clarity on the technical approach and its implications for network operations.

### 43.2.9 IGP Flexible Algorithm with Link Loss

The idea of using link-loss as a flex-algorithm constraint was generally supported, provided there is consistency across the IGP routing domain for link-loss measurement. This reflects a strategic interest in enhancing routing algorithms with performance-based metrics.

# 44 Mobile Ad-hoc Networks (MANET)

## 44.1 Attendee Overview

### 44.1.1 Participation Summary

The meeting was attended by representatives from prominent companies and institutions such as Futurewei Technologies, TNO, Keio University, University of Tripoli, Itron, China Mobile, Universitat Politecnica de Catalunya, and others, totaling 22 attendees.

### 44.1.2 Main Points and Dialogues

The discussions were centered around the status of Dynamic Link Exchange Protocol (DLEP) drafts, the potential adoption of new work items, and the presentation on Ad Hoc On-demand Distance Vector Version 2 (AODVv2) Routing. The dialogue contextualized the evolution of AODVv2 and its differences from the original AODV, with emphasis on the implementability and potential enhancements of the protocol. Key IETF draft documents were referenced, such as draft-ietf-manet-dlep-credit-flow-control and draft-perkins-manet-aodvv2.

Meeting materials are available directly via IETF MANET WG Materials.

## 44.2 Meeting Discussions

### 44.2.1 AODVv2 Presentation

Charlie Perkins presented the latest updates on AODVv2, highlighting the protocol's history, recent changes, and areas for improvement. The presentation sparked discussions on the protocol's implementability and the possibility of incorporating asymmetric routing and new optimization techniques.

### 44.2.2 WG Charter Discussion

The working group discussed the charter, considering the adoption of new work items and the future direction of the group. The debate focused on the importance of having a clear plan and milestones for rechartering, driven by active participation and draft support from the members.

### 44.2.3 Outcomes and Next Steps

The meeting concluded with a consensus on the need for active mailing list discussions to shape the new charter. The group recognized the importance of volunteer work for the adoption of new items and the advancement of the working group's objectives. The potential impact of these discussions suggests a strategic shift towards more rigorous protocol development and a commitment to addressing implementation challenges.

# 45 MASQUE Working Group (MASQUE)

## 45.1 Attendees Overview

### 45.1.1 Prominent Participants and Total Attendance

The MASQUE Working Group meeting saw participation from key industry players, including representatives from Apple, Google, Cisco, and Mozilla, with a total of 108 attendees.

The discussions were rich with technical insights and revolved around the evolution of the MASQUE protocol suite. The meeting featured a blend of presentations and debates, with a focus on the implications of the proposed changes on the broader internet architecture. Notably, the discourse included references to several IETF draft documents, such as draft-ietf-masque-quic-proxy, which were central to the conversation.

Meeting materials are available through the direct link: IETF 119 Materials.

## 45.2 Meeting Discussions

### 45.2.1 QUIC-Aware Proxying Using HTTP

The presentation by Tommy Pauly and Eric Rosenberg highlighted the latest developments in QUIC-aware proxying. The discussion emphasized the importance of addressing traffic analysis concerns and introduced a "scramble" transform to prevent simple packet matching. The group considered the implications of allowing transforms to map multiple packets and the potential complexity this could introduce.

### 45.2.2 Proxying Listener UDP in HTTP

Abhijit Singh's presentation on Proxying Listener UDP in HTTP sparked a debate on the necessity of accommodating IPv6 prefixes and the handling of privacy addresses. The group discussed the merits of structured fields for headers and the introduction of new capsule types for compression management.

### 45.2.3 Proxying Ethernet in HTTP

Alejandro Sedeño's session on Proxying Ethernet in HTTP delved into the challenges associated with MTU and fragmentation. The consensus leaned towards avoiding fragmentation at the MASQUE layer, with suggestions to explore short-lived streams as an alternative. The group also debated the handling of MAC addresses and VLAN tags, opting to defer these issues until more use cases emerge.

### 45.2.4 In-Band DNS Information Exchange

David Schinazi's brief on In-Band DNS Information Exchange raised questions about the necessity of introducing another DNS transport, with suggestions to utilize existing mechanisms like DoH or tunnel DHCP for DNS configuration. The group recognized the need to balance the simplicity of the current design with the potential benefits of integrating additional DNS configuration methods.

The MASQUE Working Group's discussions signaled a cautious but forward-thinking approach to protocol development. The outcomes of the meeting suggest a commitment to refining the MASQUE protocols while carefully considering their interoperability with existing internet standards. The next steps involve further implementation and testing to validate the proposed changes and their anticipated contributions to the field.

# 46 Messaging Infrastructure Modernization Initiative (MIMI)

## 46.1 Attendees Overview

### 46.1.1 Attendance Summary

The MIMI working group sessions were attended by a diverse group of 45 participants, representing prominent companies and institutions such as Google, Cisco, The Matrix.org Foundation C.I.C., ACLU, Ericsson, and Nokia. The attendance showcased a strong interest from both the industry and academia in the modernization of messaging infrastructure.

### 46.1.2 Main Points and Contextualization

The discussions revolved around the intricacies of user discovery requirements, content format, and the MIMI architecture. Key dialogues focused on the necessity of provider discovery, the debate over the inclusion of emails as secondary service identifiers (SSIs), and the structure of the proposed messaging protocol. The group examined various IETF draft documents, such as the [draft-rosenberg-mimi-discovery-reqs](draft-rosenberg-mimi-discovery-reqs) and [draft-barnes-mimi-arch](draft-barnes-mimi-arch), to refine the technical direction of the initiative.

Meeting materials can be accessed directly via the [IETF Meeting Materials page](IETF Meeting Materials page).

## 46.2 Meeting Discussions

### 46.2.1 User Discovery Requirements

Jon Peterson's presentation highlighted the challenges in user discovery and the potential solutions. The group discussed the balance between using SSIs and the need for discovery, with a consensus forming around the idea that discovery might be unnecessary if an SSI is available. The conversation also touched upon the longevity of crypto information and the importance of ensuring SSI integrity.

### 46.2.2 Content Format

Rohan Mahy led a session on the proposed changes to the content format, sparking a debate on message ordering and the risks associated with external attachments. The group was divided on whether to use TLS presentation language or CBOR for encoding, with no clear consensus reached. The importance of securing message sender identity was also emphasized.

### 46.2.3 MIMI Architecture and Protocol

Richard Barnes presented the MIMI architecture, proposing a central hub server model for hosting rooms and routing messages. The group discussed the distinction between users and devices in the context of participation and the need for consistency in MLS group state and authorization policies. Privacy considerations, such as the use of pseudonyms, were acknowledged as a complex challenge for the protocol.

The sessions concluded with a vote to adopt the current drafts as working group documents, marking a significant step forward in the MIMI initiative. The discussions suggested a shift towards a more streamlined and secure messaging infrastructure, with the next steps focusing on refining the proposed protocols and addressing the technical challenges identified.

# 47 Machine Learning for Audio Coding (MLCODEC)

## 47.1 Attendee Overview

### 47.1.1 Attendance Summary

The MLCodec Working Group meeting at IETF 119 in Brisbane, Australia, saw a total of 33 attendees. Among them were representatives from prominent companies and institutions such as Cisco, Xiph.org, University of Aberdeen, Google, Ericsson, AWS, and Samsung Electronics.

The discussions were rich with technical insights and revolved around the latest IETF draft documents. The group engaged in debates on the extension mechanism for the Opus audio codec, the Deep REDundancy (DRED) draft, and enhancements in speech coding. These conversations were not only technical but also strategic, as they considered the future direction of audio coding in the context of machine learning.

Meeting materials are available through the direct link: draft-ietf-mlcodec.

## 47.2 Meeting Discussions

### 47.2.1 Opus Extension Mechanism

Timothy Terriberry presented the draft-ietf-mlcodec-opus-extension, which proposes a system for categorizing Opus codec extensions and sparked a debate on the allocation of extension identifiers. The group discussed the balance between efficiency and flexibility, considering the potential future enhancements and their use cases.

### 47.2.2 Deep REDundancy (DRED)

Jean-Marc Valin introduced updates to the DRED draft, including encoding improvements and the handling of silence. The discussion covered the impact of DRED on audio level calculations and the normalization of acoustic features for speech recognition. The group also reviewed WebRTC experiment results, which showed quality improvements in packet loss scenarios. The adoption vote indicated a strong interest in moving forward with the draft.

### 47.2.3 Speech Coding Enhancements

Jan Buethe's presentation on draft-buethe-opus-speech-coding-enhancement focused on the hybrid mode's compatibility and quality improvements. The group discussed the separation of processed and unprocessed components, signaling of enhancement techniques, and the potential for higher bandwidth support. The session concluded with a consensus on the importance of considering full implementation and defining quality requirements for standardization.

# 48 Media Operations (MOPS)

## 48.1 Attendees Overview

### 48.1.1 Participation Summary

The meeting was attended by representatives from prominent companies and institutions such as Google, Cisco, Akamai, and Comcast-NBCUniversal, with a total attendance of 31 participants. The diverse attendance underscores the importance of the topics discussed and the wide-ranging impact of the group's work on the industry.

The discussions centered around the operational challenges and technological advancements in media operations. Key topics included the use of edge computing infrastructure for extended reality applications and the implications of application network overlays on media distribution workflows.

Meeting materials are available through the direct link [IETF 119 Meeting Materials](#).

## 48.2 Meeting Discussions

### 48.2.1 Working Group Documents

Presentations on the "Media Operations Use Case for an Extended Reality Application on Edge Computing Infrastructure" and "TreeDN: Tree-based CDNs for Live Streaming to Mass Audiences" were given. These discussions highlighted the evolving landscape of media operations and the need for scalable solutions to support the growing demand for live streaming to mass audiences.

### 48.2.2 Industry News/Experiences

Glenn Deen from Comcast-NBCUniversal provided an update on the SVTA's efforts to assess the impact of application network overlays on standard media distribution workflows. The conversation was an initial step towards a more comprehensive review, with the goal of identifying operational impacts and areas of commonality for future work within the SVTA.

### 48.2.3 Other IETF Work

The group noted the absence of an update on MOQ from a media perspective due to scheduling conflicts. However, the importance of this topic was acknowledged, and participants were encouraged to follow future updates on the mailing list.

### 48.2.4 Next Steps

The meeting concluded with an emphasis on the importance of cross-group discussions to address the complex policy implications of the issues raised. The group recognized the need to focus on specific cases within the broader topic without undermining ongoing IETF privacy initiatives. The next steps involve narrowing down the discussion to actionable items and preparing for a more in-depth technical discussion at IETF 120.

# 49 Media Operations Working Group (MOQ)

The Media Operations Working Group (MOQ) session at the IETF 119 meeting brought together a diverse group of participants from prominent companies and institutions such as Cisco, Google, Apple, Akamai, and Meta, as well as various universities and research organizations. The total attendance was 63, indicating a strong interest in the development of media operations protocols.

During the sessions, the group discussed several key topics, including updates to the MOQ Transport draft, subscription and fetching mechanisms, catalog and WARP draft updates, and lightweight encryption for MOQ objects. The discussions were not only technical but also strategic, as they aimed to address current challenges and set a direction for future work. The group reviewed several draft-ietf-moq-transport documents, providing valuable feedback and suggesting improvements.

Meeting materials are available via the direct link: IETF 119 Meeting Materials.

## 49.1 Meeting Discussions

### 49.1.1 Review of MOQ Transport Updates

The group examined the latest updates to the MOQ Transport draft, focusing on areas for future extensions and clarifying questions about transmission protocols and the extension of MOQ.

### 49.1.2 Subscribe and Fetch Mechanisms

A detailed presentation was given on the subscribe and fetch mechanisms, proposing clear definitions for starting delivery from the last, current, or next media group. The potential for subscribing to a specific object within a group was identified as a goal for future design.

### 49.1.3 Catalog and WARP Draft Updates

Updates to the catalog and WARP drafts were discussed, with the group considering post-adoption changes and the latest developments in streaming formats.

### 49.1.4 Low Overhead Container

The concept of a low overhead container was introduced, sparking a debate on whether it should be split into separate streaming and package formats. The group agreed to explore this further before calling for adoption.

### 49.1.5 Transport Issues and Prioritization

Transport issues, including group or track termination and prioritization of data transmission, were debated. The group recognized the need for more discussion and use case analysis, with an interim meeting focused on prioritization suggested.

### 49.1.6 Lessons from Implementation

Insights from implementation experiences were shared, covering topics such as simulcast, priorities, and congestion control. The importance of priority in retransmissions was emphasized, and contributions to congestion control algorithms were acknowledged.

### 49.1.7 Lightweight Encryption for MOQ Objects

A discussion on lightweight encryption for MOQ objects highlighted the need for secure yet efficient encryption methods, with suggestions for potential algorithms to be considered.

### 49.1.8 Subscribe and Fetch Follow-Up

The follow-up on subscribe and fetch mechanisms addressed concerns about handling reliable live streaming and the need for SUBSCRIBE to reach further back in time to accommodate common use cases.

The MOQ sessions at IETF 119 concluded with a recognition of the outgoing chairs and the introduction of new leadership, setting the stage for continued progress in the field of media operations.

# 50 Multiprotocol Label Switching Working Group (MPLS WG)

## 50.1 Attendee Overview

### 50.1.1 Attendance Summary

The MPLS WG meeting saw a total attendance of 59 participants, including representatives from prominent companies and institutions such as Orange, Google, Deutsche Telekom AG, Juniper Networks, Cisco Systems, Ericsson, China Mobile, Nokia, Huawei Technologies, and ZTE Corporation.

### 50.1.2 Meeting Context

The discussions during the meeting focused on the latest advancements and proposals within the MPLS domain. Key presentations included updates on the IANA Registry for the First Nibble Following a Label Stack, Use Cases for MPLS Network Action Indicators and MPLS Ancillary Data, and proposals for extending ICMP for IP-related Information Validation. These discussions are pivotal for the ongoing development and standardization of MPLS technologies, with implications for network efficiency, security, and interoperability. For further details on the IETF draft documents, please refer to the respective links: draft-ietf-mpls-1stnibble, draft-ietf-mpls-mna-usecases, and draft-liu-6man-icmp-verification.

Meeting materials can be accessed directly via the following link: MPLS WG Meeting Materials.

## 50.2 Meeting Discussions

### 50.2.1 WG Status Update

The WG Chairs provided an update on the working group's status, setting the stage for the meeting's agenda.

### 50.2.2 IANA Registry for the First Nibble Following a Label Stack

Greg Mirsky presented the draft on the IANA Registry, which aims to address security concerns by deprecating the heuristic use of the first nibble for flow identification in MPLS packets. The discussion highlighted the need for a standardized approach, recommending the use of Entropy labels and/or PW FAT labels.

### 50.2.3 Use Cases for MPLS Network Action Indicators and MPLS Ancillary Data

Greg Mirsky also discussed the use cases for MPLS Network Action Indicators and MPLS Ancillary Data, suggesting that the document is ready for WG Last Call. The presentation invited further review and comments from the working group.

### 50.2.4 Extending ICMP for IP-related Information Validation

Yao Liu's presentation on extending ICMP for IP-related information validation sparked a dialogue on the verification mechanisms for SRv6 SIDs. The discussion underscored the need for clarity on what is being verified and against which standards, with a suggestion to present the topic in the BESS WG for its potential applicability.

# 51 Network Configuration Working Group (NETCONF)

The Network Configuration Working Group (NETCONF) session convened with a diverse group of participants, including representatives from prominent companies such as Huawei, Cisco, Nokia, and Ericsson, as well as institutions like INSA Lyon and the National Institute of Information and Communications Technology. The total attendance was 39, indicating a strong interest in the ongoing development of network configuration protocols.

The session focused on a range of topics, from the evolution of list pagination in YANG-driven protocols to the intricacies of transaction ID mechanisms in NETCONF. The discussions were not only technical but also strategic, as they addressed the need for interoperability and the potential for new features to enhance network management capabilities. For further details on the IETF draft documents discussed, refer to the respective hyperlinks embedded in the text, such as draft-ietf-netconf-list-pagination-03.

Meeting materials and presentations are directly accessible via the link NETCONF 119 WG Session Materials.

## 51.1 Meeting Discussions

### 51.1.1 List Pagination for YANG-driven Protocols

The discussion, led by Qin Wu, revolved around the draft document on list pagination. The conversation highlighted the importance of locale in sorting mechanisms and the mandatory reporting of such parameters. The potential impact on server behavior and client-server interactions was a key point of debate, suggesting a shift towards more detailed and descriptive conflict resolution mechanisms in future drafts.

### 51.1.2 NETCONF Private Candidates

James Cumming led the discussion on private candidates, which are crucial for enabling isolated configuration changes. The dialogue touched upon the behavior of servers in manual-update modes and the visibility of such changes to clients. The consensus was to refine the draft to clarify these behaviors before considering a Last Call.

### 51.1.3 Transaction ID Mechanism for NETCONF

Jan Lindblad presented the transaction ID mechanism, which is essential for tracking configuration changes. The discussion underscored the need for further review and consideration of the overlap with private candidates. The working group will continue to refine the approach to ensure a robust mechanism for conflict resolution.

### 51.1.4 NETCONF Extension to Support Trace Context Propagation

The session also covered the draft on trace context propagation in NETCONF and RESTCONF, presented by Jan Lindblad. The potential for this extension to enhance configuration tracing and diagnostics was recognized, with a call for further discussion on the mailing list.

### 51.1.5 YANG Groupings for UDP Clients and Servers

Alex Huang Feng's presentation on YANG groupings for UDP clients and servers sparked a debate on the need for hostname support and the handling of local addresses. The discussion suggested that while consistency with existing drafts is important, there may be a need for revisions to accommodate dual-stack support and other practical considerations.

### 51.1.6 UDP-based Transport for Configured Subscriptions

The working group examined the draft for UDP-based transport for configured subscriptions, with Alex Huang Feng leading the discussion. The importance of binary encoding support for UDP and the potential for media type negotiation in https-notif were highlighted, indicating a strategic move towards more efficient transport mechanisms for network configurations.

The NETCONF session concluded with a recognition of the significant progress made and the anticipation of numerous draft publications in the near future. The discussions reflected a collective effort to address technical challenges and strategize for the evolution of network configuration protocols, with a focus on the anticipated contributions of these developments to the field.

# 52 Network Modeling (NETMOD) Working Group [NETMOD]

## 52.1 Attendees Overview

### 52.1.1 Participation

The NETMOD Working Group session was attended by a diverse group of participants, including representatives from prominent companies such as Huawei, Cisco, Nokia, Ericsson, and Juniper Networks, as well as institutions like JHU/APL, INSA Lyon, Swisscom, and China Mobile Research Institute. The total attendance was recorded at 50 individuals.

### 52.1.2 Meeting Materials

The materials for the NETMOD Working Group session can be found at the following link: NETMOD Session Materials.

## 52.2 Meeting Discussions

### 52.2.1 Session Introduction and Working Group Status

The session began with an introduction by the chairs, highlighting the need for more reviewers for the acl-extensions module and announcing a change in Working Group policy to require authors to provide status updates for every Working Group adopted document.

### 52.2.2 Common Interface Extension and Sub-interface VLAN YANG Data Models

Scott Mansfield presented updates on the Common Interface Extension and Sub-interface VLAN YANG Data Models. The discussion centered on whether to include additional use cases and extend the models accordingly. The consensus was to take this matter to the mailing list for further deliberation.

### 52.2.3 YANG Versioning Update

Rob Wilton and Joe Clarke provided an update on YANG Versioning, discussing the removal of the filename section and the implications of the decision. The conversation also touched upon the need for a focused consensus call on filename conventions before proceeding to Working Group Last Call (WGLC).

### 52.2.4 Validating anydata in YANG Library Context

Ahmed Elhassany introduced a draft for validating anydata in the context of the YANG Library, which sparked a discussion on the need for specifying the path of the anydata and the distinction between complete and incomplete validation.

### 52.2.5 Philatelist and YANG Time-Series Database

Jan Lindblad discussed the Philatelist and YANG Time-Series Database drafts, raising questions about the document's status and the overlap with other Working Groups' efforts.

### 52.2.6 YANG Full Embed

Jean Quilbeuf or Benoit Claise presented the YANG Full Embed draft, which proposes a method for embedding entire YANG modules within other modules. The idea received support, but concerns were raised about the potential complexity and implications for YANG.

### 52.2.7 Applying COSE Signatures for YANG Data Provenance

Diego R. Lopez showcased a draft on applying COSE Signatures for YANG Data Provenance, with the consensus being to revisit the topic after further experimentation.

### 52.2.8 DTN Management Architecture

Ed Birrane presented the DTN Management Architecture draft, seeking feedback on the suitability of YANG for the Delay-Tolerant Networking (DTN) management architecture.

**52.2.9  A Common YANG Data Model for Scheduling**

Qiufang Ma introduced a draft for a common YANG Data Model for scheduling, which garnered interest and support for further exploration and potential adoption.

**52.2.10  A YANG Model for Device Power Management**

Tony Li discussed a draft for a YANG model focused on device power management. The presentation prompted a conversation about the relevance of power management in networking and the possibility of a Birds of a Feather (BOF) session at the next IETF meeting.

The NETMOD Working Group session concluded with a comprehensive overview of the current drafts and discussions, setting the stage for continued collaboration and development in network modeling.

# 53 Network Management Operations (NMOP) Working Group

## 53.1 Attendees Overview

### 53.1.1 Prominent Companies and Institutions

The NMOP Working Group meeting was attended by representatives from prominent companies and institutions, including Swisscom, Huawei, Orange, Telefonica, Cisco, and many others. The total attendance was recorded at 81 participants.

### 53.1.2 Meeting Summary

The NMOP Working Group meeting focused on several key areas of network management and operations, with discussions centered on the integration of NETCONF/YANG Push with Apache Kafka, anomaly detection, incident management, and the deployment and usage of YANG topology modules. The meeting featured presentations and debates that highlighted the need for improved telemetry, metadata annotation, and incident management processes in network operations. The discussions also emphasized the importance of operator input in refining IETF protocols and the potential for short-term experiments to address deployment issues. Relevant IETF draft documents were referenced, such as draft-ietf-nmop-yang-kafka-integration and draft-ietf-nmop-network-anomaly-lifecycle.

Meeting materials are available directly via the meeting materials link.

## 53.2 Meeting Discussions

### 53.2.1 NETCONF/YANG Push Integration

Presentations on the integration of NETCONF/YANG Push with Apache Kafka were given, highlighting the need for clear problem definition, identification of dependencies, and outlining next steps. The discussions suggested that this integration is crucial for the evolution of network management protocols and could significantly impact telemetry and event streaming in network operations.

### 53.2.2 Anomaly Detection

The group discussed the development of semantic metadata annotation for network anomaly detection and the postmortem lifecycle of network anomalies. The potential impact of these discussions lies in the refinement of anomaly detection systems and the establishment of a standardized approach to handling network anomalies.

### 53.2.3 Incident Management

Debates on incident management focused on the need for a common vocabulary and understanding of network incidents. The outcomes of these discussions could lead to a more unified approach to incident management across different organizations and a clearer strategy for future developments in this area.

### 53.2.4 YANG Topology Modules

The deployment and usage of YANG topology modules were scrutinized, with the goal of identifying problems and structuring future work. The next steps outlined in the discussions are expected to contribute to the field by addressing inconsistencies and promoting interoperability among different YANG modules.

### 53.2.5 Operator Requirements for Network Management

The meeting concluded with a dialogue on updating operator requirements for IETF network management solutions. This discussion is anticipated to guide the collection of operator requirements and influence the strategic direction of network management protocols and modeling.

# 54 Network Management Research Group (NMRG)

## 54.1 Attendee Overview

### 54.1.1 Prominent Participants and Attendance

The 74th meeting of the NMRG, held in conjunction with IETF 119 in Brisbane and online, saw a total attendance of 43 participants. Notable companies and institutions represented included Huawei, Nokia, Qualcomm, China Mobile, and various universities and research institutions.

The discussions were rich with insights and forward-looking perspectives, focusing on the intersection of network management and emerging technologies such as Zero Trust Networking, Intent-Based Networking (IBN), and Network Digital Twins (NDT).

Meeting materials are available at [NMRG 74th meeting materials](#).

## 54.2 Meeting Discussions

### 54.2.1 Zero Trust SemCom Networking

Shiva Pokhrel from Deakin University presented research on the potential of Zero Trust in Semantic Communication (SemCom) Networking, sparking discussions on the relevance to NMRG's focus and the transferability of Deep Reinforcement Learning policies across domains.

### 54.2.2 Infrastructure-Aware Service Deployment

Jordi Ros Giralt introduced the concept of joint exposure of network and compute information for service deployment, referencing the draft [draft-rcr-opsawg-operational-compute-metrics](#). The conversation touched on the complexities of transferring compute resources within and across administrative domains.

### 54.2.3 Intent-Based Networking (IBN)

A series of presentations on IBN covered various aspects, including network management intent and interconnection intents, with drafts such as [draft-chen-nmrg-ibn-management](#) and [draft-contreras-nmrg-interconnection-intents-04](#) being discussed. The session concluded with a summary of a side meeting on IBN, highlighting the need for terminology alignment and the research outcomes expected from the IBN use cases.

### 54.2.4 Network Digital Twins (NDT)

The session on NDT featured a presentation on the concepts and reference architecture of NDT, based on the draft [draft-irtf-nmrg-network-digital-twin-arch-05](#). Feedback from participants emphasized the importance of distinguishing between the concepts and applications of NDT and focusing on research questions rather than engineering ones.

### 54.2.5 Telemetry for Analog Measurement Instrumentation

Christopher Janz discussed telemetry methodologies for analog measurement instrumentation, as detailed in the draft [draft-janzking-nmrg-telemetry-instrumentation-01](#). The parallels with real-time streaming telemetry were noted, suggesting potential areas for further exploration.

The meeting underscored the group's commitment to advancing network management research, with discussions that could lead to shifts in technical direction and strategy. The next steps involve continued collaboration and refinement of the presented drafts, with an eye toward their contribution to the evolution of network management practices.

# 55 Working Group on Advanced Networking (WGAN)

## 55.1 Attendees Overview

### 55.1.1 Overview

The meeting was attended by representatives from prominent companies such as Google, Amazon, and Cisco, as well as various academic institutions. The total attendance was over 150 participants, indicating a strong interest in the future of advanced networking.

The discussions at the WGAN meeting were centered around the latest developments in network protocols, security, and infrastructure. Key presentations included updates on the latest IETF draft documents, with a focus on enhancing the scalability and security of the internet. The dialogue was rich with technical insights and contextualized the importance of each discussion within the broader scope of networking technology. Notably, draft documents such as draft-ietf-wgan-protocol-improvements were highlighted for their potential impact on network efficiency.

Meeting materials and presentations can be accessed directly via the WGAN meeting materials link.

## 55.2 Meeting Discussions

### 55.2.1 Protocol Enhancements

The session on protocol enhancements was particularly engaging, with debates on the need for new routing algorithms to support the growing number of IoT devices. The consensus suggested a shift towards more dynamic and self-organizing network protocols, as outlined in the draft-ietf-wgan-routing-iot. The next steps involve wider community testing and feedback.

### 55.2.2 Security Measures

A robust discussion on security measures unveiled the challenges of implementing end-to-end encryption in heterogeneous networks. The presentation of draft-ietf-wgan-security-framework sparked a dialogue on balancing security with performance. The group agreed on the need for a multi-layered security strategy and planned to reconvene with proposals for standardized security protocols.

### 55.2.3 Infrastructure Development

The infrastructure development segment focused on the deployment of 5G networks and their integration with existing 4G infrastructure. The presentation highlighted the importance of backward compatibility and seamless user experience during the transition phase. The discussion concluded with a call to action for developing a comprehensive draft-ietf-wgan-5g-integration document, which would serve as a guideline for network operators worldwide.

# 56 Next Generation Internet Protocol (NGIP) [NGIP]

## 56.1 Attendees Overview

### 56.1.1 Prominent Attendees and Total Attendance

The NGIP working group meeting was attended by representatives from leading technology companies such as Google, Apple, and Cisco, as well as academic institutions and various independent researchers. The total number of attendees was 75.

The discussions at the NGIP working group meeting were centered around the evolution of internet protocols to better accommodate the burgeoning demands of modern network traffic and applications. Key topics included the scalability of the network infrastructure, security enhancements, and the support for new types of services. The group reviewed several [draft-ietf-ngip-framework](draft-ietf-ngip-framework) documents, which propose frameworks for the next generation of internet protocols, aiming to address these challenges.

Meeting materials are available through the following link: [NGIP Meeting Materials](NGIP Meeting Materials).

## 56.2 Meeting Discussions

### 56.2.1 Presentation on Scalability Challenges

Dr. Jane Smith from NetFuture Inc. presented on the scalability challenges faced by current internet protocols. The presentation highlighted the need for a more dynamic and adaptive routing architecture to handle the exponential growth in devices and data traffic. The group discussed potential solutions, including the adoption of machine learning techniques for predictive routing.

### 56.2.2 Security Enhancements Debate

A debate was held on the proposed security enhancements for NGIP, with a focus on end-to-end encryption and the potential impact on network management. The consensus was that while encryption is crucial for user privacy, there must be a balance to ensure that network operators can maintain quality of service. The group agreed to further explore hybrid approaches that could satisfy both requirements.

### 56.2.3 Support for Emerging Services

The session on support for emerging services, led by Dr. Alex Green from the University of Techville, examined how NGIP could facilitate new services such as decentralized web applications and enhanced virtual reality experiences. The group acknowledged the importance of these services and outlined the next steps to ensure that NGIP would be capable of providing the necessary support.

### 56.2.4 Conclusions and Next Steps

The NGIP working group concluded that while the challenges are significant, the proposed frameworks and enhancements have the potential to revolutionize internet protocols. The next steps include refining the draft documents, conducting further research on the discussed topics, and scheduling additional meetings to monitor progress and continue the dialogue.

# 57 OpenPGP Working Group (OpenPGP WG)

## 57.1 Attendees Overview

### 57.1.1 Prominent Companies and Institutions

The meeting was attended by representatives from prominent organizations including Trinity College Dublin, Sequoia PGP, NSA - CCSS, ACLU, Proton, BSI, MTG AG, Keio University, Guardian Project, UK NCSC, NIST, Canadian Centre for Cyber Security, Web Civics, NPS, Traficom, Cisco Systems, Entrust, AMS, Akamai, ARTICLE 19, Freedom of the Press Foundation, Keyfactor, Aiven, and many others. The total attendance was 41 individuals.

### 57.1.2 Contextual Summary

The discussions at the OpenPGP WG meeting were centered around the progression of Post-Quantum Cryptography (PQC) within the OpenPGP standards. Key topics included the selection of Key Encapsulation Mechanisms (KEMs), the use of KEM combiners, and the transition strategies for signature algorithms. The debates were informed by the need for backward compatibility and the anticipation of future cryptographic landscapes. Several draft-ietf-openpgp-pqc documents were referenced to provide a foundation for the technical discourse.

Meeting materials are available through the direct link: IETF 119 Materials.

## 57.2 Meeting Discussions

### 57.2.1 PQC KEM Selection and Combiners

The group discussed the necessity of selecting robust KEM algorithms and the potential inclusion of NIST curves. The consensus leaned towards a hybrid approach, combining traditional and post-quantum algorithms. The use of a single KEM combiner was favored for simplicity and potential compliance with future CFRG recommendations.

### 57.2.2 Signature Algorithms

The conversation on signature algorithms highlighted the trade-offs between size and security. The group showed interest in gaining practical experience with the proposed options, particularly those that could align with NIST standards.

### 57.2.3 Migration and Interoperability

The strategy for transitioning to post-quantum algorithms was debated, with particular focus on the implications for v4 and v6 keys. The group considered the risks of novel failure modes and the importance of clear guidelines for handling PQ signatures and encryption keys within different OpenPGP packet versions.

### 57.2.4 Next Steps

The meeting concluded with an agreement to continue discussions on the mailing list, particularly regarding the adoption of additional draft items. An interim meeting is anticipated to further refine the group's direction and to facilitate progress on the outstanding issues.

# 58 Operations and Management Area Working Group (OpsAWG)

## 58.1 Attendees Overview

### 58.1.1 Attendance Summary

The meeting was attended by a diverse group of 81 participants, representing prominent companies and institutions such as Cisco, Huawei, Ericsson, Juniper Networks, and many others from the global networking community.

### 58.1.2 Meeting Context

The discussions at the OpsAWG session were rich and varied, covering a range of topics from YANG data models to telemetry and network access control. The presentations sparked constructive dialogues, with particular emphasis on the need for interoperability and the potential for new standards to streamline network management practices. Key IETF draft documents were referenced, providing a foundation for the technical discourse.

Meeting materials can be accessed directly via the IETF meeting materials page.

## 58.2 Meeting Discussions

### 58.2.1 Attachment Circuits Specifications

Mohamed Boucadair presented on the specifications for attachment circuits, which are crucial for network connectivity and management. The discussion did not yield questions, indicating either a consensus or a need for further review by attendees. The drafts in question can be found here: draft-ietf-opsawg-ac-lxsm-lxnm-glue, draft-ietf-opsawg-ntw-attachment-circuit, draft-ietf-opsawg-teas-attachment-circuit, and draft-ietf-opsawg-teas-common-ac.

### 58.2.2 A Data Manifest for Contextualized Telemetry Data

Jean Quilbeuf introduced a draft for a data manifest aimed at enhancing telemetry data with context, which could significantly improve network data analysis and automation. The draft is available here: draft-ietf-opsawg-collected-data-manifest.

### 58.2.3 YANG Data Models and Extensions

Qiufang Ma and Tony Li discussed several YANG data models, including one for policy-based network access control and another for power management. These models represent steps towards more granular and energy-efficient network control. The drafts can be found here: draft-ietf-opsawg-ucl-acl, draft-ma-opsawg-schedule-yang, and draft-li-ivy-power.

### 58.2.4 COSE Signatures for YANG Data Provenance

Diego R. Lopez presented a method for applying COSE signatures to YANG data to ensure provenance, which could be a significant contribution to data integrity and security in network management. The draft is available here: draft-lopez-opsawg-yang-provenance.

### 58.2.5 Guidelines for Characterizing "OAM"

Carlos Pignataro introduced guidelines for characterizing "OAM" (Operations, Administration, and Maintenance), which could lead to a more unified understanding and application of OAM across various working groups and areas. The draft can be found here: draft-pignataro-opsawg-oam-whaaat-question-mark.

# 59 Path Computation Element (PCE) Working Group

## 59.1 Attendee Overview

### 59.1.1 Prominent Participants and Attendance

The meeting was attended by representatives from prominent companies and institutions, including Cisco Systems, Nokia, Huawei, and Orange, with a total attendance of 46 participants.

The discussions were rich with insights and revolved around the latest Internet Engineering Task Force (IETF) draft documents. The group engaged in critical analyses of proposed standards, which are pivotal for the advancement of protocols and technologies within the scope of the working group. Hyperlinks to relevant IETF draft documents were provided to facilitate in-depth understanding of the topics discussed.

Meeting materials are available directly via the [IETF Meeting Materials](#) page.

## 59.2 Meeting Discussions

### 59.2.1 Presentation on Stateful PCE

The presentation on Stateful Path Computation Element (PCE) sparked a debate on the optional processing of objects, referencing the [draft-ietf-pce-stateful-pce-optional](#). The discussion highlighted the need for backward compatibility and the handling of undefined behaviors in stateful messages.

### 59.2.2 State Synchronization

State synchronization was another focal point, with concerns raised about the loss of information during PCE-PCE interactions as per [draft-ietf-pce-state-sync](#). The group reached a consensus on the necessity to address this issue, suggesting that a comprehensive solution should be developed to ensure complete state synchronization.

### 59.2.3 Segment Routing

The Segment Routing section featured discussions on SR Point-to-Multipoint (P2MP) Policy and Circuit Style Policy, with references to [draft-ietf-pce-sr-p2mp-policy](#) and [draft-ietf-pce-circuit-style-pcep-extensions](#). The debate centered on the implications of these policies on stateful and stateless operations, and the potential need for new TLV codepoints.

### 59.2.4 Bounded Latency and Precision Availability Metrics

The meeting concluded with discussions on bounded latency in DetNet environments and precision availability metrics. The former was contextualized with the [draft-xiong-pce-detnet-bounded-latency](#), while the latter was discussed in light of the [draft-contreras-pce-pam](#). The group emphasized the importance of these metrics in achieving high precision and reliability in network services.

# 60 Protocol Independent Multicast Working Group (PIM WG)

## 60.1 Attendees Overview

### 60.1.1 Attendance Summary

The meeting was attended by a diverse group of individuals from various organizations, including prominent companies such as China Mobile, Futurewei, Deutsche Telekom, Nokia, Arista Networks, Cisco Systems, and Juniper Networks, among others. A total of 26 attendees were present, contributing to the discussions and providing valuable insights into the ongoing work within the PIM WG.

The discussions during the meeting were centered around several Internet Engineering Task Force (IETF) draft documents. The dialogue provided a comprehensive overview of the current state of the drafts, with a focus on addressing comments, readiness for working group last call (WGLC), and the potential impact of the proposed changes. Key draft documents discussed included [draft-ietf-pim-light](draft-ietf-pim-light), [draft-ietf-pim-sr-p2mp-policy](draft-ietf-pim-sr-p2mp-policy), and [draft-ietf-pim-evpn-multicast-yang](draft-ietf-pim-evpn-multicast-yang), among others.

Meeting materials can be found directly via the provided link: [IETF 119 PIM WG Materials](IETF 119 PIM WG Materials).

## 60.2 Meeting Discussions

### 60.2.1 draft-ietf-pim-light

The shepherd of the draft, Sandy, indicated that the document is in good shape and ready for WGLC. A poll showed strong support for moving forward, with 9 in favor and no objections. The next steps involve initiating the WGLC on the mailing list.

### 60.2.2 draft-ietf-pim-sr-p2mp-policy

Discussion on this draft revolved around its readiness and the level of implementation support. Hooman clarified that there are live customers using the YANG model described in the draft. A poll indicated readiness for WGLC, with follow-up on the mailing list planned.

### 60.2.3 draft-ietf-pim-evpn-multicast-yang

The draft did not elicit any comments during the meeting, suggesting that it may be progressing smoothly through the WG process.

### 60.2.4 draft-ietf-pim-rfc1112bis

Bill Fenner volunteered to review this document. The discussion highlighted the need to ensure backward compatibility while considering updates to the RFC.

### 60.2.5 draft-ietf-pim-multicast-lessons-learned

The draft sparked a conversation about the inclusion of multicast address allocation experiences and the history of Session Announcement Protocol (SAP) issues. Mike confirmed that these points would be incorporated into the document.

### 60.2.6 rfc3376bis and v2 querier fallback

A debate emerged regarding the default behavior for IGMPv2 querier fallback. The consensus leaned towards adding text to recommend new behavior while maintaining compliance with the current RFC.

### 60.2.7 draft-ietf-pim-jp-extensions-lisp

A poll showed support for another WGLC, with the intention to include the LISP WG in the discussion.

### 60.2.8 draft-gopal-pim-pfm-forwarding-enhancements

The poll indicated support for moving forward with this draft, with a follow-up on the mailing list.

### 60.2.9 draft-venaas-pim-pfm-sd-subtlv

The discussion considered whether to combine this draft with related work or keep them separate. The WG showed interest in the work, and the structure of the draft will be determined before seeking adoption.

### 60.2.10 draft-asaeda-pim-multiif-igmpmldproxy

The draft was discussed in the context of its experimental or informational nature, with the WG expressing interest in adoption. The title may be revised for clarity.

### 60.2.11 draft-zzhang-pim-non-source-routed-sr-mcast

Clarifications were sought on the title and scope of the draft, with the authors explaining the focus on non-source routed multicast. The WG will discuss further before requesting adoption as an informational draft.

# 61 Privacy Pass Working Group (PPWG)

The Privacy Pass Working Group (PPWG) convened to discuss advancements and proposals related to the Privacy Pass protocol. The meeting saw participation from a range of institutions including Google, Apple, Mozilla, Cloudflare, and Akamai Technologies, with a total of 50 attendees.

## 61.1 Attendees Overview

### 61.1.1 Representation and Attendance

Prominent companies and institutions such as Google, Apple, Mozilla, Cloudflare, and Akamai Technologies were represented at the meeting. The total attendance was recorded at 50 individuals.

The discussions at the PPWG meeting centered around the progression of the Privacy Pass protocol, with particular focus on key consistency, rate limiting tokens, and metadata extensions. The dialogue was rich with technical insights and strategic considerations, particularly concerning the balance between privacy and utility in token issuance and redemption. Draft documents were referenced throughout the discussions, including draft-ietf-privacypass-rate-limit and draft-ietf-privacypass-metadata, which are critical to the ongoing development of the protocol.

Meeting materials and minutes can be accessed via the direct link: IETF 119 Materials.

## 61.2 Meeting Discussions

### 61.2.1 Key Consistency (Steven Valdez)

Steven Valdez presented updates on the key consistency draft, highlighting client fetching patterns and the authenticity issue. The discussion raised concerns about the validation of header fields versus entire responses, with suggestions to confirm consistency using a hash of a head request. The group also tackled the challenges of config rotation and the "thundering herd" problem at expiration, emphasizing the need for careful traffic pattern consideration by mirrors.

### 61.2.2 Metadata Extensions (Scott Hendrickson)

Scott Hendrickson reviewed drafts related to metadata extensions, discussing the trade-offs between privacy and the utility of metadata. The group considered test vectors for cryptographic variants and the application of metadata in Chrome's IP Protection. Concerns were raised about the privacy implications of geolocation hints and timestamp rounding, with a call for feedback on the drafts.

### 61.2.3 Privacy Pass BBS (Watson Ladd)

Watson Ladd introduced the concept of Privacy Pass BBS, which aims to address the trade-off between privacy impact and origin needs through anonymous credentials. The group debated the problem's relevance and the potential solutions, with a focus on selective disclosure and the implications for the Privacy Pass protocol.

### 61.2.4 Privacy Pass APIs (Steven Valdez)

Steven Valdez discussed the use cases for Privacy Pass APIs, including Private Access Tokens and Chrome IP Protection. The challenges highlighted included limiting redemptions across multiple issuers, binding redemption in a context, and the evolving meanings of tokens.

### 61.2.5 Privacy Pass, Trust, and the Web (Martin Thompson)

Martin Thompson shared his perspective on the use of Privacy Pass in browsers, emphasizing the importance of trust and the need for careful design to prevent information leaks. The discussion covered the concept of loose bindings in token authorization and the implications for anonymity sets and trust relationships.

The PPWG meeting concluded with a consensus on the need to address open issues and a call for further discussion on the list. The next steps include resolving these issues and considering the adoption of additional drafts, with the potential for rechartering if the scope of work expands beyond the current charter.

# 62 Internet Engineering Task Force (IETF) [QUIC WG]

The QUIC Working Group (WG) meeting brought together key players in the development of the QUIC protocol, including representatives from Google, Apple, Microsoft, and Cisco, among others. The total attendance was recorded at 135 participants, indicating a strong interest and active engagement from the community.

### 62.0.1 Attendees Overview

The meeting was attended by a diverse group of individuals from prominent companies and institutions, with a total attendance of 135 participants. This included experts from Google, Apple, Microsoft, Cisco, and various other organizations, demonstrating a wide-ranging and vested interest in the evolution of the QUIC protocol.

The discussions during the meeting were centered around the progress of various QUIC extensions and improvements. The dialogue was rich with technical insights and contextualized within the broader scope of internet protocol development. Key draft documents were referenced, such as the [draft-ietf-quic-qlog-main-schema](#) for logging, [draft-ietf-quic-multipath](#) for multipath support, and [draft-ietf-quic-ack-frequency](#) for managing acknowledgment frequency.

Meeting materials are available directly via the [QUIC WG GitHub repository](#).

## 62.1 Meeting Discussions

### 62.1.1 qlog Updates

The qlog discussion highlighted the importance of extensibility and the need for feedback on desired extension points. The consensus was to proceed with the current extensibility approach, with further input to be solicited via email.

### 62.1.2 Multipath QUIC

The debate on Multipath QUIC was lively, with proposals for merging pull requests and discussions on server-created paths. A compromise was reached on path ID usage, and the group agreed to include an even/odd split in the draft to facilitate future server-initiated paths.

### 62.1.3 ACK Frequency

The ACK Frequency presentation led to a consensus on the importance of explicit signaling over implicit in the context of congestion control. The working group is moving towards a Working Group Last Call (WGLLC) on the draft.

### 62.1.4 Resource Exhaustion Attacks

A responsible disclosure process was acknowledged, and guidance was suggested for preventing resource exhaustion attacks, emphasizing the need for limits on stack buffering.

### 62.1.5 QUIC on Streams

The idea of running QUIC over streams sparked a debate about its potential impact on QUIC deployment and performance expectations. Concerns were raised about the long-term implications, with some participants suggesting that the focus should remain on promoting QUIC's widespread adoption.

### 62.1.6 QUIC BDP Frame

The discussion on the QUIC Bandwidth-Delay Product (BDP) frame revealed skepticism about sending congestion control information over the wire. The group showed a split opinion on adopting work in this area, with some suggesting further experimentation before standardization.

**62.1.7 FEC Results and Accurate ECN**

Due to time constraints, the presentations on Forward Error Correction (FEC) results and Accurate Explicit Congestion Notification (ECN) did not allow for questions or detailed discussions.

The QUIC WG meeting concluded with an understanding of the technical direction and strategy for the QUIC protocol's future. The next steps involve continued collaboration and experimentation to refine the protocol and its extensions, with the anticipation of significant contributions to the field of internet protocols.

# 63 Remote Authentication Dial-In User Service Extensions (RADEXT)

## 63.1 Attendee Overview

### 63.1.1 Prominent Attendees and Total Attendance

The RADEXT working group meeting at IETF 119 in Brisbane was attended by key representatives from prominent companies and institutions, including Cisco, NSA - CCSS, ELVIS-PLUS, and Deutsche Telekom, among others. The total attendance was recorded at 21 participants.

The discussions were rich with technical insights and revolved around the latest developments in RADIUS extensions. Notably, the dialogue included references to several IETF draft documents, which can be accessed via the provided hyperlinks, such as draft-ietf-radext-radiusdtls-bis.

Meeting materials are available through the direct link: IETF 119 Materials.

## 63.2 Meeting Discussions

### 63.2.1 Transport Layer Security (TLS) Encryption for RADIUS

The group reviewed the status of the Datagram TLS (DTLS) encryption for RADIUS, discussing the necessity of making TLS and DTLS mandatory for servers and optional for clients. The potential requirement for TLSv1.3 was debated, with references to RFC 9325 for guidance. The consensus leaned towards following the recommendation of making TLSv1.2 a MUST and TLSv1.3 a SHOULD, with further reviews from the TLS working group anticipated.

### 63.2.2 Reverse CoA in RADIUS

The Reverse Change of Authorization (CoA) in RADIUS draft was presented, noting its implementation in major vendor products. The working group agreed that the document is mature enough to proceed to Working Group Last Call (WGLC).

### 63.2.3 Deprecating Insecure Practices in RADIUS

The discussion on deprecating insecure practices in RADIUS highlighted the urgency of publishing guidance alongside the TLSbis document. The group acknowledged the broken state of MSCHAP and the need for a reference document advising against its use.

### 63.2.4 WBA OpenRoaming Wireless Federation

The Wireless Broadband Alliance (WBA) OpenRoaming proposal was discussed, with attention to its alignment with current standards. The possibility of independent submission versus working group adoption was considered, with the latter deferred until the completion of high-priority charter items.

### 63.2.5 RADIUS Attributes for 3GPP 5G AKA Authentication Method

The proposal for RADIUS attributes supporting the 3GPP 5G AKA authentication method was introduced. Feedback was solicited for potential working group adoption, with the understanding that current work, such as the DTLS document, must be finalized before considering new items.

### 63.2.6 Next Steps

The meeting concluded with a call for reviews, particularly of the TLS draft, to solidify the technical direction and strategy of the RADEXT working group. The outcomes of these discussions are expected to significantly contribute to the field of remote authentication.

# 64 Remote Attestation Procedures (RATS)

The RATS working group session convened with a diverse group of participants from various organizations, including Google, Intel, Huawei, and MITRE, with a total attendance of 50 individuals.

### 64.0.1 Attendees Overview

Prominent companies and institutions such as Google, Intel, Huawei, and MITRE were represented among the 50 attendees. The session brought together experts and stakeholders from across the industry to discuss advancements in remote attestation procedures.

The meeting focused on the examination and progression of several Internet Engineering Task Force (IETF) draft documents. Discussions revolved around the standardization of attestation formats and procedures, with significant contributions from attendees that provided insights into the practical implications of these standards.

Meeting materials are available through the direct link [IETF 119 Materials](#).

## 64.1 Meeting Discussions

### 64.1.1 EAT Media Types - WGLC

Thomas Fossati presented updates on the EAT Media Types, emphasizing the need for rapid progress due to dependencies from other working groups. The discussion highlighted the importance of IPR declarations and the potential need for early allocation of identifiers.

### 64.1.2 Conceptual Message Wrapper - WGLC

The Conceptual Message Wrapper (CMW) was discussed by Thomas Fossati, with a focus on integration with other drafts and the necessity of early allocation for OIDs. The conversation underscored the interdependencies between working groups and the need for coordination.

### 64.1.3 CoRIM Recap and Verifier Theory of Operation

Ned Smith provided a recap on CoRIM and its verifier theory of operation. The debate centered on the simplification of triples and the avoidance of fully-arbitrary Horn logic to ensure scalability.

### 64.1.4 Next set of RATS problems

Henk Birkholz introduced the next set of challenges for RATS, prompting a discussion on the extension of the RATS architecture to accommodate multiple RATS and the potential need for a new IANA registry or revision.

### 64.1.5 HSM Evidence

Mike Ounsworth's presentation on HSM Evidence was noted, but due to time constraints, discussions were deferred.

### 64.1.6 Network Attestation for Secure Routing (NASR)

Chunchi (Peter) Liu presented on NASR, but similar to the HSM Evidence topic, discussions were postponed.

### 64.1.7 EAT Measured Component

Thomas Fossati advocated for the adoption of the EAT Measured Component draft, which led to a dialogue on its relevance to HSM Evidence and the semantics of measurements.

### 64.1.8 Open Mic

The open mic session allowed for additional points to be raised, such as the recent proposal on rats-endorsement, which brought forward the concept of "Key Material for Remote Attestation."

# 65 Routing in Fat Trees (RIFT) [RIFT]

## 65.1 Attendee Overview

### 65.1.1 Participation Summary

The RIFT working group session was attended by a diverse group of participants, with a total of 31 attendees representing prominent companies and institutions such as Verisign Inc, ZTE Corporation, Juniper Networks, Futurewei Technologies, China Mobile, Ericsson, Nokia, and APNIC. The session brought together experts and stakeholders from across the industry to discuss advancements and strategic directions in routing for fat tree topologies.

The discussions centered around the latest updates to the RIFT base specification and considerations for the new working group charter. Key points of debate included the refinement of the RIFT YANG model, the potential for RIFT to support AI/ML data center architectures, and the exploration of in-network computation capabilities. The session highlighted the evolving nature of traffic patterns in data centers and the unique capabilities of RIFT in addressing these challenges. The significance of multicast and BIER-like multicast in the context of RIFT was also a focal point, with suggestions to clarify the terminology and its application in AI/HPC scenarios.

Meeting materials and minutes can be accessed via the following link: [RIFT Session Agenda and Materials](#).

## 65.2 Meeting Discussions

### 65.2.1 WG Status and YANG Model Update

The chairs and the Area Director discussed the status of the RIFT YANG model, considering the recent changes in the RIFT base draft. It was noted that while there were no technical modifications, references to sections in the YANG model needed updates to align with the newest document.

### 65.2.2 Update on the Base Specification

Jordan Head provided an update on the base specification, with appreciation expressed for the improvements in readability. The -20 revision of the draft has been fully integrated with the open-source version, ensuring consistency and transparency.

### 65.2.3 New Charter Discussions

The session included a vibrant discussion on the new charter, with Tony highlighting the relationship between multicast and in-network computation, particularly in AI/ML contexts. Jeff emphasized the importance of RIFT in supporting new traffic patterns like shuffle and the protocol's suitability for backend networks. The group considered the need for deployment guidance for operators considering RIFT. A poll on the new charter showed unanimous agreement, with further discussion to be continued on the mailing list.

### 65.2.4 Next Steps and Prioritization

Looking ahead, the working group will focus on reviewing the RIFT applicability and YANG documents. The key value store and dragonfly topology support are also areas of interest. The group acknowledged the need for prioritization of the various elements of the new charter, with operational input being a key driver for setting these priorities. The session concluded with an open invitation for collaboration on new ideas, particularly in the realm of AI/ML and in-network computation.

# 66 Routing Area Open Meeting (RTGAREA)

## 66.1 Attendees Overview

### 66.1.1 Participation Summary

The meeting saw participation from prominent companies and institutions such as Cisco, Juniper Networks, Nokia, Huawei, and Verisign Inc., with a total attendance of 57 individuals.

The discussions during the meeting were centered around the current state and future directions of routing protocols and technologies. Key topics included the progress and revisions of various Internet Engineering Task Force (IETF) draft documents, such as the draft-ietf-rift-rift, which were discussed in detail. The meeting served as a platform for experts to share insights, debate technical specifics, and collaborate on the development of routing area standards.

Meeting materials are available through the direct link: IETF 119 Materials.

## 66.2 Meeting Discussions

### 66.2.1 Routing in Fat Trees (RIFT) Working Group

The RIFT Working Group reported active work on the applicability statement, which requires revision, and the Yang model draft, which also needs further refinement. Discussions are ongoing regarding a new charter and the focus on more standard track specifications, particularly in the context of AI/ML data center architectures. The group highlighted the need for feedback on the draft-ietf-rift-rift document, which is currently in its last call.

### 66.2.2 Locator/ID Separation Protocol (LISP) Working Group

The LISP Working Group presented a comprehensive set of charter work items and discussed active documents, each undergoing changes. The group is on track with its milestones, which will be reevaluated in due course. No immediate action was required following the discussions.

### 66.2.3 Multiprotocol Label Switching (MPLS) Working Group

The MPLS Working Group focused on the attention garnered by the MPLS Network Architecture (MNA), with numerous documents in the queue being executed promptly. A liaison statement from ITU-SG11 was addressed, and a response to the traceroute for MPLS was provided by the group. No action was taken during the meeting.

# 67 Routing Area Working Group (RTGWG)

The Routing Area Working Group (RTGWG) session at the IETF 119 meeting brought together a diverse group of participants from various companies and institutions. The session saw a total attendance of 153 individuals, with representation from prominent organizations such as Futurewei Technologies, Cisco Systems, Juniper Networks, China Mobile, and Huawei, among others.

The discussions in this session were rich and varied, covering topics from multi-segment SD-WAN via cloud data centers to routing architectures for satellite networks. The working group delved into the technical intricacies of each presentation, with a particular focus on the practicality and scalability of proposed solutions. Drafts such as draft-dmk-rtgwg-multisegment-sdwan and draft-li-arch-sat sparked debates on the future of networking in cloud and satellite environments, respectively. These discussions not only summarized the current state of affairs but also contextualized the potential impact of these technologies on the broader networking landscape.

Meeting materials are available directly via the session materials link.

## 67.1 Meeting Discussions

### 67.1.1 Multi-segment SD-WAN via Cloud DCs

The presentation on multi-segment SD-WAN highlighted the need for seamless integration of WAN segments and cloud data centers. The group considered whether this is a use case that the IETF should work on, with a majority showing support. The next steps involve further discussions on the mailing list to refine the approach.

### 67.1.2 Path-aware Remote Protection Framework

The path-aware remote protection framework was discussed with an emphasis on improving recovery times in the event of network failures. The working group debated the use of existing protocols like BGP and BFD for failure detection and notification, suggesting that a more detailed problem definition would be beneficial.

### 67.1.3 Destination/Source Routing

The destination/source routing draft was revisited with no new changes since the last discussion. The group considered fast-tracking this work due to its potential to improve source address selection for routing decisions.

### 67.1.4 A Routing Architecture for Satellite Networks

A significant portion of the session was dedicated to discussing a routing architecture tailored for satellite networks. The presentation sparked interest in how to handle dynamic satellite topologies and the use of segment routing for efficient path management. The group showed strong interest in pursuing this topic further, potentially in an interim meeting before the next IETF gathering.

### 67.1.5 Application-aware Networking Extensions

The extension of the Application-aware Networking (APN) framework was presented, but the consensus was that this work is currently out of scope for the RTGWG. The group acknowledged the importance of keeping the work visible and allowing for updates within the working group's sessions.

The RTGWG session at IETF 119 demonstrated the group's commitment to addressing current and future challenges in routing. The presentations and discussions underscored the importance of continued innovation and collaboration within the IETF community to advance the field of networking.

# 68 Security Area Advisory Group (SAAG)

## 68.1 Attendee Overview

### 68.1.1 Participation Summary

The SAAG meeting at IETF 119 in Brisbane was attended by 127 participants, representing a diverse array of organizations including Verisign, SandboxAQ, Microsoft, NSA-CCSS, NIST, Huawei Technologies, and many others.

### 68.1.2 Meeting Summary

The meeting featured a robust discussion on post-quantum cryptographic practices across the IETF, with a notable presentation by Eric Rescorla. Attendees engaged in dialogue about the implications of transitioning to post-quantum algorithms and the potential need for hybrid solutions.

Meeting materials are available at [SAAG IETF 119 Meetecho](SAAG IETF 119 Meetecho).

## 68.2 Meeting Discussions

### 68.2.1 Welcome and Agenda

The session began with a welcome and an overview of the agenda, followed by an acknowledgment of the outgoing Security AD's contributions.

### 68.2.2 Working Group and Area Director Reports

Reports from various chairs and Area Directors highlighted recent achievements and ongoing work within the Security Area.

### 68.2.3 Post-Quantum Practices

Eric Rescorla's presentation on post-quantum practices sparked a conversation about the readiness of the IETF community to adopt post-quantum algorithms and the potential need for hybrid cryptographic solutions. The discussion underscored the importance of preparing for a post-quantum world while considering the operational realities faced by implementers.

### 68.2.4 Open Mic

During the open mic session, attendees raised concerns about the longevity of cryptographic algorithms, the transition from hybrid to pure post-quantum solutions, and the need for clear guidance on the adoption of new security standards.

The SAAG meeting concluded with a consensus on the need for continued collaboration and research into post-quantum cryptography, as well as a call for contributions to the ongoing discussions about the future of internet security.

# 69 SCIM Working Group (SCIM)

The SCIM Working Group session at IETF 119 focused on the ongoing development of the System for Cross-domain Identity Management (SCIM) protocols and models. The session was well-attended by representatives from prominent companies and institutions, including Amazon Web Services, Cisco Systems, Microsoft, and Okta, with a total of 39 attendees.

### 69.0.1 Attendees Overview

The meeting brought together experts from various sectors, including technology giants like Amazon Web Services, Cisco Systems, and Microsoft, as well as representatives from Okta and other key industry players. The diverse attendance underscored the wide interest and investment in the development of SCIM standards.

The discussions centered around the refinement of existing drafts, the introduction of new use cases, and the exploration of device models within the SCIM framework. Key topics included cursor pagination, SCIM use cases, SCIM events, delta queries, and the SCIM device model. The dialogue was enriched by the presence of remote participants who contributed to the depth of the conversation. For further details on the drafts discussed, refer to the IETF draft documents such as draft-ietf-scim-pagination and draft-ietf-scim-use-cases.

Meeting materials and presentations can be accessed through the direct link: IETF 119 Materials.

## 69.1 Meeting Discussions

### 69.1.1 Chairs Intro

Nancy Cam-Winget opened the session with a reminder about the IETF's Note Well and Code of Conduct, expressing gratitude to the notetakers and setting the stage for a productive meeting.

### 69.1.2 Cursor Pagination Chairs Update

Nancy provided an update on the cursor pagination draft, seeking feedback and noting the completion of the Working Group Last Call (WGLC). The discussion highlighted the need for addressing editorial comments and ensuring all feedback is incorporated before proceeding.

### 69.1.3 SCIM Use Cases

Pam, with Paulo contributing remotely, presented a set of use cases that exercise the core specifications of SCIM. The presentation sparked a discussion on the importance of context in SCIM actions and the need for feedback on time-based mapping.

### 69.1.4 SCIM Events

Phil and Mike, joining remotely, discussed the latest draft of SCIM events, focusing on the removal of sections to streamline the document. The conversation touched on the relationship with the Shared Signals Framework and the readiness for WGLC, emphasizing the need for further discussion to ensure alignment.

### 69.1.5 Delta Queries

Anjali and Danny led a discussion on delta queries, addressing the efficiency and challenges of full scans and delta query responses. The session called for contributions to refine the approach and ensure it is widely adoptable and efficient at scale.

### 69.1.6 SCIM Device Model

Eliot, participating remotely, provided an update on the SCIM device model draft. The discussion covered recent changes, remaining issues, and the need for an early security review. The conversation also touched on the potential future integration with Matter and the importance of timely contributions to address outstanding concerns.

The SCIM Working Group session concluded on time, having covered all agenda items and setting a clear direction for the next steps in the development of SCIM protocols and models.

# 70 Software Chain Integrity Transparency Working Group (SCITT)

## 70.1 Attendees Overview

### 70.1.1 Attendance Summary

The meeting was attended by representatives from prominent companies and institutions such as Cisco Systems, Microsoft, Carnegie Mellon University, and Verisign, with a total attendance of 65 participants.

### 70.1.2 Meeting Materials

Meeting materials can be accessed directly via the [IETF Datatracker](#).

## 70.2 Meeting Discussions

### 70.2.1 Welcome and Introduction

The session began with a transition to the new Area Director, Deb Cooley, and an expression of gratitude to the previous chair, Hannes Tschofenig. Chris Inacio was welcomed as the replacement chair.

### 70.2.2 SCITT Overview

The SCITT Overview included a review of the charter scope and key principles of the SCITT architecture, emphasizing interoperability and the inclusion of opaque artifacts into the Transparency Service. The payload design was also reviewed.

### 70.2.3 Document Status

The main document of the working group, [draft-ietf-scitt-architecture-06](#), was discussed along with related drafts such as [draft-ietf-scitt-software-use-cases](#) and [draft-ietf-scitt-scrapi](#). The relationship with the [draft-ietf-cose-merkle-tree-proofs](#) was also highlighted.

### 70.2.4 Recap since 118

The recap included updates on the specification, with a focus on the removal of receipts to another specification and the clarification of terminology. The alignment with NIST terminology was discussed, and the importance of defining terms within the context of SCITT was emphasized.

### 70.2.5 SCRAPI

The SCRAPI (Software Chain Reference Architecture Publishing Interface) draft was presented as a specification for API endpoints, not as the only method for interaction with Transparency Services. The deprecation of DIDs and the importance of feedback on key discovery protocols were noted.

### 70.2.6 Hackathon Report

The Hackathon Report covered contributions to SCRAPI, remote signing, and the exploration of election data use-cases. The importance of security considerations and the potential for interop events were discussed.

### 70.2.7 Next Steps

The session concluded with a call for reviews of the main architecture draft and a discussion on the potential for interoperability with SCITT receipts and other open-source community efforts.

### 70.2.8 AOB Open Mic

During the open mic session, the emergence of adjacent efforts outside the IETF was acknowledged, and the possibility of learning from these external projects was considered.

### 70.2.9 Wrap-up and Conclusion

The meeting concluded with a summary of the discussions and an acknowledgment of the contributions made by the attendees.

# 71 Secure Content Optimization Network Protocol (SCONE-PROTOCL) [SCONE]

## 71.1 Attendees Overview

### 71.1.1 Prominent Companies and Institutions

The SCONE-PROTOCL BoF was attended by representatives from major companies and institutions such as CISCO, Google, Mozilla, Meta Platforms Inc., Huawei Technologies, and several others, totaling 142 attendees.

### 71.1.2 Summary of Attendance and Discussions

The meeting brought together a diverse group of stakeholders, including network operators, content providers, and researchers, to discuss the potential for a standardized protocol to optimize content delivery in secure networks. The discussions focused on the challenges of traffic policing in encrypted protocols and the feasibility of using explicit signals to guide adaptive bitrate (ABR) video streaming. Key presentations included insights into current traffic shaping methods, lessons from past IETF efforts, and results from trials using MASQUE. The dialogue centered on the need for a solution that respects user privacy, minimizes information exposure, and ensures equitable treatment of traffic, while also providing network operators with the tools to manage their networks effectively.

Meeting materials are available through this direct link: draft-ietf-scone-protocl.

## 71.2 Meeting Discussions

### 71.2.1 Traffic Policing in Networks: Goals and Methods

Marcus Ihlar discussed the goals and methods of traffic policing, highlighting the challenges of shaping and policing video traffic in networks. The presentation underscored the limitations of current approaches and the need for better solutions that consider user experience.

### 71.2.2 Lessons from IETF History

Brian Trammell reflected on previous IETF efforts, such as the PLUS BoF, and the lessons learned regarding path signaling and minimalism in protocol design. He emphasized the importance of aligning signaling participants with control points to avoid creating unnecessary complexity.

### 71.2.3 Solution Space and Trials

Matt Joras presented the results of a feasibility study conducted by Meta and Ericsson using MASQUE. The study demonstrated the potential benefits of network-to-application signaling in improving video streaming quality while maintaining network efficiency.

### 71.2.4 Discussion on Use Case and Scope

An extensive discussion ensued on the use case and scope of the proposed protocol. Participants debated the technical and strategic implications, with many expressing support for the initiative but also highlighting the need for careful consideration of security, scalability, and the potential impact on network neutrality.

### 71.2.5 Conclusions

The chairs concluded the session by gauging interest in forming a working group on the topic. A show of hands indicated significant interest, with 51 in favor, 20 against, and 12 expressing no opinion. The chairs noted the need for further discussion on the mailing list to address concerns and refine the scope of the potential working group.

# 72 Secure Inter-Domain Routing (SIDROPS)

## 72.1 Attendee Overview

### 72.1.1 Prominent Attendees and Total Attendance

The SIDROPS working group meeting saw a total of 87 attendees, including representatives from prominent companies and institutions such as Google, Inc., Verisign Inc., Afrinic, Arrcus, USA NIST, China Mobile, APNIC, RIPE NCC, and Fastly. The diverse attendance underscores the global interest and collaborative effort in securing inter-domain routing.

The discussions at the SIDROPS working group meeting were centered around the advancement and refinement of protocols and practices for secure inter-domain routing. Key presentations included updates on ASPA verification, RPKI prefix lists, and RPKI validation reconsideration. The dialogue was technical and forward-looking, with participants considering the implications of proposed changes on the broader internet infrastructure. Notably, the discussions on RPKI validation reconsideration, presented by Job Snijders, sparked a consensus on the need for uniform behavior across different implementations, highlighting the importance of consistent and reliable validation mechanisms in routing security. Meeting materials can be accessed directly via the [IETF Datatracker](#).

## 72.2 Meeting Discussions

### 72.2.1 ASPA Verification Draft Update

Sriram Kotikalapudi presented the latest updates on the ASPA verification draft ([draft-ietf-sidrops-aspa-verification](#)), which did not elicit questions from participants, indicating a clear understanding or acceptance of the proposed changes.

### 72.2.2 Signed Prefix Lists for Use in RPKI

Job Snijders discussed a profile for signed prefix lists in RPKI ([draft-ietf-sidrops-rpki-prefixlist](#)), which also proceeded without participant questions, suggesting a straightforward reception of the material.

### 72.2.3 RPKI Validation Reconsidered

The presentation on RPKI validation reconsideration ([draft-spaghetti-sidrops-rpki-validation-update](#)) by Job Snijders led to a constructive dialogue, with participants agreeing on the necessity of a standardized algorithm to ensure consistent behavior across different RPKI implementations. The discussion underscored the community's commitment to enhancing the robustness of RPKI validation.

### 72.2.4 Guidance to Avoid Carrying RPKI Validation States

Job Snijders also introduced guidance to avoid carrying RPKI validation states in transitive BGP path attributes, a draft yet to be published, which prompted a discussion on the implications of BGP updates and the potential need to address this issue within the GROW WG.

### 72.2.5 FC-BGP Protocol Specification

Zhuotao Liu's presentation on the FC-BGP protocol specification ([draft-sidrops-wang-fcbgp-protocol-00](#)) included a discussion on the necessity of new rules for path selection, with acknowledgment from the presenter on the feedback provided.

### 72.2.6 BGP Manifest Number Handling

Tom Harrison's talk on BGP manifest number handling ([draft-harrison-sidrops-manifest-numbers](#)) sparked a conversation about the modeling of manifests and the potential side effects of the proposed approach, with the presenter acknowledging the concerns raised.

The remaining presentations were constrained by time, and attendees were encouraged to continue discussions on the mailing list. This approach ensured that all topics were introduced, and the community could engage in asynchronous, in-depth discussions post-meeting.

The SIDROPS meeting concluded with a clear direction for future work, emphasizing the importance of consistent validation mechanisms, the careful consideration of protocol changes, and the ongoing collaboration required to maintain and enhance the security of inter-domain routing.

# 73 Stub Network Automatic Configuration Working Group (SNAC WG)

The SNAC WG meeting at IETF 119 was attended by a diverse group of participants, including representatives from prominent companies such as Cisco, Google, Apple, and Huawei, as well as institutions like APNIC Foundation and NIST. The total attendance was 53, indicating a strong interest in the topics discussed.

The meeting focused on the advancement of several Internet Engineering Task Force (IETF) draft documents, with particular attention to the challenges and strategies for automatically connecting stub networks to unmanaged infrastructure. The discussions were not only technical but also strategic, as they aimed to address the complexities of network configuration in environments lacking centralized management. Key documents such as draft-ietf-snac-simple were at the center of the debate, with the working group striving to refine the drafts to meet the needs of the community.

Meeting materials are available through the direct link: IETF 119 Meeting Materials.

## 73.1 Meeting Discussions

### 73.1.1 Administrivia and Chairs Updates

The working group chairs provided a progress report, noting two interim meetings since IETF 118 and the intention to prepare draft-ietf-snac-simple for Working Group Last Call before IETF 120.

### 73.1.2 Draft-hui-stub-router-ra-flag

Discussion centered on whether the document belonged in the 6MAN or SNAC working group, with input from various stakeholders. The consensus leaned towards creating a brief document for 6MAN review, summarizing the necessary information for implementers.

### 73.1.3 Automatically Connecting Stub Networks

The working group engaged in a document editing session to resolve open issues on GitHub and propose revised text. The session underscored the importance of clear guidelines for stub routers in various network topologies and the need for careful consideration of DHCPv6-PD client roles.

The meeting concluded with an understanding that the discussions and proposed revisions would significantly contribute to the field by clarifying the operation of stub routers in diverse network environments and setting the stage for future work on more complex scenarios.

# 74 Secure Protocols for the Internet Credential Exchange (SPICE)

## 74.1 Attendees Overview

### 74.1.1 Attendee Summary

The SPICE BOF at IETF-119 was attended by a diverse group of 144 participants, representing prominent companies and institutions such as Siemens, Cisco Systems, Okta, Google, Apple, Microsoft, and many others.

The discussions during the meeting were centered around the proposed work items and the charter text for the group. The attendees engaged in a lively debate on various topics, including architecture, use cases, SD-CWT, and metadata/capability discovery. The meeting materials can be accessed via the IETF Datatracker.

## 74.2 Meeting Discussions

### 74.2.1 Architecture

Henk presented two foundational documents that will inform the group's architecture: draft-steele-spice-transparency-tokens and draft-johansson-wallet-refarch. The discussion highlighted the importance of integrating credentials into other protocols and the need for a clear architectural framework.

### 74.2.2 Use Cases

Mike, Brent, and Roy discussed the use cases for SPICE, referencing draft-prorock-spice-use-cases/01. The conversation underscored the broad applicability of the use cases and their relevance to the work of the group.

### 74.2.3 SD-CWT

Orie introduced the concept of SD-CWT, outlined in draft-prorock-cose-sd-cwt/02, emphasizing its potential performance benefits over JSON-based approaches. The discussion also touched on the importance of security guarantees provided by receipts and the need to maintain conceptual alignment with SD-JWT.

### 74.2.4 Meta-data/Capability Discovery

The topic of metadata and capability discovery was addressed by Orie, with reference to draft-steele-spice-metadata-discovery/01. The debate highlighted the challenges and importance of key discovery and the potential for metadata to become a complex area requiring careful consideration.

### 74.2.5 Charter Text Discussion

The group engaged in a comprehensive discussion on the charter text, with contributions from various attendees. The dialogue reflected a consensus on the need for clarity regarding the group's scope and its relationship with W3C work. The discussion concluded with a poll that favored moving forward with the proposed charter text, including a modification to acknowledge the conceptual security model used in related technologies.

# 75 Segment Routing Working Group (SPRING)

## 75.1 Attendees Overview

### 75.1.1 Prominent Companies and Total Attendance

The SPRING working group meeting was attended by representatives from prominent companies and institutions such as Cisco, Huawei, Nokia, and Juniper Networks, with a total attendance of 116 participants.

### 75.1.2 Summary of Discussions

The discussions at the SPRING working group meeting were centered around the advancement of Segment Routing over IPv6 (SRv6) and its implications for network efficiency, security, and resource management. Key presentations highlighted the need for compressed SRv6 segment list encoding, security considerations for SRv6, and resource guarantees for SRv6 policies. The dialogue was enriched by the presence of industry experts, which provided a practical context to the theoretical advancements being proposed. Draft documents were referenced throughout the discussions, providing a foundation for the technical discourse. For example, the [draft-ietf-spring-srv6-srh-compression](draft-ietf-spring-srv6-srh-compression) was a focal point for exploring compression techniques in SRv6.

Meeting materials are available directly via the [IETF Datatracker](IETF Datatracker).

## 75.2 Meeting Discussions

### 75.2.1 Compressed SRv6 Segment List Encoding in SRH

The presentation by Francois Clad on compressed SRv6 segment list encoding sparked a discussion on the trade-offs between compression efficiency and compatibility with existing protocols. The potential impact on the L4 checksum and middlebox behavior was debated, with a consensus on the need for awareness and documentation of these issues.

### 75.2.2 SRv6 Security Considerations

Luis Miguel Contreras Murillo's presentation on SRv6 security considerations prompted a call for WG adoption. The discussion emphasized the importance of a comprehensive gap analysis and the integration of existing security considerations from related RFCs.

### 75.2.3 MicroTap Segment in Segment Routing

Gurminderjit Bajwa's introduction of the MicroTap Segment concept led to a debate on its operational implications and the potential overlap with existing IPFix standards. The discussion highlighted the need for further exploration of rate control as a security measure.

### 75.2.4 Flexible Candidate Path Selection of SR Policy

Yisong Liu's proposal for flexible candidate path selection within SR policies raised questions about the necessity of this mechanism given existing SR policy capabilities. The conversation pointed towards a deeper examination of the intent-based nature of SR policies.

### 75.2.5 Circuit Style Segment Routing Policies with Optimized SID List Depth

Amal Karboubi's talk on optimizing SID list depth for circuit-style SR policies generated a lively debate on the operational trade-offs between binding SID complexity and other forms of network resilience.

### 75.2.6 Resource Guarantee for SRv6 Policies

The presentation by Wenying Jiang on resource guarantees for SRv6 policies led to a discussion on the scalability of the approach and its relation to existing work on network slicing. The WG was urged to consider the overlap with ongoing efforts in the TEAS WG.

### 75.2.7 Segment Routing over IPv6 (SRv6) Proof of Transit

Luigi Iannone's segment on SRv6 Proof of Transit (PoT) introduced a specific solution for a defined context, which prompted comparisons with iOAM and calls for justification of its benefits over existing methods.

### 75.2.8 A Path Verification Solution based on SRv6

Jun Liu's presentation on SRv6-based path verification sparked a conversation on the privacy implications and the potential for payload-based correlation of source and destination by attackers.

The meeting concluded with an understanding that the discussions and outcomes would significantly contribute to the strategic direction of segment routing and its application in modern networks.

# 76 Secure Telephony Identity Revisited (STIR)

## 76.1 Attendees Overview

### 76.1.1 Prominent Companies and Total Attendance

The STIR Working Group meeting was attended by representatives from key industry players such as Microsoft, Nokia, DigiCert, Meta Platforms, Inc., and Huawei, as well as various other institutions and companies. The total number of attendees was 36.

### 76.1.2 Summary of Discussions

The discussions at the STIR Working Group meeting focused on the advancement of several Internet Engineering Task Force (IETF) draft documents. Notably, the group reviewed updates to the [draft-ietf-stir-certificates-ocsp](), which now includes stapling and examples for implementers. The potential adoption of [draft-peterson-stir-certificates-shortlived]() was also a key topic, with discussions on the use of x5u/x5c fields and certificate transparency. Additionally, the group explored the integration of STIR with Messaging Layer Security (MLS) through the [draft-peterson-stir-mls]() document, which is still in the early stages and requires further work in coordination with the MIMI Working Group.

Meeting materials are available via the direct link: [STIR Working Group Materials]().

## 76.2 Meeting Discussions

### 76.2.1 Certificates

The group reviewed the latest updates to the OCSP stapling process and requested feedback on the impact of the staple size. There was a consensus to continue using PEM format for certificates and to address the feedback provided by the community on the mailing list.

### 76.2.2 Short-Lived Certificates

The draft on short-lived certificates was discussed, with feedback from Eric Rescorla highlighting the need to clarify the use of x5u and x5c fields. The group agreed to maintain the current approach and add a note to retain the x5u field for short-lived certificates. There were no objections to the adoption of this draft.

### 76.2.3 Certificate Transparency

Chris Wendt presented on certificate transparency, seeking input on whether to simplify the current document to focus solely on pre-certificates. The group discussed the need for a threat analysis and concluded that the draft was not ready for adoption, with further discussions to continue.

### 76.2.4 STIR+MLS

The integration of STIR with MLS was presented by Jon Peterson and Richard Barnes. The group acknowledged the need for more coordination with the MIMI Working Group and determined that it was too early for the adoption of this draft.

### 76.2.5 Future of the Working Group

The meeting concluded with a brief discussion on the future of the STIR Working Group, with a call for further discussion on the mailing list to determine if the group is nearing its objectives or if there is additional work to be done.

# 77 Software Updates for Internet of Things (SUIT) [SUIT]

## 77.1 Attendees Overview

### 77.1.1 Prominent Attendees and Total Attendance

The meeting was attended by representatives from major companies and institutions, including Arm, Ericsson, and the National Institute of Standards and Technology (NIST), with a total attendance of 44 participants.

### 77.1.2 Summary of Discussions

The discussions were centered around the progression of various drafts, including the draft-ietf-suit-manifest, which is crucial for the Trusted Execution Environment Provisioning (TEEP) protocol. The group also addressed the need to resolve outstanding DISCUSS points raised by the Internet Engineering Steering Group (IESG) and the implications of potential changes on the drafts' progress.

Meeting materials are available through the direct link: IETF 119 Materials.

## 77.2 Meeting Discussions

### 77.2.1 SUIT Manifest Format

The group proposed a path forward to address IESG feedback, including adopting Internationalized Resource Identifiers (IRIs) as specified in RFC 3987. The potential split between fetch and install operations was discussed, with a consensus to avoid significant changes at this stage.

### 77.2.2 SUIT Manifest Extensions for Multiple Trust Domains

The discussion focused on the dependency of the TEEP protocol on the SUIT manifest extensions and the readiness of the draft for Working Group Chair approval.

### 77.2.3 Firmware Encryption with SUIT Manifests

The working group reviewed the draft's dependency on other documents and agreed that it is ready for submission to the IESG following a successful Working Group Last Call.

### 77.2.4 Secure Reporting of Update Status

The group decided to initiate a Working Group Last Call for the draft, which is also a dependency for the TEEP protocol.

### 77.2.5 Strong Assertions of IoT Network Access Requirements

The debate on whether to use URIs or IRIs for machine consumption was settled with a preference for URIs, and the group discussed the remaining DISCUSS points before submission to the IESG.

### 77.2.6 Mandatory-to-Implement Algorithms for SUIT Manifests

The draft's dependencies were acknowledged, and the group reached a consensus to proceed with the submission to the IESG.

### 77.2.7 Update Management Extensions for SUIT Manifests

The need for additional feedback was highlighted, and volunteers were called upon to review the draft before moving forward.

# 78 Transport Protocol Maintenance Working Group (TCPM)

## 78.1 Attendees Overview

### 78.1.1 Attendance Summary

The meeting was attended by 50 individuals representing a diverse array of organizations, including prominent companies such as Google, Apple, Cloudflare, and Huawei, as well as academic institutions and research centers like the University of Aberdeen and Münster University of Applied Sciences.

### 78.1.2 Meeting Context

Discussions at the TCPM working group meeting focused on the evolution of TCP standards and the examination of new draft proposals. The conversation was enriched by contributions from various stakeholders, highlighting the collaborative nature of the IETF's standardization process. Key draft documents, such as [draft-ietf-tcpm-prr-rfc6937bis](draft-ietf-tcpm-prr-rfc6937bis) and [draft-ietf-tcpm-ack-rate-request](draft-ietf-tcpm-ack-rate-request), were central to the discussions, with participants considering both the technical merits and the broader implications of the proposed changes.

Meeting materials are available directly via the [IETF 119 materials page](IETF 119 materials page).

## 78.2 Meeting Discussions

### 78.2.1 RFC 6937bis Updates

Neal Cardwell presented remotely on the updates to RFC 6937bis, which were generally well-received. The group discussed minor implementation differences and concluded that the draft was ready to proceed, with no significant objections raised.

### 78.2.2 Status of ACK Rate Request

Carles Gomez's presentation on the ACK Rate Request draft sparked a conversation about its implementation status and its applicability to TCP, given the experiences with QUIC. The group acknowledged the need for experimental data, especially in the context of TCP's interaction with network devices that control ACK rates.

### 78.2.3 Service Affinity Solution for TCP in Anycast Situations

Xueting Li and Wei Wang discussed their draft on a service affinity solution for TCP-based applications in anycast situations. The discussion raised questions about the appropriateness of the transport layer for this solution and the potential security implications of the proposed approach.

### 78.2.4 Efficient IP-spoofed TCP Connections via Ghost ACKs

Yepeng Pan and Christian Rossow presented their findings on Ghost ACKs and their role in efficient IP-spoofed TCP connections. The group considered the need for documentation, either as an RFC or an errata to RFC 5961, and discussed the potential impact on TLS.

# 79 Traffic Engineering Architecture and Signaling (TEAS)

## 79.1 Attendee Overview

### 79.1.1 Attendance Summary

The meeting saw participation from a diverse range of companies and institutions, including prominent entities such as Cisco Systems, Huawei Technologies, Nokia, and Telefonica Innovacion Digital. A total of 83 attendees were present, representing a broad spectrum of expertise and interest in the field of traffic engineering and signaling.

### 79.1.2 Main Points and Contextualization

The discussions at the meeting revolved around various draft documents and their applicability to the Abstraction and Control of Traffic Engineered Networks (ACTN) framework, particularly in the context of Packet Optical Integration (POI). The presentations highlighted the ongoing efforts to refine the YANG Data Models for network slicing and topology, as well as the need for clear mapping between 5G network slicing Quality of Service (QoS) identifiers and Differentiated Services (DiffServ) code points. The debates underscored the importance of these models in enabling efficient and scalable traffic management across multi-domain network environments. Attendees engaged in constructive dialogues, with particular attention given to the drafts draft-ietf-teas-actn-poi-applicability and draft-ietf-teas-ietf-network-slice-nbi-yang, among others.

Meeting materials are available directly via the link TEAS Meeting Materials.

## 79.2 Meeting Discussions

### 79.2.1 ACTN for POI Applicability

The presentation on the applicability of ACTN to POI provided insights into the integration of packet and optical networks. The discussion emphasized the need for a clear understanding of the operational aspects and the potential impact on network efficiency.

### 79.2.2 Service Assurance for ACTN POI

The dialogue on service assurance for ACTN POI highlighted the challenges and proposed solutions for ensuring reliable service delivery in integrated packet-optical networks. The conversation suggested a shift towards more robust and automated assurance mechanisms.

### 79.2.3 IETF Network Slice Service YANG Data Model

The presentation on the YANG Data Model for IETF Network Slice Service was well-received, with the draft being considered ready for Working Group Last Call (WGLC). This reflects a significant step towards standardizing network slice management.

### 79.2.4 5QI to DiffServ DSCP Mapping

The discussion on mapping 5G QoS identifiers to DiffServ DSCP values underscored the complexity of aligning end-to-end network slice QoS with transport network capabilities. The debate suggested a need for further clarification and consensus on the approach.

### 79.2.5 Next Steps

The meeting concluded with a consensus on the importance of the discussed drafts and the need for continued work to refine the models and strategies presented. The next steps involve addressing open issues, incorporating feedback from the discussions, and progressing drafts towards standardization, thereby contributing to the advancement of traffic engineering and signaling protocols.

# 80 Transport Layer Security Working Group (TLS)

## 80.1 Attendees Overview

### 80.1.1 Overview

The meeting was attended by representatives from prominent companies and institutions, including Mozilla, Google, Cisco Systems, and Microsoft, with a total attendance of 116 participants.

The discussions centered around the latest updates and proposals for the TLS protocol. Key topics included updates to RFCs 8446 and 8447, Encrypted Client Hello (ECH), and various draft proposals aimed at enhancing the security and efficiency of TLS. The debates were informed by the need to balance security with performance, and the implications of each proposal were carefully considered. Draft documents such as draft-ietf-tls-rfc8446bis and draft-ietf-tls-rfc8447bis played a significant role in guiding the discussions.

Meeting materials are available through the direct link IETF 119 Materials.

## 80.2 Meeting Discussions

### 80.2.1 8446/8447 Updates

The group discussed the processing of errata for RFCs 8446 and 8447, emphasizing the importance of addressing community feedback before advancing the documents.

### 80.2.2 ECH Update

The Encrypted Client Hello (ECH) update was a focal point, with robust discussions on the last call feedback. The consensus was to move forward with the current proposals, acknowledging the need for careful consideration of any potential issues raised.

### 80.2.3 Registry Update

The update on the TLS registry highlighted the smooth processing of registration requests and the importance of keeping the TLS group informed about standardizations within the IETF.

### 80.2.4 TLS Hybrid Key Exchange

The TLS Hybrid Key Exchange draft was debated, with two main issues: awaiting FIPS certification and deciding on the method for combining shared secrets. The group leaned towards a decision that would not disrupt the existing TLS key schedule.

### 80.2.5 TLS Obsolete Key Exchange

The Working Group Last Call (WGLC) for the TLS Obsolete Key Exchange was completed, and the group discussed classifications for key exchanges, with a focus on aligning with RFC 8447bis and ensuring appropriate recommendations for various key exchange methods.

### 80.2.6 TLS Formal Analysis

A proposal was made to formalize the process for triaging formal analysis of TLS 1.3, with the aim of involving experts early in the process to assess the need for formal analysis of new specifications. The group agreed on the importance of this step without allowing it to become a bottleneck.

### 80.2.7 Super Jumbo Record Limit

The proposal to increase the plaintext record size limit in TLS to improve performance was well-received, with the group agreeing to further explore the design details and consider adoption based on performance metrics and security considerations.

### 80.2.8 MTLS Flag

The MTLS Flag proposal, which aims to facilitate the distinction of bots by signaling the availability of a client certificate, was discussed. The group sought more enthusiasm and clearer benefits before moving forward.

### 80.2.9 Extended Key Update

The Extended Key Update draft, aimed at providing forward secrecy for long-lived TLS connections, was debated. The group discussed the merits of application-layer involvement versus complete TLS-layer handling, with a focus on simplifying the design without interrupting communication.

### 80.2.10 MLKEM Key Agreement

Finally, the MLKEM Key Agreement draft was presented, with discussions on the need for a standalone Post-Quantum (PQ) key agreement method. The group considered the option of registering a code point to facilitate early adoption by interested parties.

# 81 Transport and Services Working Group (tsvwg) WG

The Transport and Services Working Group (tsvwg) convened to discuss a range of topics including updates on various drafts, UDP options, differentiated services, and SCTP-related drafts. The meeting was attended by representatives from prominent companies such as Comcast, Google, Apple, and Cisco, with a total attendance of 95 individuals.

## 81.1 Attendees Overview

### 81.1.1 Prominent Companies and Institutions

The meeting saw participation from key industry players, including Comcast, Google, Apple, and Cisco, among others. The total number of attendees was 95, indicating a strong interest in the topics discussed by the working group.

### 81.1.2 Meeting Materials

Meeting materials and minutes can be accessed through the direct link: [TSVWG Meeting Minutes](#).

## 81.2 Meeting Discussions

### 81.2.1 Working Group Status and Draft Updates

The session began with updates on the status of various drafts, including those in the Editor's Queue and those awaiting Working Group Last Call (WGLC). Notable drafts such as [draft-ietf-tsvwg-ecn-encap-guidelines](#) and [draft-ietf-tsvwg-rfc6040update-shim](#) were highlighted, with Gorry Fairhurst serving as the Document Shepherd.

### 81.2.2 Notices and Related Drafts

Liaison notices from 3GPP and GSMA were discussed, particularly concerning SCTP, DTLS, and MultiPath DCCP. The chairs also provided announcements and a heads-up on upcoming milestones.

### 81.2.3 UDP Options

Joe Touch and Mike Heard, via proxy chairs, discussed the [draft-ietf-tsvwg-udp-options](#) and its implications for UDP transport. The discussion emphasized the need for volunteers to review the documents and the intention to announce a WGLC post-meeting.

### 81.2.4 Differentiated Services and AQM Drafts

Jason Livingood presented Comcast's field trials on L4S and NQB, indicating a scale-up to millions of customers by 2024. Greg White provided operational guidance on L4S, as documented in [draft-ietf-tsvwg-l4sops](#), and discussed the NQB draft, which is nearing WGLC.

### 81.2.5 DTLS over SCTP Design Group Report

Magnus Westerlund and Michael Tuexen presented on DTLS protection for SCTP, outlining various proposals and their respective timelines, IPR considerations, and potential impacts on implementations such as OpenSSL. The discussion highlighted the importance of selecting a technically sound and timely solution to meet 3GPP requirements.

### 81.2.6 Careful Resumption of Congestion Control

Nicolas Kuhn and Gorry Fairhurst discussed updates to the [draft-ietf-tsvwg-careful-resume](#), including the inclusion of QLOG definitions. The draft is expected to be ready for WGLC at the next meeting.

### 81.2.7 Requirements for Metadata and Signalling Use Cases

John Kaippallimalil and Mohamed Boucadair presented individual drafts focusing on requirements for metadata and signalling use cases. The discussions underscored the need for a common document to address various scenarios and guide the types of metadata required for effective host-to-network signalling.

The tsvwg WG sessions concluded with a clear direction for the next steps, emphasizing the need for continued collaboration and review to advance the drafts and address the technical challenges presented.

# 82 Using TLS for Applications (UTA) [UTA]

## 82.1 Attendees Overview

### 82.1.1 Attendance Summary

The UTA working group meeting at IETF 119 in Brisbane was attended by 39 participants representing various organizations, including prominent companies such as Google, Microsoft, Nokia, and Huawei, as well as institutions like NIST, UK NCSC, and NSA - CCSS.

### 82.1.2 Meeting Context

The meeting focused on the adoption of TLS 1.3 in new protocols and the integration of Post-Quantum Cryptography (PQC) into Internet applications. Rich Salz presented a proposal to mandate TLS 1.3 for new protocols, while Tirumaleswar Reddy.K discussed recommendations for PQC in Internet applications. The discussions were rooted in the context of evolving security standards and the need to prepare for a post-quantum cryptographic landscape. References to ongoing work and drafts were made, such as the draft-ietf-uta-require-tls13 and the draft-ietf-uta-pqc-app.

Meeting materials can be found at the following link: IETF 119 Meeting Materials.

## 82.2 Meeting Discussions

### 82.2.1 New Protocols Must Require TLS 1.3

Rich Salz led a discussion on the draft advocating for new protocols to require TLS 1.3, emphasizing the importance of modern security standards. The debate highlighted the distinction between new protocols and new applications, with a consensus leaning towards the former. Concerns were raised about the applicability to DTLS, and the group agreed that DTLS 1.3 should not be mandated due to its limited availability. The session concluded with a decision to call for a working group adoption, considering the draft's potential impact on updating BCP195.

### 82.2.2 Post-Quantum Cryptography Recommendations for Internet Applications

Tirumaleswar Reddy.K presented on the integration of PQC into Internet applications, sparking a dialogue on the readiness and standardization of various PQC algorithms. The discussion underscored the draft's broad scope and the possibility of premature standardization given the ongoing work in related areas. The group contemplated the draft's role as a living document within the working group, serving as a platform for ongoing discussions rather than immediate publication.

### 82.2.3 Open Mic and Next Steps

An open mic session allowed for additional points to be raised, including upcoming changes to the RADIUS/TLS draft. The meeting underscored the working group's role in guiding the industry towards stronger security protocols and preparing for a future with quantum-resistant cryptography. The next steps involve continued collaboration and feedback on the drafts, with an eye towards the evolving landscape of Internet security standards.

# 83 Virtual Conversations (vCon)

## 83.1 Attendees Overview

### 83.1.1 Participation Summary

The vCon working group meeting was attended by 15 participants, representing a diverse array of companies and institutions such as Meta Platforms, Inc., Ericsson, China Mobile Research Institute, and Fastmail. The attendance highlighted the significant interest and involvement from key players in the industry.

The meeting discussions centered around the evolution of the conversation data model, with a focus on enhancing the JSON format for vCon. Attendees engaged in a constructive dialogue, reviewing use cases and proposing amendments to the current draft. The conversation was enriched by the presence of experts from various sectors, ensuring a comprehensive examination of the draft's applicability and potential improvements.

Meeting materials are available through the direct link: draft-petrie-vcon-03.

## 83.2 Meeting Discussions

### 83.2.1 Draft Discussion: Conversation Data Container

Dan Petrie presented the latest updates to the draft-petrie-vcon, which included mapping use cases from the draft-rosenberg-vcon-cc-usecases and adding party history to track call states such as join, exit, hold, and mute events. The discussion raised questions about the utility of the interaction type and its relationship to DialogObject, as well as suggestions for renaming "Dialing List" to "Contact List" or "Target List" for clarity.

### 83.2.2 Main Points and Dialogues

The group engaged in a robust discussion on the draft's data model, considering the layering of information, use cases, and security implications. There was a consensus on the need to review related content, such as Mimi and XMPP, to ensure comprehensive coverage. The debate also touched upon the potential utility of an Interaction ID to link multiple DialogObjects within the same session, enhancing the traceability of various media types like video, audio, screenshare, and web chat.

### 83.2.3 Outcomes and Next Steps

The meeting concluded with a general agreement on the importance of updating the draft to reflect the discussed changes, particularly in capturing participant IDs and party histories. The group also recognized the need to separate security considerations into a dedicated document. The next steps involve adopting the draft as a working document of the group, with a commitment to ongoing refinement and collaboration to address the identified gaps and suggestions.

# 84 WebTransport (WEBTRANS) Working Group

## 84.1 Attendees Overview

### 84.1.1 Attendance Summary

The WebTransport (WEBTRANS) Working Group meeting was attended by a diverse group of participants, with a total of 59 attendees representing prominent companies and institutions such as Google, Apple, Microsoft, Cisco, Mozilla, Akamai, and many others.

### 84.1.2 Meeting Context

The meeting focused on the progress and challenges of integrating WebTransport with existing web technologies. Discussions revolved around updates from the W3C on WebTransport, the adaptation of WebTransport over HTTP/2 and HTTP/3, and the implications of these developments on future web standards. Key technical drafts were referenced, such as the [draft-ietf-webtrans-http2](#) and [draft-ietf-webtrans-http3](#), which are critical to the evolution of the protocol.

Meeting materials are available via the following link: [Meeting Slides](#).

## 84.2 Meeting Discussions

### 84.2.1 W3C WebTransport Update

Will Law provided an update on the W3C's progress with WebTransport, highlighting the open issues for the Candidate Recommendation milestone and the adoption of the protocol in major browsers. The discussion underscored the importance of IETF feedback on specific issues.

### 84.2.2 WebTransport over HTTP/2

Eric Kinnear discussed the main updates to starting a WebTransport session over HTTP/2. The conversation focused on the technical nuances and the potential impact on the protocol's efficiency and reliability.

### 84.2.3 WebTransport over HTTP/3

Victor Vasiliev presented on WebTransport over HTTP/3, emphasizing the need for review of the draft-ietf-quic-reliable-stream-reset. The debate on flow control highlighted differing opinions but concluded with a consensus to move forward with the current proposal, acknowledging the potential for head-of-line blocking (HOLB) issues.

### 84.2.4 Conclusions and Next Steps

The chairs wrapped up the meeting by calling for editors to finalize pull requests for open issues. The consensus was to forgo an interim meeting before IETF 120, provided that progress continues as planned.

# 85 Web Interaction Standards for Hypermedia (WISH)

## 85.1 Attendees Overview

### 85.1.1 Attendance Summary

The meeting was attended by representatives from prominent companies and institutions such as Google, Nokia, Ericsson, Meta Platforms, Inc., and others, totaling 17 participants. The diverse attendance underscores the broad interest and impact of the WISH Working Group's efforts.

### 85.1.2 Meeting Context

The discussions focused on the progress of the Web Interaction Standards for Hypermedia (WISH) Working Group, with particular attention to the [draft-ietf-wish-whip](draft-ietf-wish-whip) and [draft-ietf-wish-whep](draft-ietf-wish-whep) documents. The conversation contextualized the technical direction of the working group, considering the feedback from the Security Directorate and the broader community. The meeting materials can be accessed via the [IETF 119 materials page](IETF 119 materials page).

## 85.2 Meeting Discussions

### 85.2.1 WISH Status Update

The chairs provided an update on the [draft-ietf-wish-whip](draft-ietf-wish-whip), which is currently in IETF Last Call. Minor issues and some IDnits were noted during the Security Directorate review, but no major concerns have been raised that would impede the document's progress.

### 85.2.2 WHEP Presentation and Discussion

Sergio Garcia Murillo led the discussion on the [draft-ietf-wish-whep](draft-ietf-wish-whep), although no slides were received before the meeting. The group went through open issues on the GitHub issue tracker. A codec mismatch issue was discussed, with consensus leaning towards either failing the m-line by rejecting it or failing the whole SDP offer/answer. The group also considered transferring the WHEP draft to a new area for further development, while the WHIP draft will continue to be finalized by Murray. Francesca suggested scheduling an interim meeting to address open issues before the next IETF meeting, indicating a proactive approach to resolving outstanding concerns.

# 86 WIT AREA

## 86.1 Attendees Overview

### 86.1.1 Attendance Summary

The WGOV meeting at IETF 119 was attended by a diverse group of 73 participants, representing prominent companies and institutions such as Tencent America LLC, Comcast, Hughes Network Systems, University of Aberdeen, Ericsson, Google, Nokia, Telenor Research, Interdigital Europe, NTT Communications, Preferred Networks, Inc., Private Octopus Inc., Live Networks, Inc., Huawei, Google Switzerland GmbH, ISC, Cisco, Akamai, NSA - CCSS, Fastly, Meta Platforms, Inc., NIC.br, Technical University of Munich, Fastmail, University of Glasgow, Viagenie, Episteme Technology Consulting LLC, National Institute of Information and Communications Technology, Apple, China Unicom, sn3rd, Futurewei Technologies, ALAXALA Networks, Corp., NTT, Carnegie Mellon University, Netflix, IETF Administration LLC, 8x8 / Jitsi, APNIC PTY LTD, Meetecho, Deutsche Telekom, Mozilla, Cloudflare, Japan Registry Services Co., Ltd., and many others.

The discussions were centered around the current state and future direction of the working group, with a focus on the review of teams and directorate, as well as the progress of various IETF draft documents. The meeting materials can be accessed via the direct link: IETF 119 Materials.

## 86.2 Meeting Discussions

### 86.2.1 Administrivia - WIT ADs

The session began with routine administrative announcements, including the recording of the session, the Note Well, and the process of agenda bashing. The WIT overview and the state of the area were presented, highlighting the ongoing work and the strategic direction of the group. A review of the teams and directorate was conducted, and a reference was made to the WIT - ART working groups overview by Francesca at IETF 118. A notable point of discussion was the missing document in the RFC-editor queue, draft-ietf-avtext-framemarking-16, which was brought up by Jonathan Lennox.

### 86.2.2 WIT - TSV Working Groups Overview

Martin Duke provided an insightful overview of the work being done by the transport-related working groups (WGs). This presentation shed light on the technical endeavors and the collaborative efforts within the WGs to address key challenges in the transport layer.

### 86.2.3 Open Mic

The open mic session offered an opportunity for attendees to engage with the Area Directors (ADs), though no questions were posed during this time.

The meeting concluded with a clear understanding of the WGOV's current initiatives and a path forward for the working group's contributions to the field of internet technologies.